\documentclass[12pt]{article}
\pdfoutput=1
\usepackage{amssymb}
\usepackage{amsmath}
\usepackage{graphicx}
\usepackage{hyperref}
\usepackage{color}
\usepackage{fullpage}

\numberwithin{equation}{section}

\newcommand{\eq}{\begin{equation}}
\newcommand{\eqx}{\end{equation}}
\newcommand{\eqs}{\begin{equation*}}
\newcommand{\eqsx}{\end{equation*}}
\newcommand{\eqn}{\begin{eqnarray}}
\newcommand{\eqnx}{\end{eqnarray}}
\newcommand{\eqns}{\begin{eqnarray*}}
\newcommand{\eqnsx}{\end{eqnarray*}}

\newcommand{\f}[2]{\frac{#1}{#2}}

\newcommand{\cor}[1]{\left\langle{#1}\right\rangle}
\newcommand{\ket}[1]{\left|{#1}\right\rangle}
\newcommand{\bra}[1]{\left\langle{#1}\right|}

\newcommand{\HH}{{\mathcal H}}

\DeclareMathOperator{\arctanh}{arctanh}

\renewcommand{\th}{\theta}
\newcommand{\sg}{\sigma}

\newcommand{\dl}{\delta}
\newcommand{\Dl}{\Delta}
\newcommand{\Gm}{\Gamma}
\newcommand{\al}{\alpha}

\newcommand{\om}{\omega}

\newcommand{\kap}{\kappa}

\newcommand{\eps}{\varepsilon}
\newcommand{\qqqq}{\quad\quad\quad\quad}
\newcommand{\qq}{\quad\quad}

\newcommand{\nn}{{\cal N}}

\DeclareMathOperator{\res}{Res}

\newcommand{\Nfin}{\mathbf{N}}

\newcommand{\suii}{{$su(2)$\ }}
\newcommand{\sutt}{{$su(1|1)$\ }}

\newcommand{\id}{\,I\!d}

\newcommand{\oo}[1]{{\mathcal O}\left(#1\right)}

\title{String field theory vertex from integrability}

\author{Zoltan Bajnok$^{a}$\thanks{e-mail: {\tt bajnok.zoltan@wigner.mta.hu}},\ \  
Romuald A. Janik$^{b}$\thanks{e-mail: {\tt romuald@th.if.uj.edu.pl}} \\ \\ 
\small 
${}^a$ MTA Lend\"ulet Holographic QFT Group\\\small
Wigner Research Centre\\\small
H-1525 Budapest 114, P.O.B. 49, Hungary\\\small
${}^b$ Institute of Physics\\\small
Jagiellonian University\\\small
ul. {\L}ojasiewicza 11, 30-348 Krak{\'o}w, Poland}

\date{}

\begin{document}

\maketitle

\begin{abstract}
We propose a framework for computing the (light cone) string field theory vertex
in the case when the string worldsheet QFT is a generic integrable theory.
The prime example and ultimate goal would be the $AdS_5 \times S^5$ superstring theory
cubic string vertex and the chief application will be to use this framework as a formulation
for $\nn=4$ SYM theory OPE coefficients valid at any coupling up to
wrapping corrections.
In this paper we propose integrability axioms for the vertex, illustrate them
on the example of the pp-wave string field theory and also uncover similar structures
in weak coupling computations of OPE coefficients. 
\end{abstract}

\vfill

\pagebreak

\section{Introduction}

The integrability properties of string theory in $AdS_5 \times S^5$ background \cite{intreview}
together with the AdS/CFT correspondence \cite{adscft1} allows for obtaining exact results 
for various observables in $\nn=4$ Super-Yang-Mills (SYM) theory for any value
of the gauge theory coupling in the planar, large $N_c$ limit.
Currently this program is very well developed for the spectral problem, namely
for the determination of the scaling dimensions of \emph{all} local operators \cite{TBA1}-\cite{QSC2}.
For other observables we have currently only partial results like various
strong and weak coupling expansions
or exact answers but restricted to some particular concrete observables like generalized
cusp Wilson loops, circular loops or for some ingredients
of scattering amplitudes. 

A class of observables for which it would be crucial to obtain a similar level
of understanding as for the scaling dimensions are the OPE coefficients
or, equivalently, the 3-point correlation functions of local operators. Namely, 
these quantities  provide  the remaining fundamental data for any conformal field
theory (CFT). Indeed, higher point functions do not carry any independent dynamical
content and can be reduced to scaling dimensions, OPE coefficients and
conformal blocks determined by conformal symmetry alone.

On the string side of the AdS/CFT correspondence these quantities are also
interesting for their own sake,  namely the AdS/CFT string diagram corresponding
to a 3-point function can be interpreted as a three string interaction.
In fact, the first wave of interest in OPE coefficients of (unprotected) operators
in $\nn=4$ SYM theory \cite{7AUTH}-\cite{alday2} came from the proposed link with the 3-string
string field theory vertex in the pp-wave \cite{BMN} string field theory (SFT) \cite{SV}-\cite{SHIMADA}.
The SFT vertex is also interesting as it is related  to the first
$1/N_c$ corrections to the string hamiltonian/scaling dimensions, too.

Unfortunately, there is practically no information on generalizing
the pp-wave SFT to the full $AdS_5 \times S^5$ case. This is not an
issue of technical or calculational complexity but rather a more fundamental one.
A unique feature of the pp-wave geometry is that, although it is curved,
the worldsheet quantum field theory of the string in an appropriate
light cone gauge reduces to free massive bosons and fermions \cite{MT}, thus allowing for
the use of mode expansions in implementing continuity conditions
for the SFT (light cone) vertex \cite{SV} similarly as for the flat space
SFT vertex \cite{GS}. For an interacting worldsheet QFT,
as is the case for the full $AdS_5 \times S^5$ geometry, we do not
have any techniques so far for finding the SFT vertex.

Thus the main goal of  the present paper is to provide a new formulation
for the problem of determining the (light cone) SFT vertex in the case
when the worldsheet theory is a generic integrable QFT, which
includes as a key special case the $AdS_5 \times S^5$ background.
We propose an integrable bootstrap formulation of the SFT vertex,
namely a set of coupled functional equations for the SFT amplitudes
understood as the value of the vertex with specific string excited
states on each of the three legs. The dependence on the concrete
background/worldsheet QFT enters through the appearance of the S-matrix
in the SFT vertex axioms.
This formulation should be valid up to exponential `wrapping corrections'.

The bootstrap approach for obtaining various physical quantities in
two dimensional integrable quantum field theories has already
a long and successful history. Basically it amounts to
implementing very general functional and analyticity properties of
the various observables and using in addition key properties of
integrability like factorized scattering~etc.

Initially, the bootstrap program was developed for determining
the scattering amplitudes (and at the same time the particle content,
hence the name bootstrap) for a theory on a two-dimensional plane \cite{ZamON}-\cite{Dorey:1996gd}.
The result is the explicit knowledge of the 2-particle scattering
S-matrix and the mass spectrum of the theory, e.g. the masses of bound states
in terms of the masses of the fundamental particles.
Subsequently this information was used to obtain the spectrum
of such a theory on a cylinder of finite size \cite{ZamTBA,Dorey:1996re}.

Since then, the bootstrap program was extended to cover theories
with integrable boundary conditions \cite{Bndry}, providing exact formulas
for reflection factors; as well as for theories with integrable defects \cite{Defect}.

A whole new field of research started when bootstrap was applied
to more fine-grained, and in a certain sense off-shell observables
such as form factors \cite{KW}-\cite{Babujian:1998uw}. Here, in contrast to ordinary scattering
amplitudes the number of incoming and outgoing particles does not need
to be balanced.  All the above developments appeared within
the context of ordinary relativistic integrable quantum field
theories and reflected the main questions of interest in such contexts.

One of the most intriguing features of the AdS/CFT correspondence is that
it provides a mapping between observables in a 4-dimensional gauge theory
and in the 2-dimensional string worldsheet quantum field theory.
As such, some natural questions in the 4-dimensional gauge theory
suggest completely novel problems/geometrical configurations
in the dual 2-dimensional integrable QFT, which were never
investigated hitherto by the relativistic integrable QFT community.

Some prime examples of such problems involve, on the classical level,
strong coupling scattering amplitudes (equivalently null polygonal Wilson loops) \cite{AM},
classical solutions with the topology of a thrice-punctured sphere
relevant for the OPE coefficients of (classical) operators at strong coupling 
\cite{JW}-\cite{KK3}.

The first application of bootstrap ideas in such a novel geometrical 
context was the very interesting work \cite{NEWBOOTSTRAP}, which provided
bootstrap equations for (excited) pentagonal scattering amplitudes relevant for
general multigluon scattering amplitudes in $\nn=4$ SYM.

\begin{figure}[t]
\hfill\includegraphics[height=4cm]{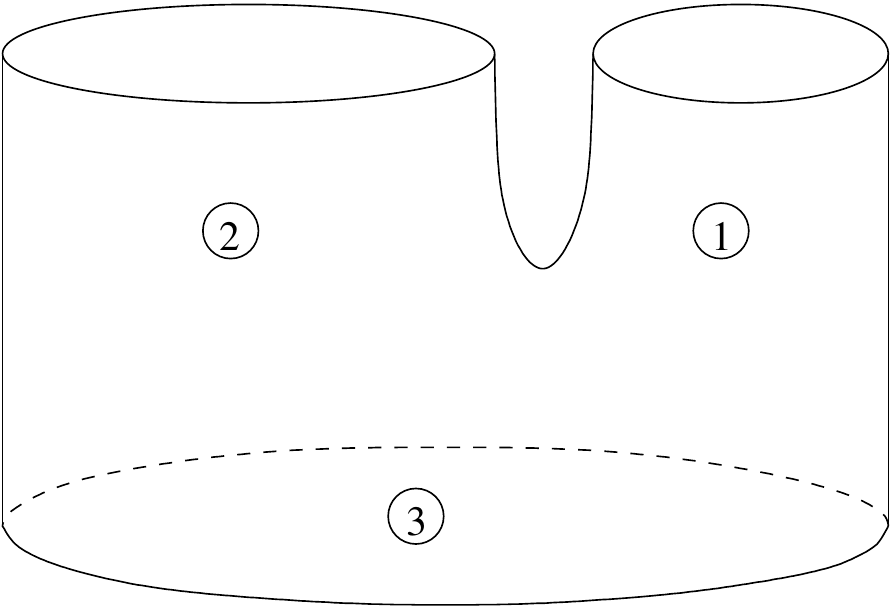}\hfill{}
\caption{The geometry of the worldsheet for the cubic light cone string field theory vertex.   
\label{fig.sftpants}}
\end{figure}

The goal of this work is to apply the bootstrap methodology to the classic
string pants diagram (see fig.~\ref{fig.sftpants}) relevant for the (light cone) SFT vertex.

In the following section, which is still a continuation
of the introduction, we  provide an explanation of the main ideas and
motivations behind our approach and then give an outline of the remaining parts
of the paper.  

\section{Insight from the spectral problem and form factors}

The spectral problem for an integrable quantum field theory is defined as finding
the energy levels of the theory defined on a cylinder of arbitrary size.

As a first step in solving this problem, one passes to the same theory
but defined in infinite volume -- on the whole two-dimensional plane.
There we have well defined asymptotic states so we can consider the S-matrix.
What is crucial, however, is that only in this setting we have at our disposal
analyticity properties of the S-matrix, especially crossing.
Thus one first solves the theory in infinite volume by
implementing the symmetries of the problem, solving the Yang-Baxter
equation together with unitarity and crossing, and determining any
remaining possible CDD factors. At this stage one obtains the exact analytical 
form of the S-matrix. This procedure is commonly called the S-matrix bootstrap.

In the second step, one considers the same theory defined on a large cylinder
of circumference $L$. A multiparticle
state on the cylinder can be considered just as a quantum mechanical multiparticle state
parametrized by the particles momenta $\{p_i\}$. These momenta are quantized
by the Asymptotic Bethe Ansatz quantization condition\footnote{With appropriate nested Bethe Ansatz
structure in case of nondiagonal S-matrices.}
\eq
e^{i \Phi_k(\{p_i\})}\equiv e^{ip_k L} \prod_{i \neq k} S(p_k,p_i)=1
\label{ABA}
\eqx
which essentially amounts to the single valuedness condition for the wave function.
The energy of the relevant state is then given by the sum of the particles' energies
\eq
E=\sum_{k=1}^N E(p_k)
\label{BAEnergy}
\eqx

As we decrease the size of the cylinder, quantum-field-theoretical virtual effects
become important (so-called \emph{wrapping corrections}) with the leading terms
being described by (generalized) L{\"u}scher corrections \cite{LUSCHER,KONISHI}, again in terms of infinite volume data.
These corrections give additional terms of order $e^{-mL}$ in the energy formula (\ref{BAEnergy}) 
and quantization conditions (\ref{ABA}). 
Subsequent multiple wrapping terms are much more involved (although progress has recently been made \cite{BOMBARDELLI})
but surprisingly the whole infinite set of wrapping corrections can be effectively resummed
through the so-called Thermodynamic Bethe Ansatz, which provides the exact spectrum
for any size of the cylinder. In the nondiagonal case, this last step is, however, quite involved 
(this is especially true in the $AdS_5\times S^5$ setup \cite{TBA1}-\cite{QSC2}).

In the above description we would like to emphasize two points. Firstly, the neccessity
of having an infinite volume description in order to formulate functional equations for the S-matrix.
Here the existence of crossing invariance is of particular importance.
Secondly, the simplicity of the finite volume answer as long as we neglect
the exponential wrapping corrections $\sim e^{-mL}$.
The obtained answer is valid for any value of the coupling in the integrable QFT.
Ultimately we would like to have a similar framework for the OPE coefficients.

A suggestion has been made for the use of form factors in this context 
\cite{NORDITATALK}-\cite{FFUS}.

Form factors are expectation values of a local operator on the worldsheet sandwiched
between multiparticle \emph{in} and \emph{out} states.
\eq
{}_{out}\!\cor{\th'_1,\ldots,\th'_m| \oo{0} | \th_1,\ldots,\th_n}_{in}
\eqx
In infinite volume one may use crossing to put all particles into the \emph{in} state
\eq
\cor{0 | \oo{0} |\th_1,\ldots,\th_n} \equiv F_n(\th_1,\ldots,\th_n)
\eqx
and formulate functional equations for these quantities. Assuming for simplicity
a theory with just one species of particles and no bound states, the equations
take a very transparent form:
\eqn
\label{e.ffwatson}
F_n(\th_1,\ldots,\th_i,\th_{i+1},\ldots,\th_n) &=& F_n(\th_1,\ldots,\th_{i+1},\th_{i},\ldots,\th_n) 
S(\th_i,\th_{i+1}) \\
\label{e.ffperiod}
F_n(\th_1+2\pi i,\th_2,\ldots,\th_n) &=& F_n(\th_2, \ldots,\th_n,\th_1) \\
\label{e.ffkinem}
-i \res_{\th'=\th} F_{n+2}(\th'+i\pi,\th,\th_1,\ldots,\th_n) &=&
(1-\prod_{i=1}^n S(\th,\th_i) ) F_n(\th_1,\ldots,\th_n)
\eqnx 
Equation (\ref{e.ffwatson}) can be understood as a simple consequence of the commutation
relation between Zamolodchikov-Faddeev creation operators. Equation (\ref{e.ffperiod})
is very important as it involves in a crucial way crossing properties. Last particle
with rapidity $\th_1$ gets crossed up to the \emph{out} state, then it will get crossed
back on the other side. Equation (\ref{e.ffkinem}) is the so-called kinematical singularity
axiom and is the crossed version of the fact that the form factor has a singularity
once an outgoing and an incoming particle have the same rapidity.

These axioms are the form factor counterpart of the S-matrix bootstrap and
similarly allow for an exact explicit solution.
Indeed, the form factor axioms have been solved exactly for numerous relativistic integrable 
quantum field 
theories, \cite{LYFF,shGFF} including ones with nondiagonal scattering 
\cite{Smirnov,Babujian:1998uw}.

If we again would be content with neglecting wrapping corrections, the finite volume form factors
can be expressed in a very simple way through the infinite volume ones \cite{FVFF}
\eq
\label{e.fvff}
\cor{0 | \oo{0} |\th_1,\dots,\th_n}_L = \f{1}{\sqrt{\rho_n \cdot \prod_{i<j}S(\th_i,\th_j)}} 
\cdot F_n(\th_1,\dots,\th_n)
\eqx
Here the finite volume rapidities $\th_1$, \dots , $\th_n$ are constrained to obey the Bethe Ansatz
quantization condition and  $\rho_n$ is the Gaudin norm 
\eq
\rho_n=det \left \vert \frac{\partial \Phi_k}{\partial p_j}\right \vert
\eqx
which accounts to the difference between the natural finite volume normalization and
the continuum normalization in infinite volume. Finally, the square root of the product 
of S-matrices, which is just a phase,
ensures that the finite-volume form factor is a completely symmetric function of the rapidities
in contrast to the infinite volume one which obeys (\ref{e.ffwatson}).

Thus we see a similar pattern as for the spectral problem --- functional equations in infinite volume
and a simple passage to finite volume up to wrapping corrections.

Form factors seem to be a promising framework for OPE coefficients in the special case
of so-called HHL (Heavy-Heavy-Light) \emph{diagonal} 3-point functions, where two operators 
correspond to a specific multiparticle state (with large anomalous dimension at strong coupling)
while the light operator does not carry any conserved R-charges.
In this case the strong coupling classical formula (\cite{Z,Costa} modified in \cite{FFUS})
denoted schematically by
\eq
C_{HHL} \sim \int_{Moduli} \int d^2\sg V_L[ X^I(\sg^a) ]
\eqx
coincides exactly with a classical computation of a ‘diagonal’ form factor (here we integrate over 
the moduli space of the classical 2-point correlation function solution of the \emph{Heavy} operator).
This has a distinctive pattern of finite volume dependence (a bit more complicated than (\ref{e.fvff})
due to diagonality and disconnected terms. See \cite{FFUS,FFUS2} for details).
However this hypothesis has been so far tested only at strong coupling.

The form factor formulation in the context of OPE coefficients has both significant
advantages as well as disadvantages. On the positive side, through the existence of infinite volume axioms
and simple finite volume reduction, they have the potential to work at any coupling
up to wrapping corrections. On the negative side, they are potentially applicable only
if the initial and final volume remain the same (i.e. the third `light' operator does not carry
any $J$ charge), and probably only if the two `heavy' operators are conjugate to each other.
This is not a generic situation as typically we have $J_1+J_2=J_3$ with all $J_i$'s distinct from zero.
The case $J_i=0$ is an important albeit very special case.
Another difficulty with the form factor formulation is that the three gauge theory operators
are treated very asymetrically. Two gauge theory operators are considered as external multiparticle
\emph{in} and \emph{out} states, while the third operator is represented by a specific `effective'
worldsheet vertex operator which corresponds, in the form factor language, to a particular solution
of the form factor axioms. It is for the moment \emph{a-priori} not clear how to associate the
specific solution of form factor axioms to a particular gauge theory operator/massive string state.
Of course this is still premature as currently we do not have at our disposal \emph{any} solution of the
form factor axioms in the case of the worldsheet $AdS_5 \times S^5$ string theory,
which remains an outstanding open problem.

In this paper we will pursue an alternative formulation which involves the study of
the cubic (light-cone) string field theory vertex. An $AdS_5 \times S^5$ string diagram
corresponding to a 3-point correlation function has the topology of two strings
joining into a third one and certainly involves the cubic vertex as an essential
ingredient. In fact this line of approach was widely used in the pp-wave limit
with a formula of the kind
\eq
\label{e.c123pp}
C_{123}=f(\Dl_1,\Dl_2,\Dl_3) \cdot \bra{1}\bra{2}\cor{3|V_3}_{pp-wave}
\eqx
where $\bra{i}$ represent the appropriate BMN operators described using
pp-wave string excitations and $\ket{V_3}_{pp-wave}$ is the pp-wave
cubic string vertex constructed in \cite{SV,STEFANSKI,KH,DY}.
Various concrete formulas were put forward \cite{7AUTH,DY,SHIMADA} but we do not have
currently a clear generalization of this formula to the full $AdS_5 \times S^5$
context\footnote{Some problems were recently encountered in \cite{ZayS}.}.

Leaving this issue aside,
in this paper we will concentrate on proposing an integrable approach for
computing the cubic vertex
\eq
\label{e.vertexgen}
\bra{1}\bra{2}\cor{3|V_3}_X
\eqx
which would be applicable in principle for a curved background $X$ such that the
worldsheet string QFT is integrable.
Our formulation is \emph{a-priori} restricted up to wrapping corrections relative to the sizes
of the three closed strings\footnote{Although wrapping corrections for a single string may be incorporated in this approach.}. We will provide functional equations for the above
quantities (\ref{e.vertexgen}) in a certain decompactification limit (to be defined later in the paper)
and provide a recipe for obtaining the physical finite volume version of (\ref{e.vertexgen})
along the lines of the relation (\ref{e.fvff}).

Since we will not control the overall normalization of the vertex in this paper, the function $f(\Dl_1,\Dl_2,\Dl_3)$ may be
incorporated into the vertex so the functional equations may be potentially interpreted
as functional equations directly for the OPE coefficients (although this interpretation
should be treated with care as the relation between the $AdS_5 \times S^5$ vertex
\eq
\bra{1}\bra{2}\cor{3|V_3}_{AdS_5 \times S^5}
\eqx
and the OPE coefficient $C_{123}$
may will be of a more general form than (\ref{e.c123pp})).

The chief obstacle in defining the string vertex for an interacting worldsheet theory
is that the hitherto applied constructions of the string field theory vertex
used in an essential way mode expansions of the worldsheet fields and operator
continuity conditions \cite{GS,SV}. In the interacting context we do not have such
tools at our disposal\footnote{Although expansions into Zamolodchikov-Faddeev operators
may in principle exist, they seem to be impossible to control even in the simplest
interacting contexts.} so we apply a form of an integrable bootstrap approach by isolating
a decompactification limit allowing to define functional equations incorporating
crossing and a subsequent finite volume reduction which should be straightforward
as long as we are neglecting wrapping corrections. We thus adopt the same philosophy which
was so successful both in the case of the spectral problem and for form factors in relativistic theories.

In the remaining part of the paper we will first recall some information about the
pp-wave string field theory and its exact solution, then define the decompactified vertex
and propose the string vertex functional equations. Then we will analyze these equations
in the case of the massive free boson and compare with the pp-wave results in order to
get insight into the required analyticity structure of the solution. In particular
we will show how the very nontrivial special functions appearing in the exact pp-wave solution of
\cite{LSNS} can be obtained easily from our functional equations.
Then we will proceed to define the program for the finite volume reduction and give
the string vertex axioms in the general nondiagonal case.
Finally, we will also show that some of the general properties of our axioms
can be observed in direct \emph{weak coupling} computations of OPE coefficients
in the \suii and \sutt sectors.
We relegate various technical details to the appendices.

\section{The pp-wave light cone string field theory vertex}
\label{s.ppintro}

A unique feature of the pp-wave limit of $AdS_5 \times S^5$ is that when the Green-Schwartz 
superstring action
is considered in an appropriate light-cone gauge, the worldsheet theory reduces to
a set of \emph{noninteracting} massive boson and fermion fields.
Hence in this geometry the superstring can be easily quantized exactly \cite{MT}.
Similarly, the light cone string field theory vertex can also be formulated
in a direct generalization of the well known flat space case \cite{GS} (although
there are several significant subtleties in implementing target space supersymmetry \cite{GS,STEFANSKI,RUSSO}
in order to determine the so-called `prefactor' part of the SFT vertex).

\begin{figure}
\hfill\includegraphics[height=4cm]{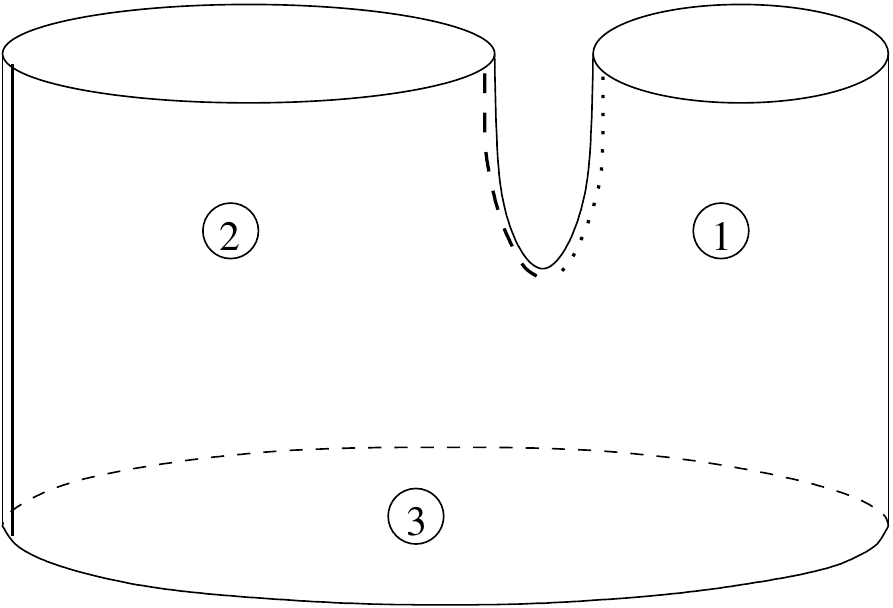}\hfill
\includegraphics[height=4cm]{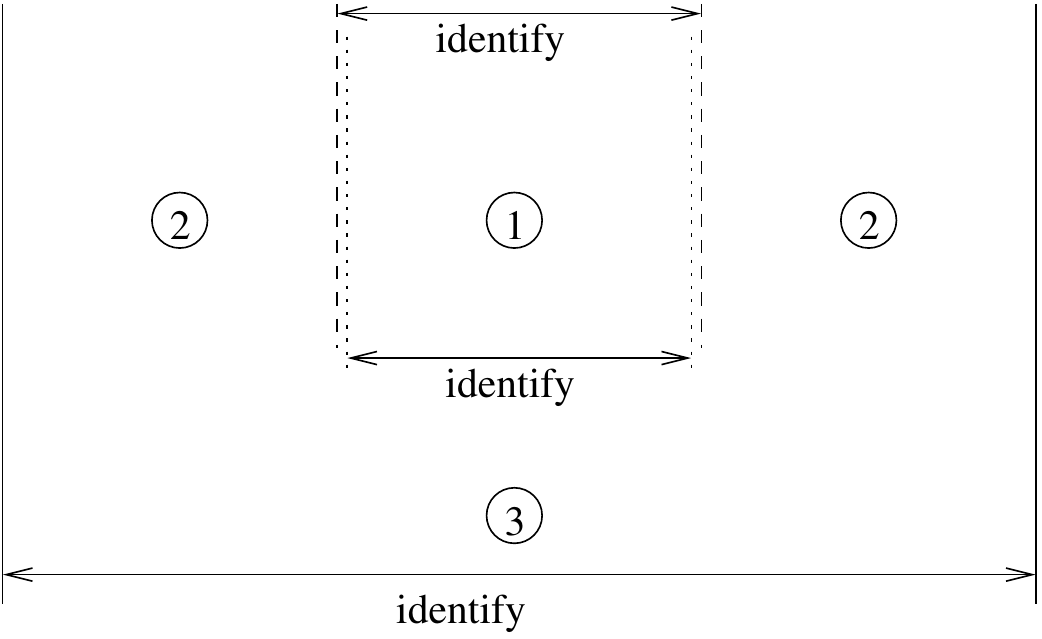}\hfill{}
\caption{The geometry of the worldsheet for the cubic light cone string field theory vertex in two different representations. The embedded left figure is flattened on the right by cutting along the various lines, which are 
identified on the right picture.   \label{fig.sft}}
\end{figure}

The light cone string field theory cubic vertex describes the splitting (or joining)
of an incoming string into two outgoing strings (see fig. \ref{fig.sft} (left)). The sizes of the strings, which
are proportional to conserved charges add up, hence we have
\eq
L_1+L_2=L_3.
\eqx 
(or $J_1+J_2=J_3$ --- we will often identify the $J$ charge with the size of the cylinder
and use one or the other notation depending on whether we want to be closer to the $AdS_5\times S^5$
string context or whether we want to emphasize a generic integrable QFT point of view).

The pp-wave vertex consists of two distinct parts. One is a universal exponential part which
follows from putting the worldsheet QFT onto the geometry shown in the right of fig.~\ref{fig.sft},
while the second part is the so-called `prefactor' which is an operator inserted at the splitting point,
and takes the form of a quadratic polynomial in creation and annihilation operators. The latter part
is required by target-space supersymmetry algebra, while the former part basically implements
just the continuity equations for the worldsheet QFT at the string splitting. Here we will concentrate
the discussion on this universal part, although our methods should be applicable also to the full vertex.

Technically, the (bosonic) universal exponential part of the vertex is obtained in the following way.
The free massive boson is expanded into cosine and sine modes in the three regions corresponding to
strings \#1, \#2 and \#3, with coefficients being the appropriate creation and annihilation operators
of the modes. Then one requires the continuity of $\phi$ and $\Pi \equiv \partial_\tau \phi$
at the string splitting time to obtain linear relations between the relevant creation and annihilation
operators:
\eq
\label{e.continuity}
\sum_{r=1}^3 \f{X^r_{nm}}{\sqrt{\om^r_m}} \left(a^{+(r)}_m- a^{(r)}_m\right) =0 \quad ;
\qquad
\sum_{r=1}^3 {\mathrm{sgn}}_r X^r_{nm} \sqrt{\om^r_m} \left(a^{+(r)}_m+ a^{(r)}_m\right) =0
\eqx 
In the above formula $a^{+(r)}_m$ is the creation operator for string $r$ with mode number $m$,
$\om^r_m$ is proportional to the energy of that mode (see below), while $X^r_{nm}$ is a purely geometric
overlap between mode $m$ on string $r$ and modes defined on the whole interval (and thus coinciding
with modes of string \#3). ${\mathrm{sgn}}_r$ is a sign which is opposite for ingoing and outgoing strings.

The above equations are implemented as operator equations acting on a state 
$\ket{V} \in \HH_1 \otimes \HH_2 \otimes \HH_3$ which represents the SFT vertex.
The simplest solution of these equations is an exponential of a quadratic form in the creation
operators:
\eq
\label{e.vertex}
\ket{V}=\exp\left\{\f{1}{2} \sum_{r,s=1}^3 \sum_{n,m} \bar{N}^{rs}_{nm}\, a^{+(r)}_n a^{+(s)}_m \right\} \ket{0}
\eqx 
The coefficients $\bar{N}^{rs}_{nm}$ are the famous Neumann coefficients\footnote{The bar in $\bar{N}^{rs}_{nm}$
comes form the fact that we are dealing here with cosine and sine modes. When we pass to modes with definite worldsheet momentum, which will be the case relevant for this paper, we will use unbarred notation.} and the problem of finding their explicit form is surprisingly
intricate. This comes from the fact that they involve finding the inverse of an infinite dimensional matrix
defined through (\ref{e.continuity}).
In the case of the pp-wave, the solution has been found in two steps. Firstly, the Neumann coefficients
where shown to obey a factorization property:
\eq
\bar{N}^{rs}_{nm} = - \f{m n \al}{1-4\mu \al K} \f{ \bar{N}^r_m \bar{N}^s_n}{\al_s \om^r_n +\al_r \om^s_m}
\eqx
where
\eq
\al_1=\f{J_1}{J_3} \quad ;\quad \al_2=\f{J_2}{J_3} \quad;\quad \al_3=-1 \quad;\quad \al=\al_1\al_2\al_3
\eqx
and $\mu$ is a parameter of the pp-wave background while $K$ and the Neumann vector $\bar{N}^r_m$
are the nontrivial quantities. Then the Neumann vectors and $K$ have been ultimately 
determined in the impressive works \cite{HSSV} and \cite{LSNS}.
The latter paper (to which we will often refer by the shorthand LSNS) provides a very explicit form 
for the exact answer which we will discuss at length in section~\ref{s.lsns}.

Before we finish this section with some comments, let us emphasize that the standard approach
to the string field theory vertex outlined above is almost impossible to generalize to the case of 
an interacting worldsheet QFT (as would be the case for $AdS_5 \times S^5$). 
In the interacting case, we do not have a workable analog of 
mode expansions hence it is extremely difficult to imagine how to implement continuity relations.
Moreover, the above formulation using integer mode numbers seems to be intrinsically
tied to a finite volume setup which makes matters even more complicated. The goal of this paper
is to find an alternative approach which bypasses these problems.

Let us now comment on various properties and features of the pp-wave SFT vertex
which will be important for our subsequent considerations.

\begin{figure}
\hfill\includegraphics[height=5cm]{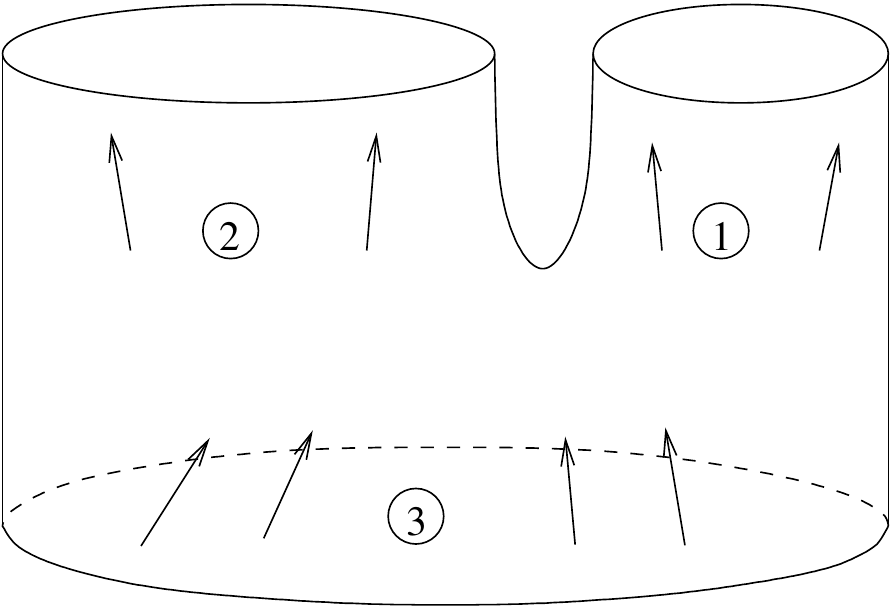}\hfill{}
\caption{The string field theory vertex with some incoming and outgoing particles in each of the three strings.\label{fig.particles}}
\end{figure}

The form of the exponential vertex (\ref{e.vertex}) provides for us a clear physical interpretation
of the Neumann coefficients. $\bar{N}^{rs}_{nm}$ is just the amplitude of a free massive scalar
theory on the pants diagram (fig.~\ref{fig.sft} (right)) with just two particles/modes -- one on string $r$
with mode number $n$, the other on string $s$ with mode number $m$, and vacuum on the remaining string(s).
The exponential form of the vertex (\ref{e.vertex}) means essentially that all amplitudes with
a higher number of particles distributed on the pants diagram are expressible in a simple way in terms
of the 2-particle ones (i.e. in terms of the Neumann coefficients).
We expect that in the interacting case the relation between the amplitudes with higher number of particles
and lower ones will be less trivial so the goal of formulating the vertex corresponds to
finding (equations for) amplitudes with all possible distributions of particles among the three strings
and \emph{not} just generalizing Neumann coefficients to the interacting case.

Let us now introduce some general notation for a generic SFT amplitude with particles with 
rapidities\footnote{For the sake of clear notation here we  parametrize the particles by
relativistic rapidities, but the definitions will go over verbatim either to a parametrization
in terms of momenta or in terms of complex $AdS$ rapidities.}
$\th_1,\ldots,\th_n$ on string \#3, $\th'_1,\ldots,\th'_m$ on string \#2 and $\th''_1,\ldots,\th''_l$ on string \#1.
We  also explicitly mark the sizes of the respective strings:
\eq
\Nfin^{3|2;1}_{L_3|L_2;L_1}\left( \th_1,\ldots,\th_n \;\biggl|\; \th'_1,\ldots,\th'_m\, ;\, \th''_1,\ldots,\th''_l \right)
\eqx
For the case of the pp-wave SFT vertex, these quantities can be directly expressible in terms
of the Neumann coefficients. Assuming for the moment the absence of the prefactor, we would have
\eq
\label{e.examples}
\Nfin^{3|2;1}_{L_3|L_2;L_1}\left(\th_1, \th_2 \;\bigl|\; \varnothing \,;\, \varnothing \right) \equiv N^{33}_{n_1 n_2}
\qqqq
\Nfin^{3|2;1}_{L_3|L_2;L_1}\left(\th_1 \;\bigl|\; \th_2 \,;\, \varnothing\right) \equiv N^{32}_{n_1 n_2}
\eqx
where $n_i$ are mode numbers corresponding to particular momenta/rapidities and the empty set $\varnothing$ just denotes
the vacuum.
A more complicated example is
\eq
\Nfin^{3|2;1}_{L_3|L_2;L_1}\left(\th_1, \th_2 \;\bigl|\; \th_3 \,;\, \th_4 \right) \equiv N^{33}_{n_1 n_2} N^{12}_{n_3 n_4}+
N^{32}_{n_1 n_3} N^{31}_{n_2 n_4}+ N^{32}_{n_2 n_3} N^{31}_{n_1 n_4}
\eqx
When we give formulas for the pp-wave case, we will alternatively use the conventional notation of Neumann coefficients, 
but always recall (\ref{e.examples}). 

Another interesting observation comes from analyzing some important parameter regimes appearing in the pp-wave case. 
$\mu$ is a parameter
which is essentially the inverse of the `t Hooft coupling. It appears in the frequency of
the appropriate mode as
\eq
\om^r_m =\sqrt{m^2+\mu^2 \al_r^2}
\eqx
For our purposes it is convenient to reformulate all formulas by trading the integer mode numbers
for physical worldsheet momenta. The momenta are given by $p=\pm 2\pi m/J_r$, and thus the frequency
becomes
\eq
\om^r_m = \f{|\al_r| J}{2\pi} \sqrt{p^2+M^2}
\eqx
where $J\equiv J_3$ and the mass of the scalar field is related to $\mu$ through
\eq
\label{e.mass}
M =\f{2\pi}{J} \mu
\eqx
In the pp-wave times, people were mostly interested in comparison with gauge theory perturbative
computations and thus concentrating on an expansion around $\mu=\infty$ in inverse powers of $\mu$.
In particular they employed simpler asymptotic versions of the Neumann coefficients
which neglected terms of the type
\eq
e^{-2\pi \mu |\al_r|} 
\eqx
It is interesting to realize that this term, when expressed in terms of the physical mass of the 
free boson (\ref{e.mass}), becomes 
\eq
e^{-M J_r}
\eqx
which is exactly the scale of wrapping corrections associated to string $r$.
In fact this nicely explains the observation made in \cite{LSNS} about the similarity
of the formulas of the leading exponential corrections to the Neumann coefficients
with Casimir energy of the free massive boson.

In the following, we will also need expressions corresponding to modes 
with definite worldsheet momentum -- thus the so-called BMN modes instead of the cosine and
sine modes used in the derivation of the Neumann coefficients.
The explicit relations are given e.g. in \cite{KH}, in particular we have
\eq
\label{e.bmn}
N^{rs}_{mn} = \f{1}{2} ( \bar{N}^{rs}_{mn} - \bar{N}^{rs}_{-m\,-n} )
\eqx
for positive mode numbers.

We will be mostly, but as it will turn out not exclusively, concentrated on the string vertex
when neglecting wrapping corrections. In this limit, the expression for $N^{rs}_{mn}$ 
no longer involves special functions but is still apparently quite cumbersome \cite{HSSV}:
\eq
N^{rs}_{mn} \propto \left[ \f{\sqrt{(\om^r_m+\mu\al_r)(\om^s_n+\mu\al_s)}}{\om^r_m+\om^s_n} -
\f{\sqrt{(\om^r_m-\mu\al_r)(\om^s_n-\mu\al_s)}}{\om^r_m+\om^s_n} \right] s_{rm}s_{sn}
\eqx
with
\eq
s_{1m}=s_{2m}=1 \quad\quad s_{3m}=-2 \sin(\pi m \al_1)
\eqx
Surprisingly enough, once we parametrize the modes by rapidities $p=M \sinh \th$,
the above expression simplifies drastically\footnote{We provide more formulas and discuss various 
intriguing features of this limit in section~\ref{s.asympt}.}:
\eq
\label{e.n33as}
N^{33}(\th,\th')_{asympt} \propto - 2 \f{\sin \f{p L_1}{2} \sin \f{p' L_1}{2}}{
\cosh\frac{\th-\th'}{2}} 
\eqx
where we extracted simple factors related to the normalization condition for the modes and
some overall constant factor. The subscript ${}_{asympt}$ denotes the fact that we neglected all exponential
wrapping corrections in $e^{-M L_1}$.

Let us make some comments on the above expression (\ref{e.n33as}).
Firstly, we see that the discrete nature of the finite volume integer modes
does not play here any important role. 
In fact the above expression is extremely simple when expressed 
in terms of infinite volume rapidities. 
The passage to finite volume amounts here\footnote{Recall that we are always neglecting exponential wrapping corrections.} 
just to evaluating the above expression (\ref{e.n33as}) for rapidities corresponding
to quantized momenta i.e. $p=M \sinh(\th)=2\pi n/L$. This is in direct
correspondence with the finite volume evaluation of form factors (\ref{e.fvff}).

Secondly, the analytic structure of this function is also
quite appealing as there is a pole at $\th=\th '+i\pi$, which is exactly the characteristic
position of the so-called kinematical singularity for form factors, with the $i\pi$ 
intimately related to crossing properties. 

Thirdly, there are nevertheless still some surprising
features of the expression (\ref{e.n33as}). Two-particle form factors typically have vanishing residue at the
kinematical pole (see (\ref{e.ffkinem})), while here\footnote{Recall from the discussion above that Neumann coefficients
can be interpreted as two particle amplitudes.} the residue is nonzero and is in fact quite bizarre.
The $\sin \f{pL}{2}$ factors are also quite surprising by themselves. They are almost of the `wrapping' type, however instead 
of being exponentially suppressed, they are oscillatory.

Moreover, if one would consider the asymptotic form of $N^{32}$
\eq
\label{e.n23as}
N^{32}(\th,\th')_{asympt}  \propto  \f{\sin\f{p L_1}{2}}{\sinh\frac{\th-\th'}{2}} 
\eqx
one would see that (\ref{e.n33as}) and (\ref{e.n23as}) are related by a surprisingly modified
form of crossing relation
\eq
\label{e.asmodcross}
N^{33}(\th, \th' - i\pi)_{asympt} = -2i \sin \f{p' L_1}{2} N^{32}( \th, \th')_{asympt}
\eqx
which, incidentally bears a striking resemblance to the modified crossing observed in \cite{MINWALLA1,MINWALLA2}.

So to conclude this section, we see that the asymptotic form of the pp-wave Neumann coefficients
very strongly suggests the existence of an infinite volume formulation based on analyticity properties
such as crossing, kinematical singularity etc. In the remaining part of the paper we will indeed provide such
a formulation and also show that the apparent modification of crossing in (\ref{e.asmodcross})
is in fact an artefact of the large volume limit and the true crossing property should be different.

\section{The decompactified string vertex and the SFT axioms} 
\label{s.decomp}

\begin{figure}
\hfill\includegraphics[height=4cm]{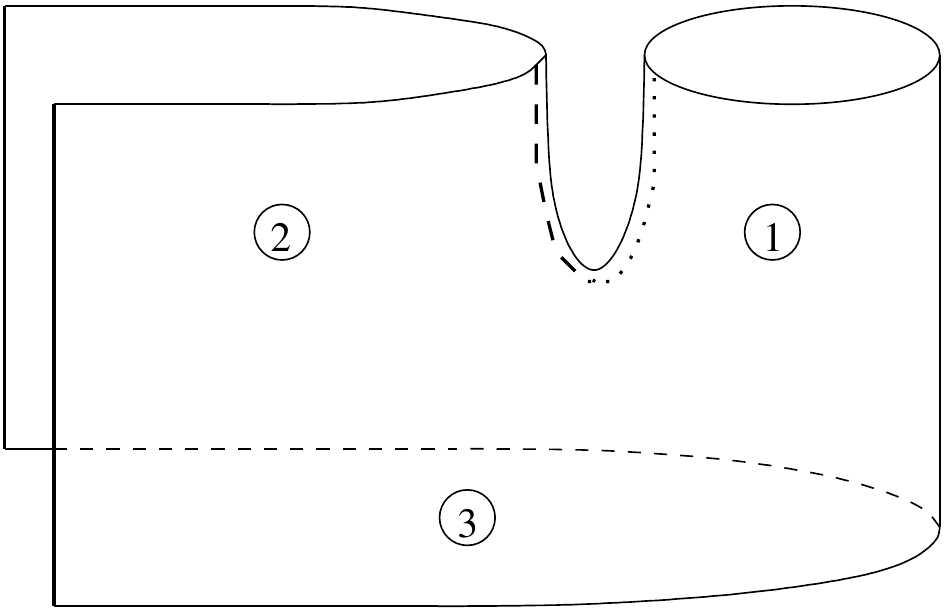}\hfill
\includegraphics[height=4cm]{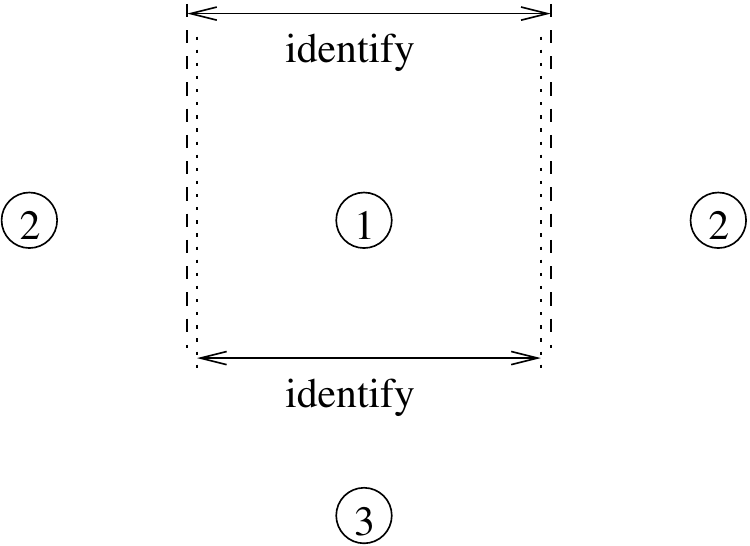}\hfill{}
\caption{The decompactified string field theory vertex in two different geometrical representations. \label{fig.decomp}}
\end{figure}

As emphasized before, in order to be able to formulate functional equations incorporating
crossing property it is crucial to define a decompactified version of the SFT vertex.
We show such a construction in fig.~\ref{fig.decomp}, where we cut strings \#2 and \#3
and extended their boundaries to infinity. The right hand side of this figure shows the
resulting pattern of identifications in the two-dimensional plane.
Here string \#1 remains of finite size $L\equiv L_1$ and there will be a nontrivial 
dependence on the dimensionless product $mL$. In particular, we expect that the decompactified
vertex amplitudes will incorporate all wrapping corrections associated with string \#1.

Of course, we could have just as well made the cut along string \#1 and \#3, leaving string \#2
at finite size. In fact in order to find the physical finite volume SFT vertex from integrability
we will advocate considering simultaneously both possibilities, solving the associated two sets
of functional equations and then \emph{requiring} that the finite volume reductions of both
solutions will coincide. We will describe this in more detail in section~\ref{s.program}.

Finally, we note that unfortunately we cannot decompactify both outgoing strings, because then 
the ingoing string \#3 would split into two disconnected pieces. In fact we will find that
the size of the leftover finite size string $L$ will play a crucial role in formulating
the SFT axioms.

\subsection{The decompactified SFT axioms}

In the following we cut string \#2  and string \#3 and extend their
boundaries to infinity as shown on the left of figure 2. On the right,
one can see the full infinite spacetime domain of string \#3 on the lower part, 
while the infinite spacetime domain of string \#2 with a missing strip of size
$L$ on the upper part of the figure. The two sides of the strip are identified
to form the space-time cylinder of string \#1. The other identification
on the figure makes the space-time for string \#2 continous, i.e. leaving 
from left to right on the left of the strip we appear immediately on the right of 
the strip. 

The aim of this section is to propose functional equations
for the amplitude with prescribed number and momenta of particles on 
the decompactified strings \#2 and \#3, while the particle content
in the compact string \#1 may be arbitrary. Since these string \#1
excitations will not enter the equations at all, we will denote them
by $\bullet$ below.

As the decompactified SFT vertex amplitude has slightly different
properties\footnote{This is exactly as for the relation between
finite volume and infinite volume form factors which differ
by a Jacobian factor and a product of S-matrices neccessary to ensure
symmetry of the finite volume one~ c.f.~(\ref{e.fvff}).} from 
the finite volume one discussed in section~\ref{s.ppintro},
we  introduce some specific notation in this case.
We thus denote the decompactified SFT vertex amplitude
by
\eq
\Nfin^{3|2}_{\bullet,L} \Bigl( \th_1,\ldots,\th_n \;\Bigl|\; \th'_1,\ldots,\th'_m\Bigr)
\eqx
which contains particles with rapidities $\{\theta_{i}\}$ in domain
\#3 and with rapidities $\{\theta_{j}'\}$ in domain \#2.

\begin{figure}[h]
\begin{center}
\includegraphics[height=7cm]{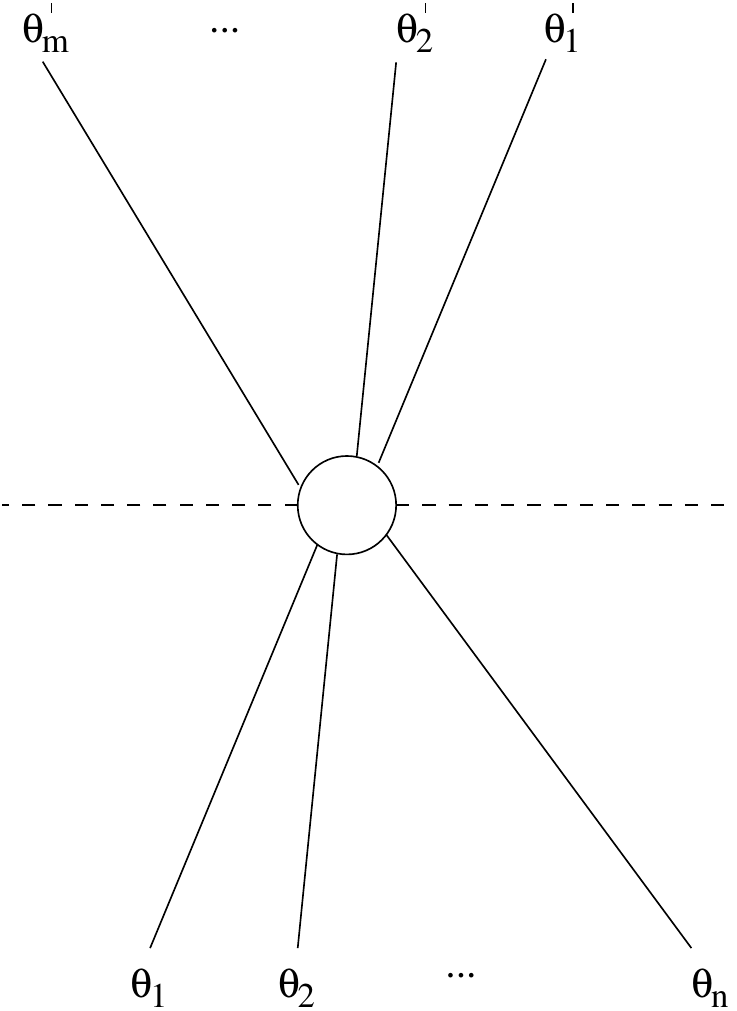}
\par\end{center}
\caption{Graphical notation for the SFT vertex. Domain \#3 is below the dashed line and contains
incoming particles $\{\theta _i\}$, while domain \#2 is above the line and contains outgoing particles
$\{\theta _i\}$. The emission of string \#1 is represented by the circle,
which can be understood as a partly nonlocal operator insertion in the form factor language.
It introduces a discontinuity
which is distributed symmetrically on the border of the two domains indicated by dashed lines.
\label{fig.nm}}
\end{figure}

The superscripts denote the noncompact ingoing and outgoing strings,
$L$ is the size of the remaining closed string \#1 and $\bullet$
denotes its specific state as well as any local operator inserted at the
string splitting point (like the prefactor in the pp-wave SFT vertex), see
Figure \ref{fig.nm} for a graphical notation.

In the following, we will sometimes suppress the lower subscript ${}_{\bullet,L}$ 
as long as it remains unchanged.
We assume that the particles scatter on each other diagonally
with the scattering matrix $S(\theta_{i},\theta_{j})$. 
We will cover the general non-diagonal case in section~\ref{s.nondiag}.
This S-matrix
does not necessarily depend on the differences of the rapidities but
satisfies unitarity
\eq
S(\theta_{i},\theta_{j})=S(\theta_{j},\theta_{i})^{-1}
\eqx
When particles pass through each other they scatter with the S-matrix,
thus their ordering is essential. States in domain \#3 are preparated
at $t=-\infty$ and contain particles with ordered rapidities $\theta_{i}>\theta_{i+1}$,
i.e. the fastest is on the leftmost. We call these states initial
states. States in domain \#2 contain ordered particles with rapidities
$\theta_{i+1}'<\theta_{i}'$, in which the fastest is the rightmost.
These states are called final states. The coefficient $\Nfin^{3|2}_{\bullet,L}$ above
describes the transition amplitude from an initial to a final state.
Clearly if there were no space deficiency, $L=0$, (and trivial operator
insertion), $\Nfin^{3|2}_{\bullet,L}$ would be nothing but the scattering matrix element,
nonvanishing only for coinciding sets of rapidities. If, however,
$L\neq0$, or there is an operator insertion the corresponding $\Nfin^{3|2}_{\bullet,L}$
is similar to a form factor: that is to a matrix element of an operator.
This operator is local for $L=0$ but is non-local for $L\neq0$.
In the following we focus on the $L\neq0$ case. This is similar to
the situation, when we analyze the form factors of an operator, which
is nonlocal with respect to the particles. Moving a particle around
the space deficiency would pick up a phase factor proportional
both to $L$ and to its momenta: $e^{ipL}$. 
This bears some similarity with the form factor
axioms with nonzero index of mutual locality \cite{Yurov:1990kv}. Note, however,
that here the analogous index is momentum dependent which is a completely novel
and unique feature of the string vertex. 
By the choice of the
bases in domains \#3 and \#2 we can freely place this nonlocality
wherever we want. To be in accordance with the pp-wave conventions we distribute
the nonlocality in an equal way on the border of domains \#3 and \#2, which 
we indicate by dashed lines on the figures.

\begin{figure}[h]
\begin{center}
\includegraphics[height=6cm]{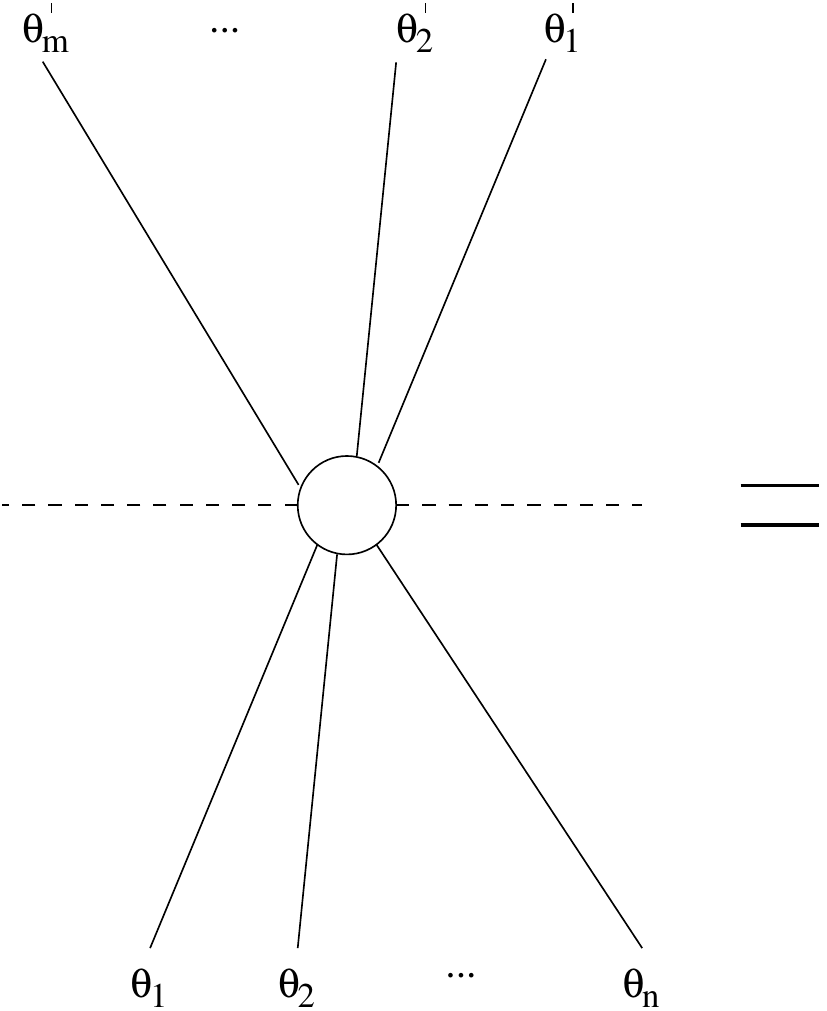}~~\includegraphics[height=6cm]{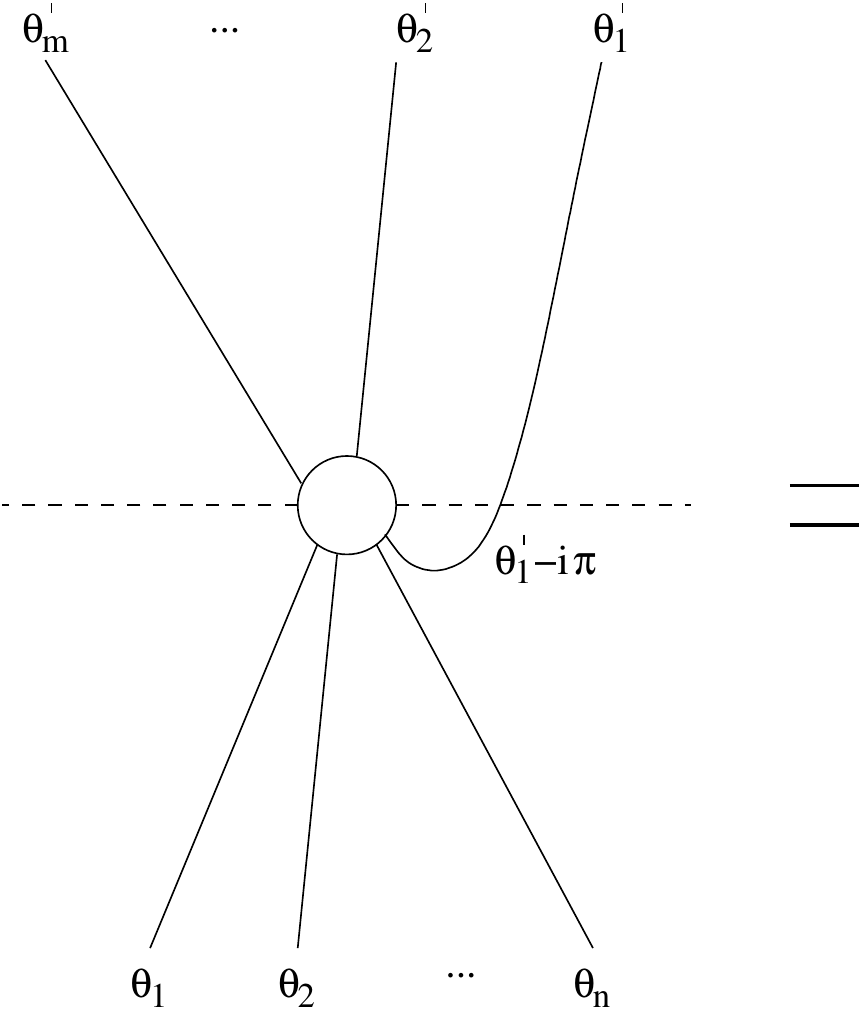}~
~\includegraphics[height=6cm]{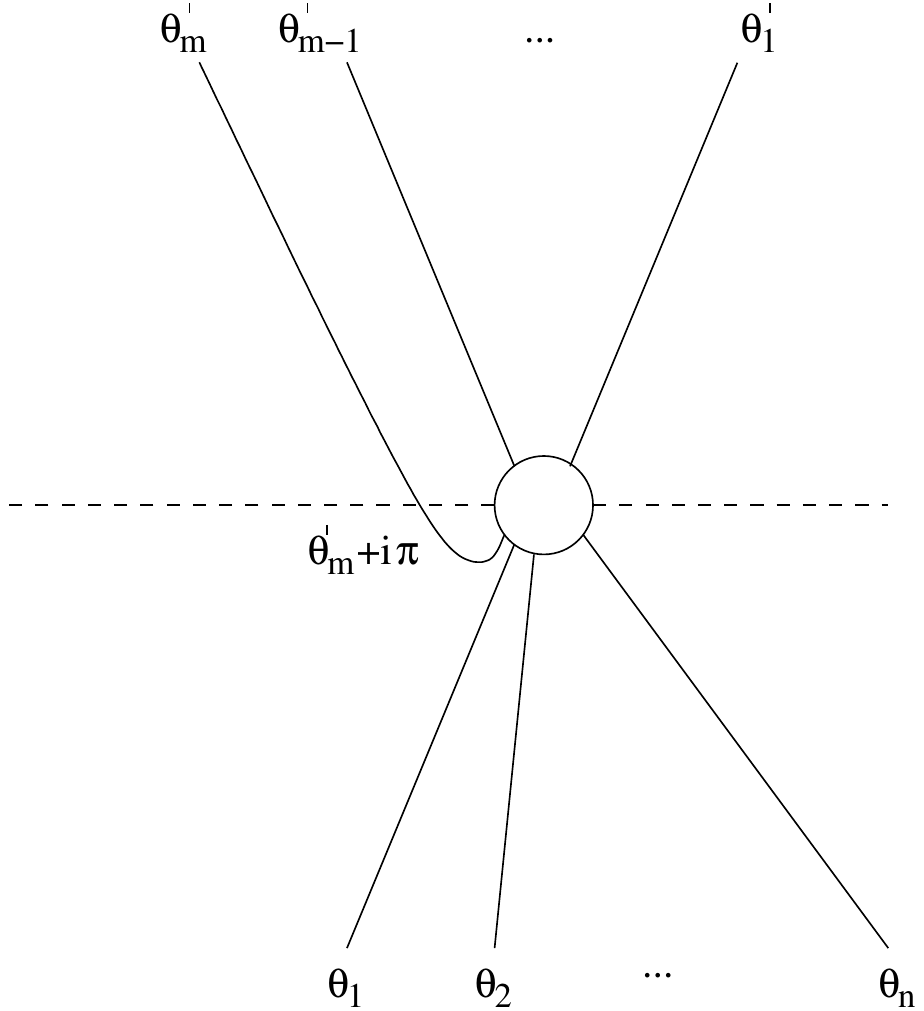}
\par\end{center}
\caption{Crossing transformations for the SFT vertex.}
\label{fig.cross}
\end{figure}

This means we define the crossing equations as 
\begin{align}
\label{e.crossi}
\Nfin^{3|2}_{\bullet,L} \bigl( \th_1,\ldots,\th_n \;\bigl|\; \th'_1,\ldots,\th'_m \bigr) &=
e^{ip(\theta_{1}')\frac{L}{2}} \,
\Nfin^{3|2}_{\bullet,L}\bigl( \th_1,\ldots,\th_n,\th'_1-i\pi \;\bigl|\; \th'_2,\ldots,\th'_m\bigr) \\
\Nfin^{3|2}_{\bullet,L}\bigl( \th_1,\ldots,\th_n \;\bigl|\; \th'_1,\ldots,\th'_m\bigr) &=
e^{-ip(\theta_{m}')\frac{L}{2}} \,
\Nfin^{3|2}_{\bullet,L}\bigl( \th'_m+i\pi, \th_1,\ldots,\th_n \;\bigl|\; \th'_1,\ldots,\th'_{m-1}\bigr)
\label{e.crossii}
\end{align}
They are represented graphically on Figure \ref{fig.cross}. 

In these equations it is understood that no rapidites in the initial and
final states coincide $\theta_{i}\neq\theta_{j}'$ as otherwise disconnected
terms can arise\footnote{Let us note that it is possible to introduce a sign in \emph{both} equations
(\ref{e.crossi}) and (\ref{e.crossii}) to accommodate different normalization conventions. This
does not change, of course, any physical content.}.  
\begin{figure}[h]
\begin{center}
\includegraphics[width=4cm]{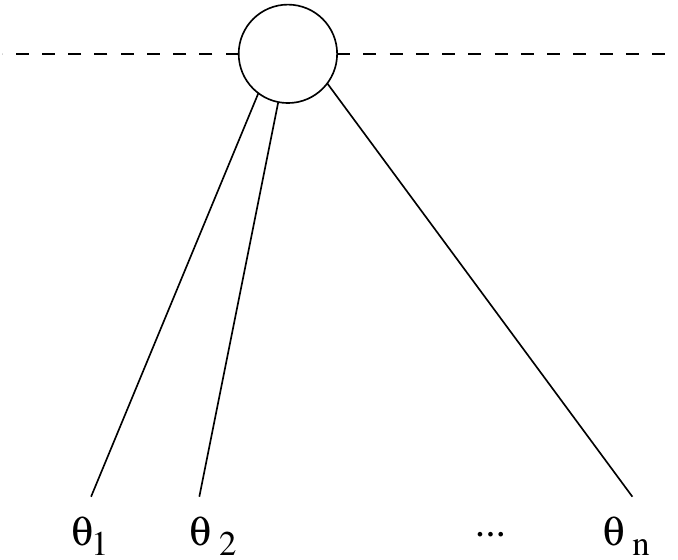}
\par\end{center}
\caption{Elementary SFT vertex.}
\label{fig.Nelem}
\end{figure}
Crossing all particles to domain \#3 we can define the elementary SFT vertex
\eq
\Nfin_{\bullet,L}(\theta_{1},\dots,\theta_{n})\equiv 
\Nfin^{3|2}_{\bullet,L}\bigl( \th_1,\ldots,\th_n \;\bigl|\varnothing \bigr)
\eqx
which obviously contains all the information\footnote{Recall that the empty set $\varnothing$ denotes the vacuum
(no particles) while $\bullet$ stands for any particle content.}. This vertex is represented
graphically on Figure \ref{fig.Nelem}.

\begin{figure}[h]
\begin{center}
\includegraphics[height=3.5cm]{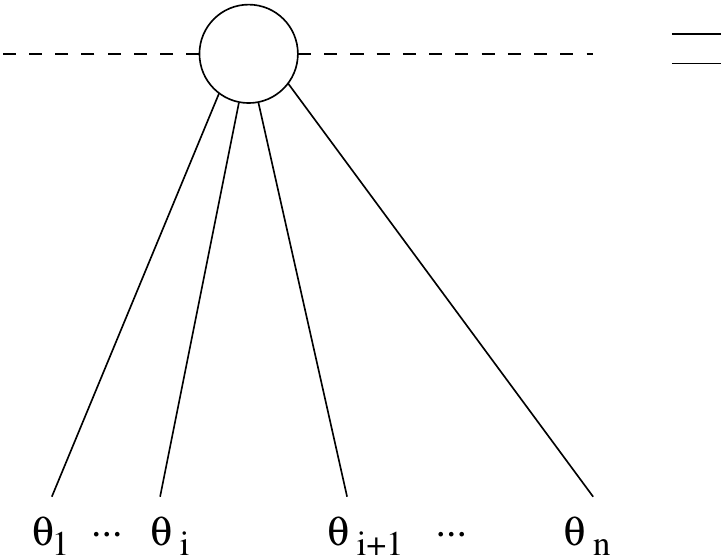}~~~~~~~\includegraphics[height=3.5cm]{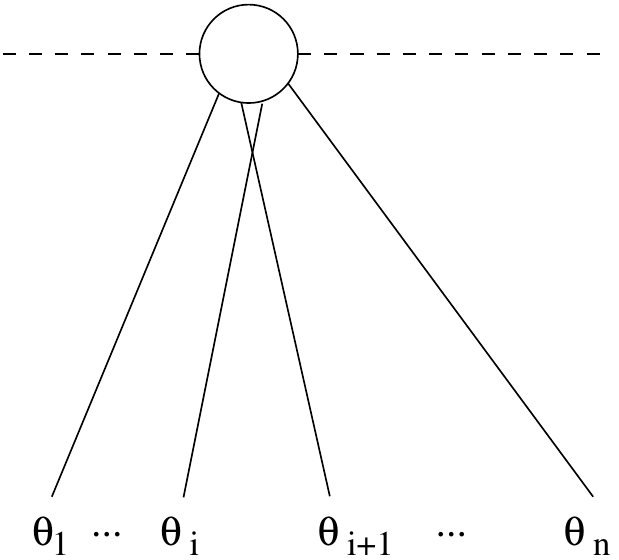}
\par\end{center}
\caption{Permutation axiom of the  SFT vertex.}
\label{fig.NPerm}
\end{figure}

 Now we formulate the
axioms it should satisfy. As the initial state is a scattering state
exchanging two neighbouring particles leads to the factor of the scattering
matrix, see Figure \ref{fig.NPerm}: 
\eq
\label{e.symmetry}
\Nfin_{\bullet,L}(\theta_{1},\dots,\theta_{i},\theta_{i+1},\dots,\theta_{n})=
S(\theta_{i},\theta_{i+1})\Nfin_{\bullet,L}(\theta_{1},\dots,\theta_{i+1},\theta_{i},\dots,\theta_{n})
\eqx

\begin{figure}[h]
\begin{center}
\includegraphics[width=5.2cm]{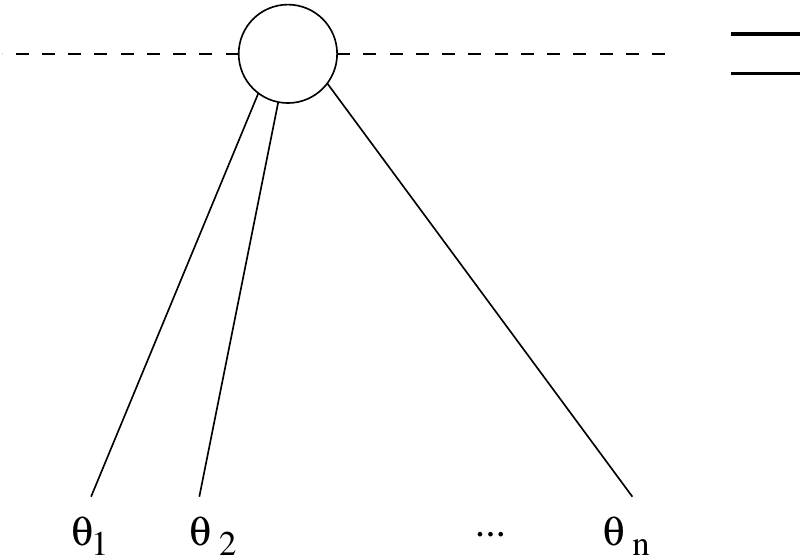}~~~~~~~\includegraphics[width=4cm]{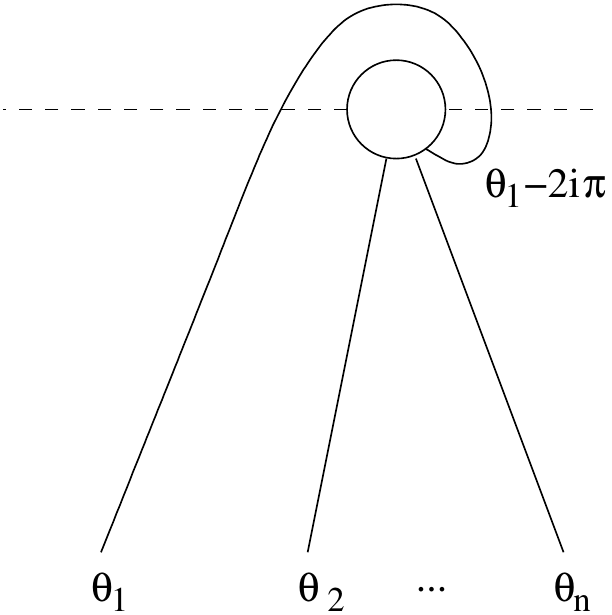}
\par\end{center}
\caption{Monodromy axiom of the  SFT vertex.}
\label{fig.Nmon}
\end{figure}

Crossing the first particle to domain \#2 and crossing back to the
last position we obtain the monodromy relation
\eq
\label{e.monodromy}
\Nfin_{\bullet,L}(\theta_{1},\theta_{2},\dots,\theta_{n})=
e^{-ip(\theta_{1})L}\, \Nfin_{\bullet,L}(\theta_{2},\dots,\theta_{n},\theta_{1}-2i\pi)
\eqx
which expresses the nonlocality of the ``operator insertion'', see Figure \ref{fig.Nmon}. Here
we used that $p(\th_1 +i\pi)=-p(\th_1) $. 

\begin{figure}[h]
\begin{center}
\hspace{-1cm}\includegraphics[width=6.5cm]{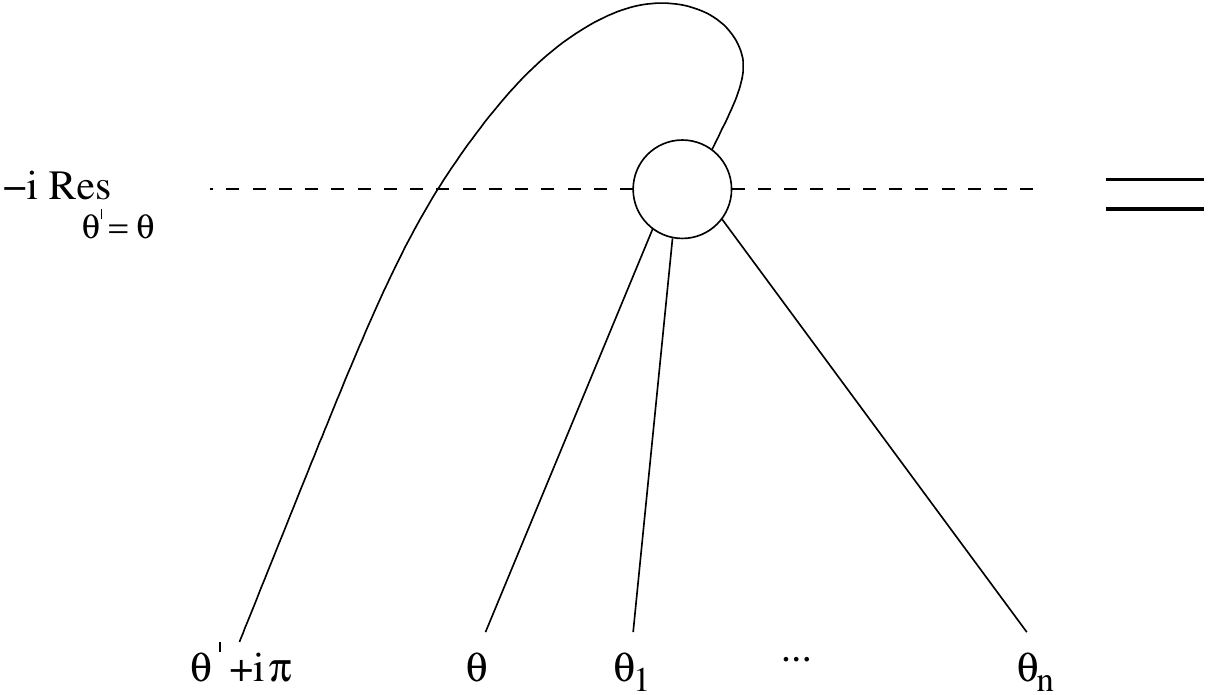}~~\includegraphics[width=5.5cm]{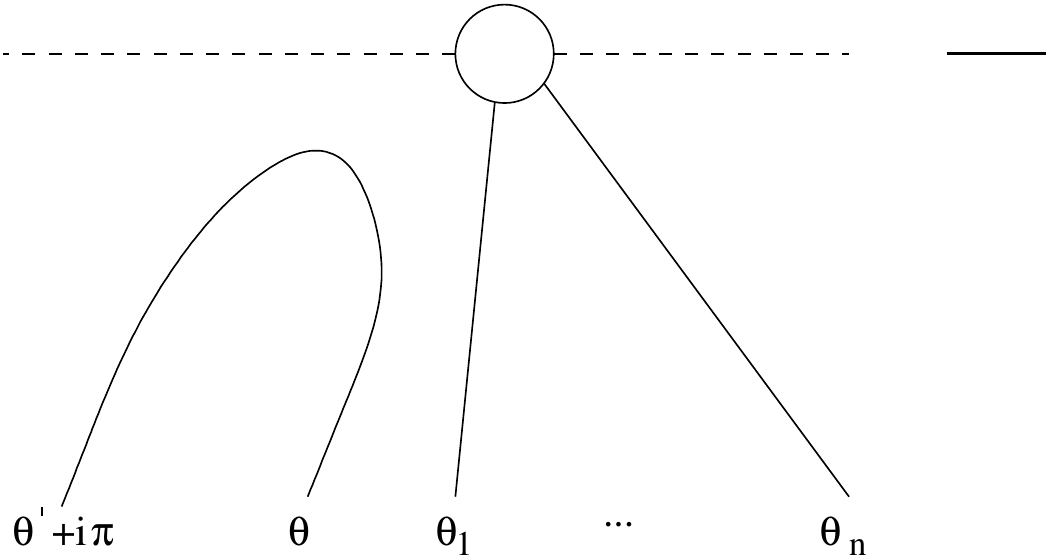}~
~\includegraphics[width=4.5cm]{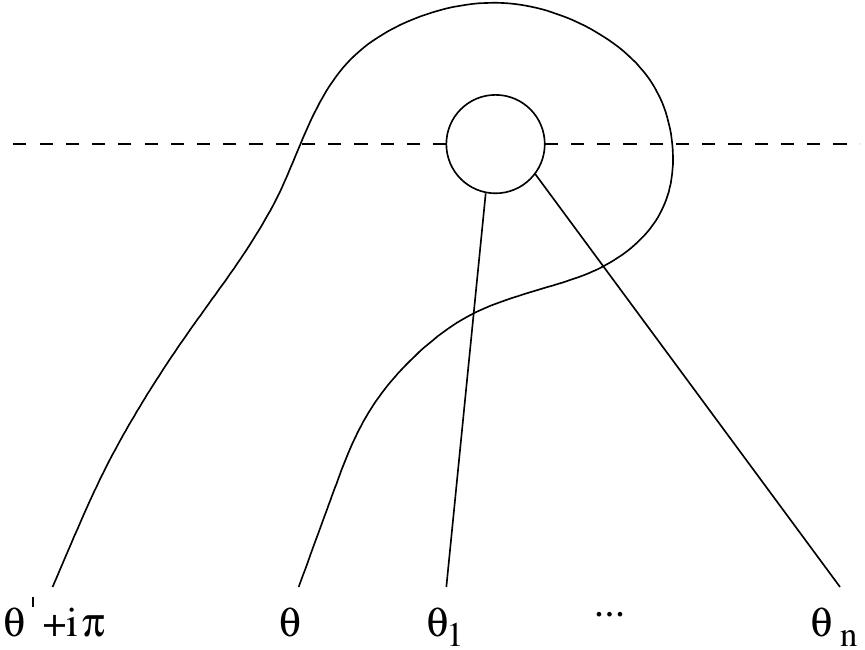} 
\par\end{center}
\caption{Kinematical singularity axiom of the  SFT vertex.}
\label{fig.kin}
\end{figure}

The above two relations
provide functional equations which enable to determine the coefficient
$\Nfin_{\bullet,L}$ once its analytical structure is known. $\Nfin_{\bullet,L}$ must be
a meromorphic function of the rapidites, whose poles have physical
origins. There are poles which have kinematical and others which have
dynamical origins. A kinematical singularity can appear whenever, after
crossing, an initial particles' rapidity coincides with a final one.
The residue of the pole is proportional to the amplitude where the
two particles are missing as: 
\eq
-i\mbox{Res}_{\theta'=\theta}\Nfin_{\bullet,L}(\theta'+i\pi,\theta,\theta_{1},\dots,\theta_{n})=
\Bigl (1-e^{ip(\theta)L}\prod_{i=1}^{n}S(\theta,\theta_{i})\Bigr )\Nfin_{\bullet,L}(\theta_{1},\dots,\theta_{n})
\label{e.kinematical}
\eqx
The proportionality factor expresses the fact that the on-shell particle
can pass the other particles and the defect on both sides as shown on Figure \ref{fig.kin}.

The dynamical singularity axiom is related to the existence of boundstates and expresses 
the SFT vertex of the boundstate in terms of that of the fundamental particles. 
As this axiom does not provide any restriction on the elementary SFT vertex we 
do not write out explicitly here, but spell out the details in the nondiagonal case in section~\ref{s.nondiag} below.

Let us finish this part by specifying the kinematical singularity axiom for the case when there are particles both in the initial and final states, as this equation will show up in the weak coupling limit of the OPE coefficients:\eqn
-i\mbox{Res}_{\theta'=\theta}\Nfin_{\bullet,L}^{3\vert2}
(\theta,\theta_{1},\dots,\theta_{n}\vert \theta_{1}',\dots,\theta_{m}',\theta' )&=&e^{-i p(\theta)L/2}
\Bigl (1-e^{ip(\theta)L}\prod_{i=1}^{n}S(\theta,\theta_{i})
\prod_{j=1}^{n}S(\theta_j',\theta)\Bigr )\times \nonumber \\
&&\qquad\qquad \Nfin_{\bullet,L}^{3\vert2}
(\theta_{1},\dots,\theta_{n}\vert \theta_{1}',\dots,\theta_{m}')
\eqnx

Finally, note that the above equations do not depend in any way on the state of the
compact string (string \#1 here). This is in fact very natural and is analogous
to the well known form factor axioms which have exactly the same
form for \emph{any} local operator. The form factor axioms do not have a unique solution, however,
and various solutions correspond to form factors of various local operators.
We expect the situation to be similar here -- the axioms for $\Nfin_{\bullet,L}(\theta_{1},\dots,\theta_{n})$
will have many solutions depending on the particle content of string \#1
and on the choice of prefactor operator in the SFT vertex inserted at the splitting point.

However, this time \emph{in contrast} to the ordinary form factor case,
we will be able to provide additional information which will severely restrict
the dependence on the string \#1 state. This will be discussed in detail in section~\ref{s.program},
where we complete the formulation of our program for the SFT vertex.

\section{The free massive boson example (or the pp-wave SFT vertex)}
\label{s.lsns}

In this section we will discuss the simplest case for which our integrable approach should work
i.e. a free massive boson. This is precisely the case of the pp-wave SFT vertex\footnote{Since the treatment of massless particles in the integrable S-matrix language
is in general quite subtle, together with the importance of wrapping, a discussion of the flat space SFT vertex of \cite{GS} would
require a lot of care.}, the Neumann
coefficients of which are known exactly.

The consideration of the pp-wave vertex is interesting for a variety of reasons. Firstly, we may check
that the proposed booststrap axioms are indeed satisfied. Secondly, we may analyze the analytical
structure of the pp-wave Neumann coefficients to put forward some `empirical' analyticity requirements
for the solutions of the bootstrap SFT vertex axioms in the general interacting case. Thirdly, we may investigate 
directly their asymptotic limit neglecting wrapping corrections, which limit turns out to have surprisingly
subtle properties. Finally, we may see how to reconstruct the exact (decompactified) pp-wave Neumann coefficients
directly from our axioms together with the analyticity assumptions mentioned above.

\subsection{A review of LSNS formulas}
\label{s.lsnsreview}

Let us start by reviewing the known exact solution for the pp-wave Neumann coefficients
as given by \cite{LSNS}. These formulas involve quite a lot of notation and new special functions $\Gamma_\mu(z)$
introduced by LSNS, whose properties and definitions we recall in Appendix~\ref{s.appGammaMu}.
We also pass here to rapidity variables instead of integer mode numbers and introduce
some modifications of the special functions -- which we denote by $\tilde{\Gamma}_{\mu}(\theta)$ --
which are more convenient for our purposes.

Recall from Section 3 that the Neumann coefficients in the
cosine-basis have the structure 
\eq
\bar{N}_{mn}^{rs}=\rho\frac{mn}{\alpha_{s}\omega_m^r+\alpha_{r}\omega_n^s}\bar{N}_{m}^{r}\bar{N}_{n}^{s}
\eqx
where we focused on the dependence on the quantization numbers $n,m$
and put the rest into the normalization constant $\rho$. The Neumann
vectors $\bar{N}_m^r$ are defined in terms of the function $f_{m}^{(r)}$
as 
\eq
\bar{N}_{m}^{r}=\sqrt{\frac{\omega_m^r}{m}}\frac{\omega_m^r+\alpha_{r}\mu}{\alpha_{r}m}f_{m}^{(r)}
\eqx
The Neumann matrix above is related to the cosine basis. The coefficients
for the sine basis can be obtained as
\eq
\bar{N}_{-m-n}^{rs}=-\frac{\omega_m^r-\alpha_{r}\mu}{m}\frac{\omega_n^s-\alpha_{s}\mu}{n}\bar{N}_{mn}^{rs}
\eqx
The Neumann coefficients in the exponential basis relevant for our considerations 
follow through (\ref{e.bmn})
\eq
N_{mn}^{rs}=\frac{1}{2}
\left(1+\frac{\omega_m^r-\alpha_{r}\mu}{m}\frac{\omega_n^s-\alpha_{s}\mu}{n}\right)\bar{N}_{mn}^{rs}
\eqx
As we explained earlier, the rapidity parametrization considerably simplifies the
formulas. Thus we express $\omega_m^r$ and the mode number $m$ in
terms of the rapidity $\theta_{m}$ as: 
\eq
\omega_m^r=\sqrt{m^{2}+\alpha_{r}^{2}\mu^{2}}=\vert\alpha_{r}\vert\mu\cosh\theta_{m}\qquad;
\qquad m={\vert\alpha_{r}\vert} \mu\sinh\theta_{m}
\eqx
In the following we introduce formulas, which are valid for any signs
of $\alpha_{r}$. However, the expressions will depend on this sign,
which we denote by ${\mathrm{sgn}}_{r}$. Using the formulas above we get 
\eq
\frac{\omega_m^r-\alpha_{r}\mu}{m}\frac{\omega_n^s-\alpha_{s}\mu}{n}=
\left(\tanh\frac{\theta_{m}}{2}\right)^{{\mathrm{sgn}}_{r}}\left(\tanh\frac{\theta_{n}}{2}\right)^{{\mathrm{sgn}}_{s}}
\eqx
and the general Neumann matrix can be written as: 
\eqn
N_{mn}^{rs}&\!\!=&\!\! \frac{\rho}{2}\frac{\mu\sinh\theta_{m}\sinh\theta_{n}}{{\mathrm{sgn}}_{s}
\cosh\theta_{m}+{\mathrm{sgn}}_{r}
\cosh\theta_{n}}\frac{1}{\alpha_s \alpha_r}
\left[1+\left(\tanh\frac{\theta_{m}}{2}\right)^{-{\mathrm{sgn}}_{r}}
\left(\tanh\frac{\theta_{n}}{2}
\right)^{-{\mathrm{sgn}}_{s}}\right]  \\ \nonumber
&&\times \sqrt{\frac{\cosh\theta_{m}}{\sinh\theta_{m}}}\sqrt{\frac{\cosh
\theta_{n}}{\sinh\theta_{n}}}f_{m}^{(r)}f_{n}^{(s)}
\eqnx
Let us spell out the details in the three distinct cases. For $\alpha_3=-1$ we have
\begin{equation}
N_{mn}^{33}=-\rho\frac{d^{(3)}(\theta_{m})d^{(3)}(\theta_{n})}{\cosh\frac{1}{2}(\theta_{m}-\theta_{n})}\quad;\qquad d^{(3)}(\theta_{m})=-\sqrt{\mu}\sinh\frac{\theta_{m}}{2}\sqrt{\frac{\cosh\theta_{m}}{\sinh\theta_{m}}}f_{m}^{(3)}\label{eq:N33}
\end{equation}
while for the other  cases with  $i=1,2$
\begin{equation}
N_{mn}^{3i}=N_{nm}^{i3}=-\rho\frac{d^{(3)}(\theta_{m})d^{(i)}(\theta_{n})}{\sinh\frac{1}{2}(\theta_{m}-\theta_{n})}\quad;\qquad d^{(i)}(\theta_{n})=\frac{\sqrt{\mu}}{\alpha_i}
\cosh\frac{\theta_{n}}{2}\sqrt{\frac{\cosh\theta_{n}}{\sinh\theta_{n}}}f_{n}^{(i)}
\end{equation}

\begin{equation}
\label{e.Niii}
N_{mn}^{ij}=\rho\frac{d^{(i)}(\theta_{m})d^{(j)}(\theta_{n})}{\cosh\frac{1}{2}(\theta_{m}-\theta_{n})}\qquad;\qquad i,j=1,2
\end{equation}
The $d^{(r)}(\theta)$ are closely related to the Neumann vectors, but are more 
convenient in the following. 
We now analyze the expressions $d^{(r)}(\theta)$ one by one by starting with $d^{(3)}(\theta)$. 
We recall from \cite{LSNS} that\footnote{There is also an extra factor $(-1)^{m+1}$ for $d^{(1)}$, which 
we choose to be $1$ to unify the notation for $d^{(1)}$ and $d^{(2)}$.} 
\eq
f_{m}^{(3)}=\frac{\sqrt{m}}{\pi}\sin(m\pi\alpha_{2})\frac{e^{\tau_{0}(\mu-\omega^3_{m})}}{\omega^3_{m}}\frac{\Gamma_{\mu\alpha_{1}}(m\alpha_{1})\Gamma_{\mu\alpha_{2}}(m\alpha_{2})}{\Gamma_{\mu}(m)}M(0^{+})
\eqx
where 
\eq
\tau_{0}=\alpha_{1}\log\alpha_{1}+\alpha_{2}\log\alpha_{2}=\alpha_{1}\log\mu\alpha_{1}+\alpha_{2}\log\mu\alpha_{2}-\log\mu
\eqx
and $\Gamma_\mu(z)$ is defined in Appendix B. We move $e^{\tau_{0}\mu}M(0^{+})$
to $\rho$, as it appears in all $f$ coefficients. Using the renormalized
deformed $\tilde{\Gamma}_\mu$ functions introduced in Appendix B, together with
the rapidity parametrization, we obtain
\eq
d^{(3)}(\theta_{m})=-\frac{\sinh\frac{\theta_{m}}{2}\sin(\mu\pi\alpha_{2}\sinh\theta_m)}{\pi\sqrt{\cosh\theta_{m}}}
\frac{\tilde{\Gamma}_{\mu\alpha_{1}}(\theta_{m})\tilde{\Gamma}_{\mu\alpha_{2}}(\theta_{m})}{\tilde{\Gamma}_{\mu}(\theta_{m})}
\eqx
After a similar manipulation on $d^{(i)}(\theta)$ for $i=1,2$ we
can turn
\eq
f_{n}^{(i)}=\frac{e^{\tau_{0}(\mu+\omega_{\frac{n}{\alpha_{i}}})}\alpha_{i}}{\omega_{\frac{n}{\alpha_{i}}}\sqrt{n}\alpha_{1}\alpha_{2}}\frac{\Gamma_{\mu}(\frac{n}{\alpha_{i}})}{\Gamma_{\alpha_{2}\mu}(\frac{\alpha_{2}n}{\alpha_{i}})\Gamma_{\alpha_{1}\mu}(\frac{\alpha_{1}n}{\alpha_{i}})}M(0^{+})
\eqx
with $ \omega_z=\sqrt{z^2+\mu^2} $ into the expression 
\eq
d^{(i)}(\theta_{n})=\frac{1}{2\mu\sqrt{\alpha_{i}}\alpha_1\alpha_{2}\sqrt{\cosh\theta_{n}}\sinh
\frac{\theta_{n}}{2}}\frac{\tilde{\Gamma}_{\mu}(\theta_{n})}{\tilde{\Gamma}_{\alpha_{2}\mu}(\theta_{n})\tilde{\Gamma}_{\alpha_{1}\mu}(\theta_{n})}
\eqx
These expressions together with equations (\ref{eq:N33})-(\ref{e.Niii}) provide
the exact finite $L_{i}$ expressions for the Neumann coefficients.

\subsection{The decompactification limit of the LSNS formulas and their analyticity properties}
\label{s.lsnsdecomp}

In order to make contact with the SFT vertex axioms introduced in section~\ref{s.decomp},
let us take the same decompactification limit, in which 
we send $L_{3},L_{2}\to\infty$ and keep $L_{1}=L_{3}-L_{2}$ finite. 
Moreover we will also keep the mass of the scalar field and the particle rapidities fixed.
This entails sending also $\mu$ and the integer mode number $m$ to $\infty$, such that 
\eq
\f{2\pi \mu}{L_3} \equiv M  \quad\quad  \f{2\pi m_i}{|\al_i| L_3} \equiv M \sinh \th
\eqx
are kept fixed. In this limit
\eq
\alpha_{1}=\frac{L_{1}}{L_{3}}\to0\quad;\qquad\alpha_{2}\to1
\eqx
while the quantities 
\eq
\mu\alpha_{1}=\frac{ML_{1}}{2\pi}\quad;\qquad m_3\alpha_{1}=
\frac{pL_{1}}{2\pi}=\frac{ML_{1}\sinh\theta}{2\pi}=\mu\alpha_{1}\sinh\theta
\eqx
stay finite. In the following formulas we will drop the subscript in $L_1$ and use the notation $L \equiv L_1$ 
as in the SFT vertex axioms of section~\ref{s.decomp}.

The key quantities appearing in the Neumann coefficient formulas (\ref{eq:N33})-(\ref{e.Niii}) now have the
following finite decompactified limits\footnote{  We choose again the factor $(-1)^{m+1}$ to be $1$.}:
\eq
d^{(3)}(\theta)=-\frac{\sinh\frac{\theta}{2}\sin \f{p L}{2}}{\pi\sqrt{\cosh\theta}} \cdot
\tilde{\Gamma}_{\f{ML}{2\pi}}(\theta)  \cdot e^{-\f{\th}{2\pi}{pL}}
\eqx
and 
\eq
d^{(2)}(\theta)=\frac{\pi}{ML \sqrt{\cosh\theta}\sinh\frac{\theta}{2}}
\cdot
\f{1}{\tilde{\Gamma}_{\f{ML}{2\pi}}(\theta)}
\cdot
e^{\f{\th}{2\pi} pL}
\eqx
where  here and from now on $p=M \sinh \th$.
We must still address, however, one minor detail. In the decompactified case, the external states are 
conventionally normalized to a Dirac delta function in rapidities, while the finite volume mode states
are normalized to Kronecker deltas in mode numbers. So we have to factor out the $1/\sqrt{\cosh\th}$
terms\footnote{Since in any case we are not controlling the overall normalization here, we absorb
any remaining $1/\sqrt{M}$ factors in the normalization $\rho$.} into the Jacobian.
This yields finally the decompactified expressions in the natural infinite volume normalization:
\begin{align}
\label{e.d3decomp}
d^{(3)}(\theta)&=-\sin \f{p L}{2} \cdot \frac{\sinh\frac{\theta}{2}}{\pi}
 \cdot
\tilde{\Gamma}_{\f{ML}{2\pi}}(\theta)  \cdot e^{-\f{\th}{2\pi}pL} \\
d^{(2)}(\theta) &=\frac{\pi}{ML \sinh\frac{\theta}{2}}
\cdot
\f{1}{\tilde{\Gamma}_{\f{ML}{2\pi}}(\theta)}
\cdot
e^{\f{\th}{2\pi} pL}
\label{e.d2decomp}
\end{align}

It is important to note that the above expressions contain an infinite set of
exponential wrapping corrections w.r.t. the size of the string \#1, i.e. terms
of the form $e^{-n ML}$. Later we will describe the asymptotic limit defined
by neglecting these exponential corrections which turns out to be surprisingly
subtle.

In appendix~\ref{s.appdecomp} we will directly formulate the continuity conditions
for the decompactified SFT vertex for the massive free scalar and check that the above
limit of the LSNS expression (\ref{e.d3decomp}) is indeed a solution. This is important to make sure 
that the puzzling terms like $\sin \f{pL}{2}$ appearing in the Neumann coefficients
indeed exist directly for the decompactified vertex and do not arise from 
some unknown subtlety in finite volume reduction.

We can now verify that the decompactified Neumann coefficients, defined through (\ref{eq:N33})-(\ref{e.Niii})
in terms of (\ref{e.d3decomp})-(\ref{e.d2decomp}) satisfy the SFT vertex axioms of section~\ref{s.decomp}. 
In the present case the symmetry (\ref{e.symmetry}) is satisfied trivially and we are left with 
checking the monodromy (\ref{e.monodromy}), crossing
(\ref{e.crossi})-(\ref{e.crossii}) and the kinematical singularity axioms (\ref{e.kinematical}).

The monodromy property is seen to be easily implemented in terms of the last factor in (\ref{e.d3decomp})
as $\tilde{\Gamma}_{\f{ML}{2\pi}}(\theta)$ is $2\pi i$-periodic. Note that the additional signs generated 
by $\sinh\frac{\theta}{2}$ get canceled by signs coming from the denominators of (\ref{eq:N33})-(\ref{e.Niii}).

Using the crossing property of the deformed gamma functions 
\eq
\tilde{\Gamma}_{\f{ML}{2\pi}}(\theta+i\pi)\sinh\theta \sin \f{p L}{2}
= -\frac{2\pi^2}{\tilde{\Gamma}_{\f{ML}{2\pi}}(\theta) ML}
\eqx
one can see the crossing relation between $d^{(3)}$ and $d^{(2)}$ is 
\eq
d^{(3)}(\theta\pm i\pi) = \mp ie^{\pm ip(\theta)\frac{L}{2}}d^{(2)}(\theta)
\eqx
Inserting this into (\ref{eq:N33})-(\ref{e.Niii}) we get in particular the crossing properties {\bf }
\begin{align}
N^{32}(\th,\th')=e^{ip(\th')L/2} N^{33}(\th,\th'-i \pi) \quad\quad
N^{22}(\th,\th')=e^{ip(\th)L/2} N^{32}(\th-i\pi,\th')
\end{align}

Finally let us consider the kinematical singularity axiom (\ref{e.kinematical}) for $N^{33}(\th,\th')$.
It is seen to be satisfied using the property
\eq
\label{e.d3kin}
d^{(3)}(\theta + i\pi) d^{(3)}(\theta ) = -\f{1}{2ML} \left(1-e^{ip L} \right)
\eqx
An analogous property for $N^{22}(\th,\th')$ follows from 
\eq
\label{e.d2kin}
d^{(2)}(\theta + i\pi) d^{(2)}(\theta ) = \f{1}{2ML} \left(1-e^{-ip L} \right)
\eqx

Let us note some important features of the analytical properties of the (decompactified) Neumann coefficients in 
the complex rapidity plane. A notable feature of the functional equations (\ref{e.d3kin}) and (\ref{e.d2kin})
is that they are (almost) identical, while the explicit solutions (\ref{e.d3decomp}) and (\ref{e.d2decomp})
are clearly quite different. The difference lies in the location of zeroes in the physical strip. Due to the
factor $\sin \f{pL}{2}$, all the zeroes of $d^{(3)}(\th)$ lie on the line of \emph{real} $\th$'s, while
in the case of $d^{(2)}(\th)$, they lie on the line $\Im m(\th)=\pi$. This directly carries over to the different
location of zeroes in the Neumann coefficients $N^{rs}(\th,\th')$. Note that the physical difference between
strings \#2 and \#3 is that string \#2 is accompanied by the emission of string \#1.
Thus the Neumann coefficient of string \#3 vanishes exactly at the rapidities which are allowed by the asymptotic BA
equations for the finite size string \#1. We do not have currently a physical understanding of this property but expect similar features to occur 
for the generic interacting case.

\subsection{Asymptotic limit}
\label{s.asympt}

Let us now describe the asymptotic large $L$ limit of the (decompactified) Neumann coefficients
or equivalently of the elementary $d^{(r)}(\th)$ functions defined in (\ref{e.d3decomp}) and (\ref{e.d2decomp}).
Recall that $L$ is the size of the third finite string, and the asymptotic limit is defined
by neglecting all exponential $e^{-ML}$ corrections. Since this corresponds exactly to neglecting
wrapping corrections, such a limit is of chief interest for the subsequent reconstruction
of the finite volume SFT vertex, which we describe in section~\ref{s.program}, and for
potential applications to OPE coefficients in $\nn=4$ SYM theory.
Incidentally, this was also exactly the relevant limit used when comparing
pp-wave SFT vertex with perturbative OPE coefficients of BMN operators.

To this end, let us quote the large $ML$ asymptotics (\ref{e.Gammamuas}) of $\tilde{\Gamma}_{\f{ML}{2\pi}}(\theta)$
which follows from the properties derived in the LSNS paper \cite{LSNS}.
\eq
\tilde{\Gamma}_{\f{ML}{2\pi}}(\theta) \sim \sqrt{\f{2\pi^2}{ML}} 
\f{e^{\f{\th}{2\pi} pL}}{\sinh \f{\th}{2}}
\eqx
It is extremely important to emphasize that the above formula holds only on an 
open subset $|\Im m( \th)|<\pi$. In particular, it does not hold on the `crossing line' $\Im m( \th)=\pi$.
A very intriguing feature of the above expression is that it has a monodromy
when $\th \to \th+2\pi i$ which is in apparent contradiction with the $2\pi i$ periodicity of 
$\tilde{\Gamma}_{\f{ML}{2\pi}}(\theta)$. Of course, there is no real contradiction
due to the fact that this asymptotic formula breaks down on the line $\Im
m( \th)=\pi$.

The above mentioned apparent monodromy has, however, very important consequences
for the behaviour of the asymptotic Neumann coefficients. It cancels exactly
the explicit monodromies in (\ref{e.d3decomp}) and (\ref{e.d2decomp}) and one obtains
\begin{align}
d^{(3)}(\th)_{asympt} &=- \sqrt{\f{2}{ML}} \cdot \sin \f{p L}{2} \\
d^{(2)}(\th)_{asympt} &= \f{1}{\sqrt{2ML}}
\end{align}
This leads to the following asymptotic Neumann coefficients\footnote{Recall that we factor out
some constant normalization.} (valid for $|\Im m(\th)|<\pi$ and\\ $|\Im m(\th')|<\pi$):
\begin{align}
\label{e.asN33}
N^{33}(\th,\th')_{asympt} &=-\f{2}{ML} \f{\sin \f{p L}{2} \sin \f{p' L}{2}}{
\cosh\frac{\th-\th'}{2}} \\
N^{32}(\th,\th')_{asympt} &=\f{1}{ML} \f{\sin\f{p L}{2}}{\sinh\frac{\th-\th'}{2}} \\
N^{22}(\th,\th')_{asympt} &=\f{1}{2ML} \f{1}{\cosh\frac{\th-\th'}{2}}
\label{e.asN22}
\end{align}
Despite the simplicity of the above expressions, one should keep in mind that 
they are in fact equivalent to the \emph{all order}
$1/\mu$ formulas in the pp-wave SFT vertex.

The asymptotic expressions (\ref{e.asN33})-(\ref{e.asN22}) are quite intriguing. Firstly, we loose the nontrivial monodromy
of the exact formulas (\ref{e.d3decomp}) and (\ref{e.d2decomp}) and obtain simple antiperiodic functions. 
We believe that this property may be
necessary in the general case for solving the consistency equation (\ref{e.consistency}) when
using the decompactified formulas for constructing the finite volume SFT vertex up to
wrapping corrections.
Secondly, if we were to extend the above asymptotic formulas by analytical continuation
to the whole complex plane, the kinematical singularity axiom and crossing property would
be modified. The case of this effective asymptotic crossing is particularly intriguing
as we get e.g.
\eq
N^{33}(\th, \th' - i\pi)_{asympt} = -2i \sin \f{p' L}{2} N^{32}( \th, \th')_{asympt}
\eqx
which bears quite striking resemblance to the recently discovered modifications
of crossing in Chern-Simons theories \cite{MINWALLA1,MINWALLA2}.

We do not want to make here any statement about the effective asymptotic crossing and
kinematical axioms in the general interacting case and leave this problem
for future investigation.

\subsection{Reconstruction of $\hat{\Gm}_\mu(\th)$ from the SFT axioms}

In the final part of this section, let us see how to reconstruct the known LSNS solution directly from solving
the SFT vertex axioms of section~\ref{s.decomp}. Firstly, we will see that  obtaining the
solution in this way is very simple, and definitely much simpler than the direct approach 
of \cite{HSSV} and \cite{LSNS}. Secondly, we will see that by themselves, the functional
equations are not restrictive enough and one needs additional input about
the analytical structure, in particular the location of zeroes, in order
to fix the solution.

Let us concentrate on the $N^{33}(\th,\th')$ Neumann coefficient
which is equal to our function $\Nfin_{\bullet,L}(\th,\th')$
with vacuum on string \#1 (i.e. $\bullet \equiv \varnothing $ here).
The functional equations in this case read
\begin{align}
\Nfin_{\varnothing,L}(\th,\th') &= \Nfin_{\varnothing,L}(\th',\th) \\
\Nfin_{\varnothing,L}(\th+2\pi i,\th') &= e^{-ipL} \Nfin_{\varnothing,L}(\th,\th') \\
\Nfin_{\varnothing,L}(\th+i\pi+\eps,\th) &= \f{i}{\eps} \bigl( 1-e^{ipL} \bigr)\Nfin_{\varnothing,L} + \oo{\eps^0}
\end{align}
and we take $\Nfin_{\varnothing,L}$ with no arguments to be equal to $1$ (i.e. we normalize
the answer w.r.t. taking the amplitude with vacuum on all three strings).
We further assume that the large real $\th$ asymptotics of the solution is bounded.

It is convenient to solve first the monodromy axiom by factoring out 
\eq
e^{-\f{\th}{2\pi} pL-\f{\th'}{2\pi} p'L}
\eqx
from $\Nfin_{\varnothing,L}(\th,\th') \equiv N^{33}(\th,\th')$. Also we may implement the kinematical singularity by
introducing a denominator $e^\th+e^{\th'}$, i.e.
\eq
N^{33}(\th,\th') =\f{e^{-\f{\th}{2\pi} p L-\f{\th'}{2\pi} p' L}}{e^\th+e^{\th'}} Q(\th,\th')
\eqx
The kinematical singularity axiom implies the following functional equation for $Q(\th,\th')$:
\eq
Q(\th+i\pi,\th)i e^{-\th} =e^{-ip \f{L}{2}}
- e^{ip \f{L}{2}} 
\eqx
Let us introduce a simple factorizable ansatz for $Q(\th,\th')$:
\eq
Q(\th,\th') = \f{2h(\th)}{1+e^{-\th}} \f{2h(\th')}{1+e^{-\th'}}
\eqx 
Then we have
\eq
\label{e.heq}
h(\th) h(\th+i\pi) =-\sinh \th \sin \f{pL}{2}
\eqx
We recover thus the functional equation for $1/\tilde{\Gamma}_{\f{ML}{2\pi}}(\theta)$ up to a normalization
factor.
It is nevertheless instructive to try to solve this directly in order to rederive the 
special function $\tilde{\Gamma}_\mu$ 
and also to understand its space of solutions. In particular,
we can verify directly that
\eq
\label{e.second}
\f{1}{\tilde{\Gamma}_{\f{ML}{2\pi}}(\theta+i\pi)} \propto \sinh \th \sin \f{pL}{2} \cdot \tilde{\Gamma}_{\f{ML}{2\pi}}(\theta)
\eqx 
is also a solution (up to an appropriate overall constant).

We can rewrite the functional equation (\ref{e.heq})
as 
\eq
h(p) h(-p)=-p \sin\f{pL}{2}
\eqx
and expand the sine into an infinite product $\sin \pi z=\pi z \prod_{n=1}^\infty \left(1-\f{z^2}{n^2}\right)$.
We can find a solution in terms of a product of elementary solutions solving
\eq
f(p) f(-p) =\left(1-\f{p^2L^2}{4\pi^2 n^2} \right)
\eqx
The right hand side has two zeroes and \emph{a-priori} we are free to distribute them either into the first or second
factor on the left hand side or into both of them. This corresponds to the choice whether the zeroes lie on the line
$\Im m ( \th)=0$ or $\Im m( \th)=\pi$. From the exact LSNS solution for $N^{33}(\th,\th')$ discussed in section~\ref{s.lsnsdecomp}, we see that both zeroes
should be on the real line. The simplest and most obvious solution $f(\th)=1-p L/(2\pi n)$ does not 
provide the right location of zeroes as it leads to zeroes lying on both of the two lines and we
need a slightly more involved factorization:
\eq
\left(\sqrt{M^2+ \f{4\pi^2 n^2}{L^2}} -E(\th) \right)\left(\sqrt{M^2+ \f{4\pi^2 n^2}{L^2}} -E(\th+i\pi) \right)
=\f{4\pi^2 n^2}{L^2}-p^2
\eqx 
where $E(\th)=M \cosh \th$. We are now free to include either of the two factors into $f(\th)$ and consequently
into $h(\th)$. If we choose the right hand factors for all $n$, the solution will not have zeroes on the real line.
The consistent choice of the left hand factor would conversely ensure that all the zeroes lie on the real axis
-- this choice will lead to the second solution (\ref{e.second}) relevant for $N^{33}(\th,\th')$.
However, we could have made different choices for any $n$ constructing many (nonphysical) solutions of (\ref{e.heq}).
This shows that the assumptions on the location of zeroes are of crucial importance.

Introducing appropriate exponential factors for convergence leads to the infinite product
representation of the $\hat{\Gamma}_\mu(\th)$ following from the formulas in Appendix~B.

\section{Axioms for the nondiagonal case}
\label{s.nondiag}

\begin{figure}[h]
\begin{centering}
\includegraphics[width=10cm]{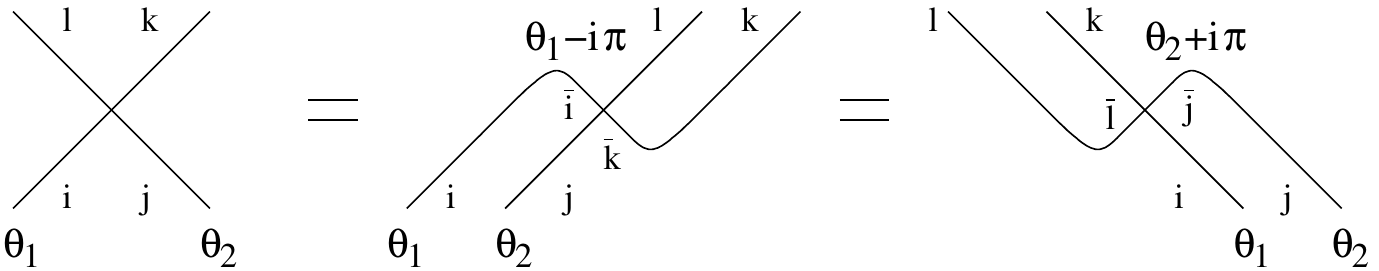}
\par\end{centering}
\caption{Crossing transformation of the scattering matrix.}
\label{fig.Scross}
\end{figure}

In this section we formulate axioms for the SFT vertex in the case
when the integrable worldsheet theory contain particles of different
types: additionally to the rapidity, $\theta,$ the particles are
characterized also by their particle type: $i$. In general, the scattering
matrix is non-diagonal, but due to integrability, the  multiparticle scatterings
factorize into two particle scatterings, and the S-matrix satisfies crossing
symmetry: 
\eq
S_{ij}^{kl}(\theta_{1},\theta_{2})=C^{k\bar{k}}S_{j\bar{k}}^{l\bar{i}}(\theta_{2},\theta_{1}-i\pi)C_{\bar{i}i}
=C^{l\bar{l}}S_{\bar{l}i}^{\bar{j}k}(\theta_{2}
+i\pi,\theta_{1})C_{\bar{j}j}
\eqx
where $C_{ij}$ is the charge conjugation matrix and its inverse is
$C^{jk}$ : $C_{ij}C^{jk}=\delta_{i}^{k}$. The crossing transformation
connects anti-particles in the initial states to particles in the
final state and vica versa. The graphical representation of the crossing
symmetry of the scattering matrix is demonstrated on Figure \ref{fig.Scross}.

\begin{figure}[h]
\begin{centering}
\includegraphics[height=7cm]{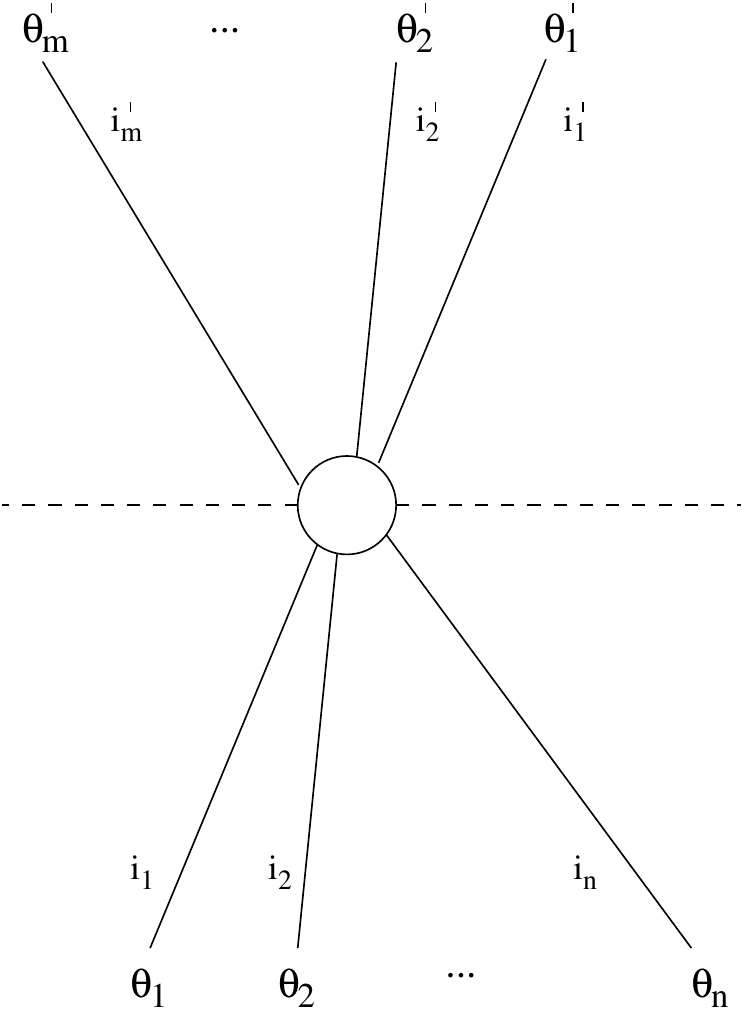}
\par\end{centering}
\caption{SFT vertex in the generic case. The initial state contains particles
with rapidities $\theta_{1},\dots,\theta_{n}$ and particle content
$i_{1},\dots,i_{n}$, while the final state, has rapidities $\theta_{1}',\dots,\theta_{m}'$
and particle content $i_{1}',\dots,i_{m}'$ . }
\label{fig.Nnondiag}
\end{figure}

The decompactified SFT vertex, additionally to the rapidities, depends
also on the types of the particles which we denote as
\eq
N_{\bullet,L}^{3\vert2}(\theta_{1},\dots,\theta_{n}\vert\theta'_{1},
\dots,\theta_{m}')_{i_{1},\dots,i_{n}}^{i_{1}',\dots,i_{m}'}
\eqx
where $\theta_{1},\dots,\theta_{n}$ are the rapidities of the initial
state in domain \#3 with particle content $i_{1},\dots,i_{n}$, while
the final state, in domain \#2, has rapidities $\theta_{1}',\dots,\theta_{m}'$
and particle content $i_{1}',\dots,i_{m}'$ . The placement of the
indices and their orderings reflect the geometry of the amplitude
as shown on Figure \ref{fig.Nnondiag}.

\begin{figure}[h]
\begin{centering}
\includegraphics[height=6cm]{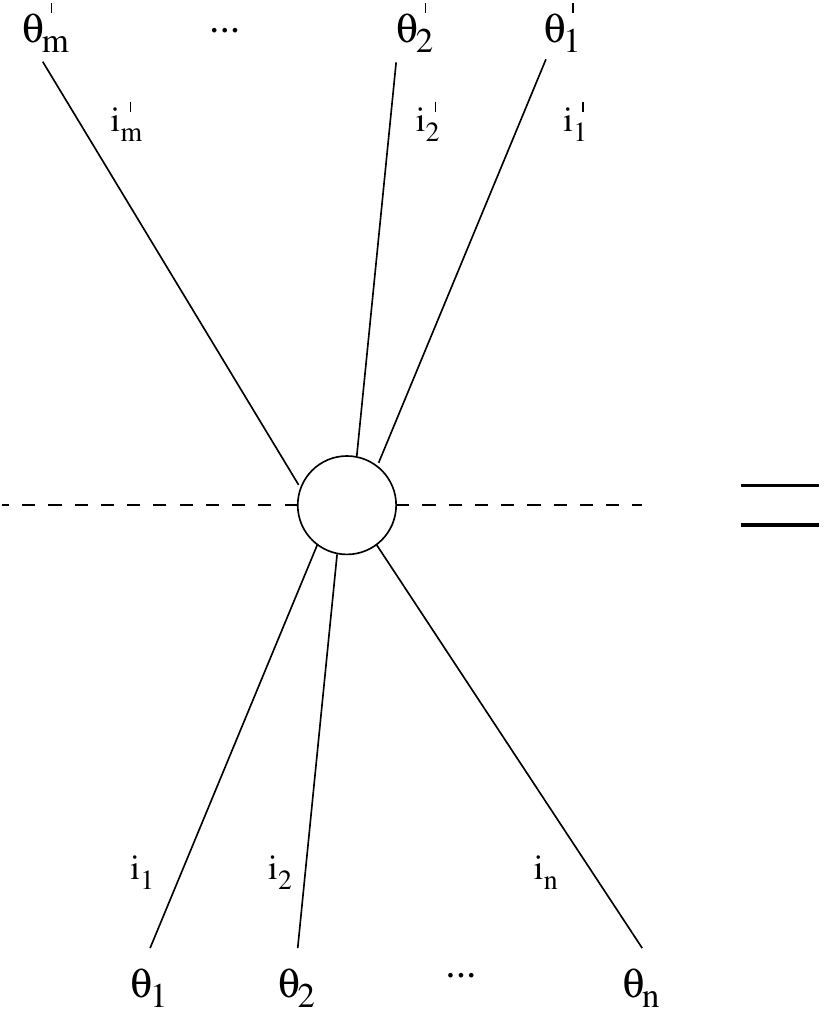}~~
\includegraphics[height=6cm]{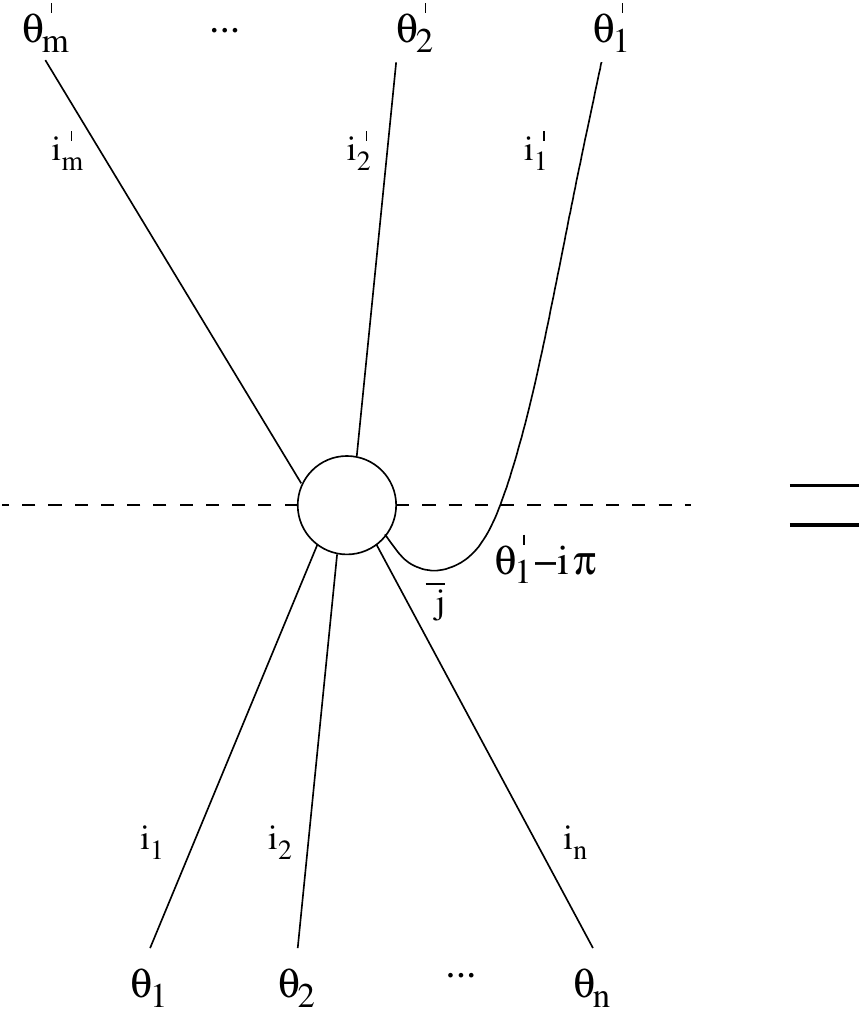}~
~\includegraphics[height=6cm]{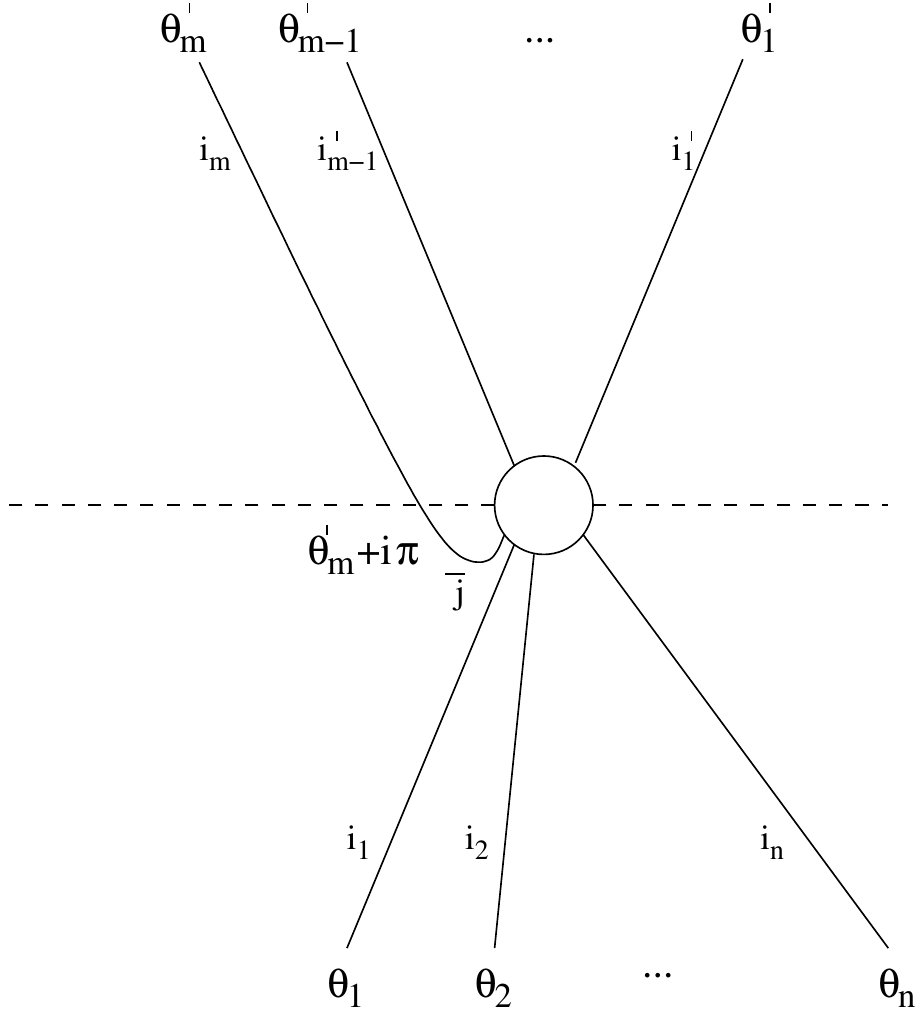}
\par\end{centering}
\caption{Graphical representation of the crossing transformation in the nondiagonal
case. Crossing a particle from outgoing to incoming comes with a charge
conjugation matrix, which replaces the particle with its antiparticle. }
\label{fig.Nndcross}
\end{figure}

\begin{figure}[h]
\begin{centering}
\includegraphics[width=5cm]{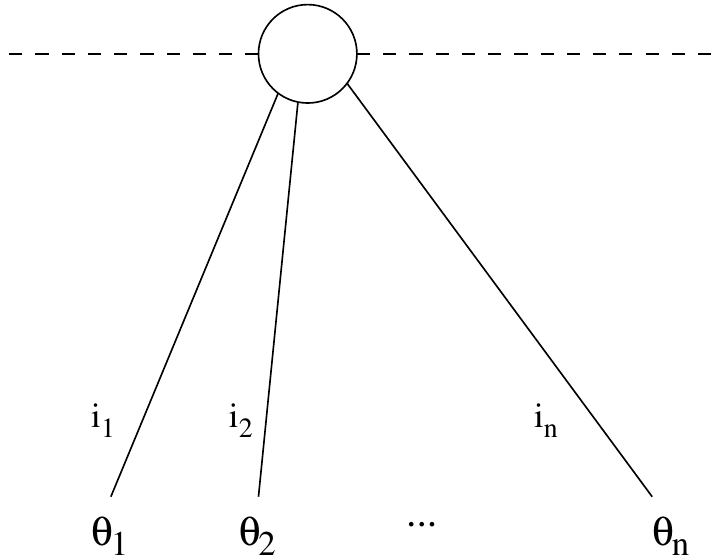}
\par\end{centering}
\caption{The elementary SFT vertex in the generic case.}
\label{fig.Nndelem}
\end{figure}

By keeping the previous convention we distribute the space discontinuity
equally on the border of domains \#3 and \#2. The space deficiency
has no effect on the particles type so the generalized crossing relations
take the form
\eq
N_{\bullet,L}^{3\vert2}(\theta_{1},\dots,\theta_{n}\vert\theta'_{1},\dots,
\theta_{m}')_{i_{1},\dots,i_{n}}^{i_{1}',\dots,i_{m}'}=
e^{ip(\theta'_{1})L/2}N_{\bullet,L}^{3\vert2}(\theta_{1},\dots,\theta_{n},
\theta'_{1}-i\pi\vert\theta'_{2},\dots,\theta_{m}')_{i_{1},\dots,i_{n},\bar{j}}^{i_{2}',
\dots,i_{m}'}C^{\bar{j}i_{1}'}
\eqx
\eq
N_{\bullet,L}^{3\vert2}(\theta_{1},\dots,\theta_{n}\vert\theta'_{1},\dots,
\theta_{m}')_{i_{1},\dots,i_{n}}^{i_{1}',\dots,i_{m}'}=
e^{-ip(\theta'_{m})L/2}N_{\bullet,L}^{3\vert2}(\theta'_{m}-i\pi,\theta_{1},
\dots,\theta_{n}\vert\theta'_{1},\dots,\theta_{m-1}')_{\bar{j},i_{1},\dots,i_{n}}^{i_{1}',
\dots,i_{m-1}'}C^{\bar{j}i_{m}'}
\eqx
Graphically they can be represented as we show on Figure \ref{fig.Nndcross}.
These crossing relations are valid if the incoming and outgoing particle
states have no overlaps, as otherwise singularities can appear. We
explain later these disconnected terms, which are related to amplitudes
with less particles. 
By crossing all particles into the initial state we can define the
elementary SFT vertex
\eq
N_{\bullet,L}(\theta_{1},\dots,\theta_{n})_{i_{1},\dots,i_{n}}=N_{\bullet,L}^{3\vert2}(\theta_{1},\dots,
\theta_{n}\vert\varnothing)_{i_{1},\dots,i_{n}}
\eqx
which we represent graphically on Figure \ref{fig.Nndelem}. 

\begin{figure}[h]
\begin{centering}
\includegraphics[height=4cm]{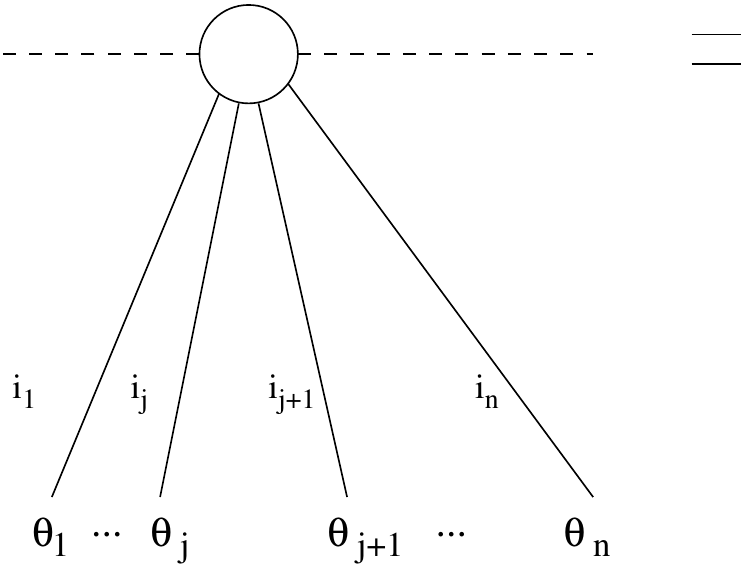}~~~~
~~~\includegraphics[height=4cm]{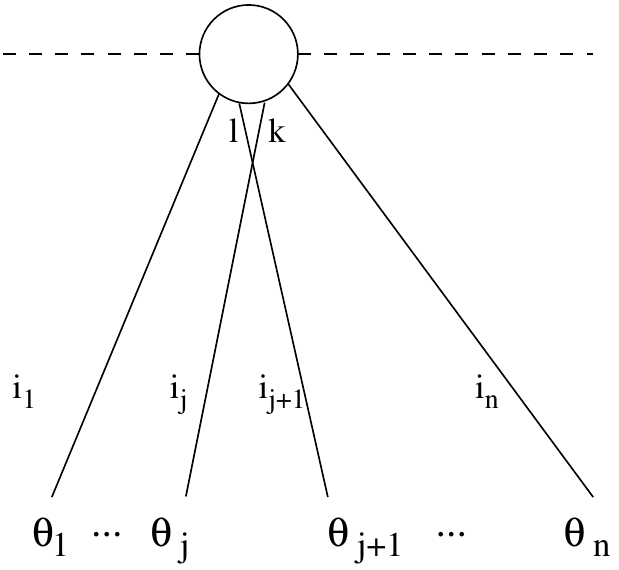}
\par\end{centering}
\caption{Permutation axiom for the SFT vertex. Exchanging two neighbouring particles
introduces an S-matrix factor.}
\label{fig.Nndperm}
\end{figure}

\begin{figure}[h]
\begin{centering}
\includegraphics[height=3.6cm]{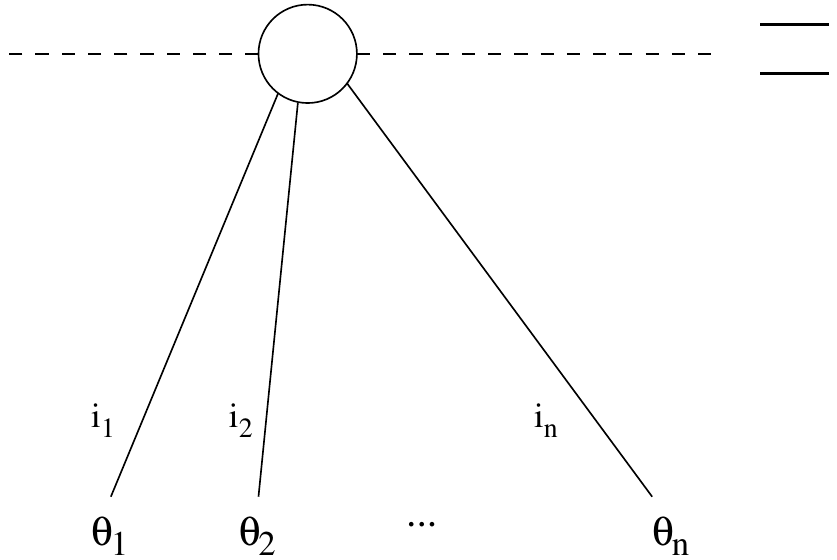} ~~~~ \includegraphics[height=4cm]{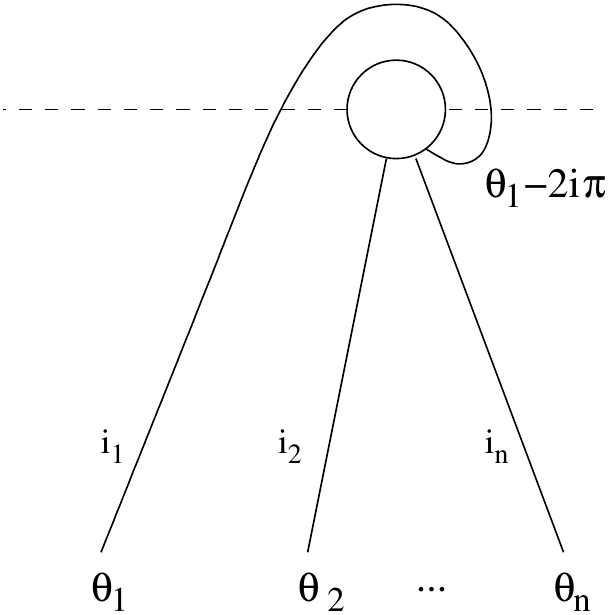}
\par\end{centering}
\caption{Monodromy property for the generic SFT vertex.}
\label{fig.Nndmon}
\end{figure}

This elementary SFT vertex satisfies several axioms. The permutation
axiom expresses that exchanging two particles comes with an S-matrix
factor:
\eq
N_{\bullet,L}(\theta_{1},\dots,\theta_{j},\theta_{j+1},\dots,\theta_{n})_{i_{1},\dots i_{j},i_{j+1},
\dots,i_{n}}=S_{i_{j}i_{j+1}}^{kl}(\theta_{j},\theta_{j+1})N_{\bullet,L}(\theta_{1},\dots,\theta_{j+1},\theta_{j},
\dots,\theta_{n})_{i_{1},\dots l,k,\dots,i_{n}}
\eqx
Graphically it takes the form shown on Figure \ref{fig.Nndperm}.
By crossing the leftmost incoming particle to an outgoing antiparticle
and crossing back again to the rightmost particle we obtain the monodromy
property
\eq
N_{\bullet,L}(\theta_{1},\dots,\theta_{n})_{i_{1},\dots,i_{n}}=e^{-ip(\theta_{1})L}N_{\bullet,L}
(\theta_{2},\dots,\theta_{n},\theta_{1}-2i\pi)_{i_{2},\dots,i_{n},i_{1}}
\eqx
which is shown on Figure \ref{fig.Nndmon}. 

\begin{figure}[h]
\begin{centering}
\includegraphics[width=6cm]{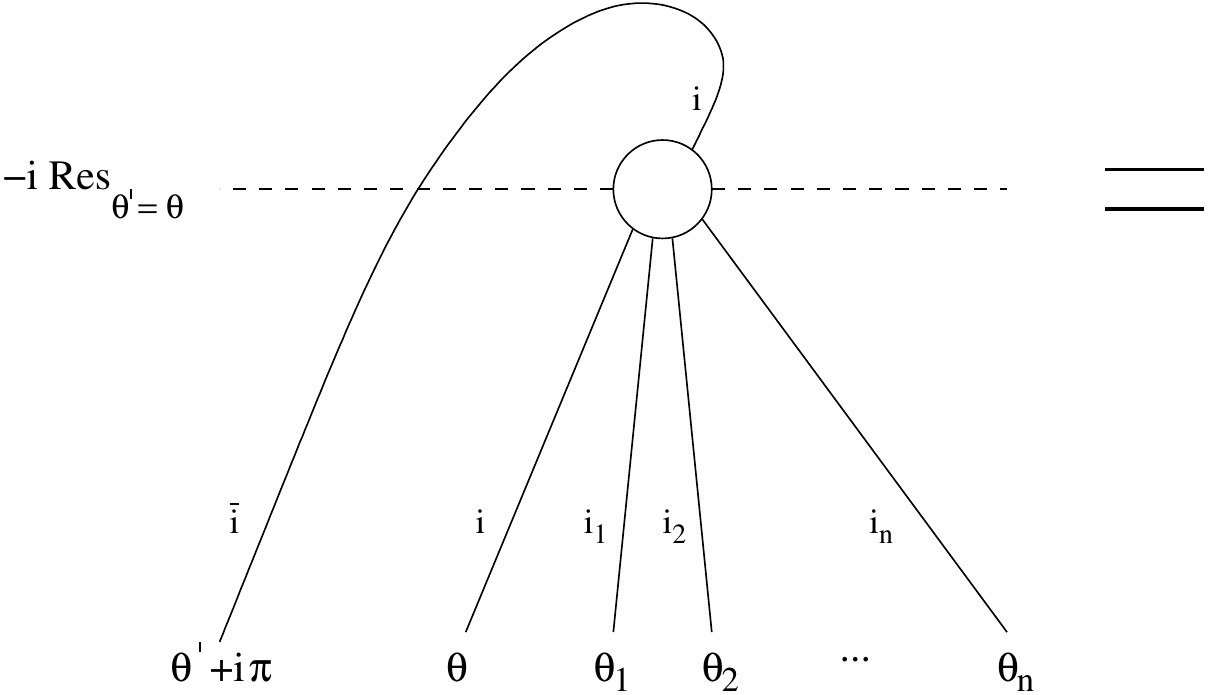}~~\includegraphics[width=5.5cm]{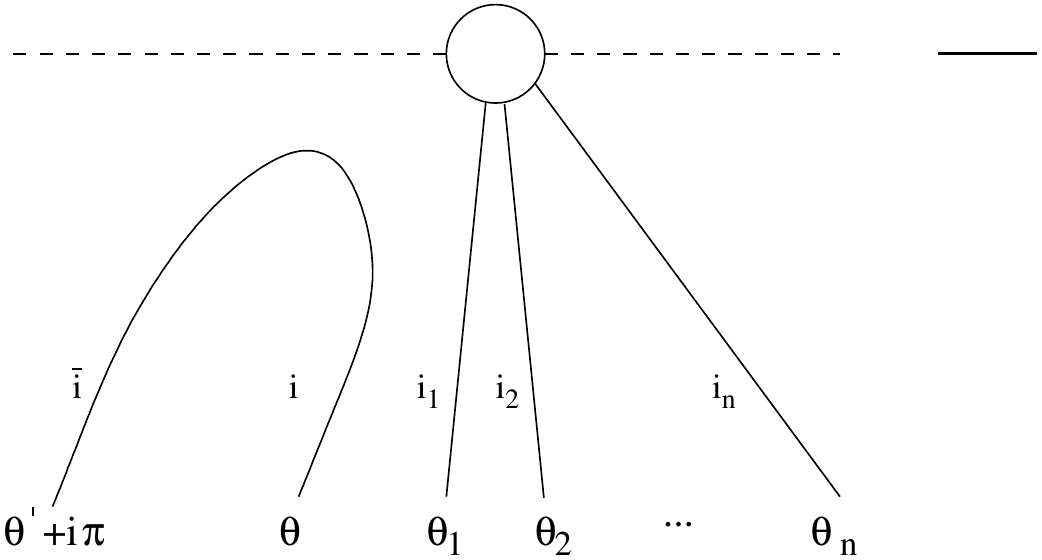}~
~\includegraphics[width=4.5cm]{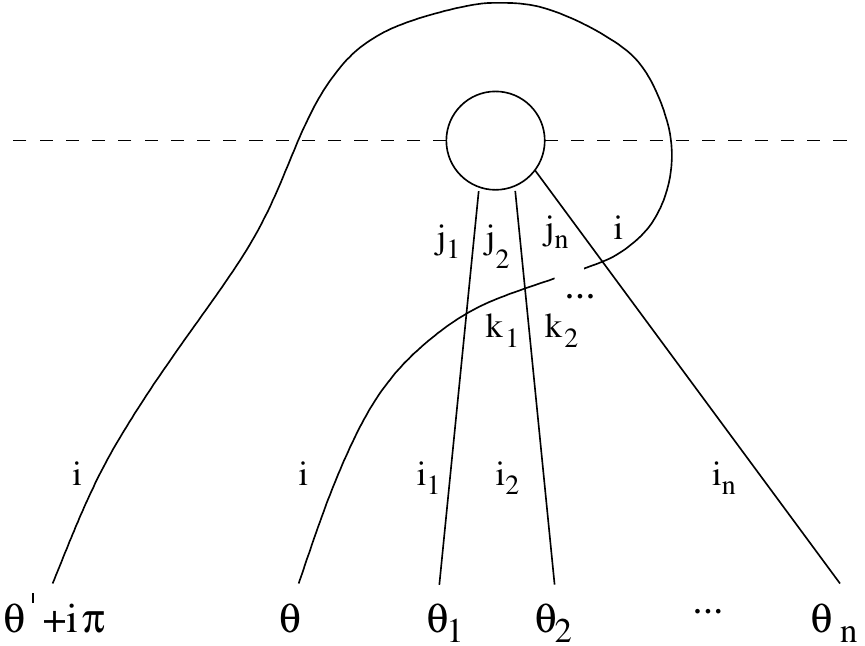}
\par\end{centering}
\caption{Kinematical singularity axiom for the generic SFT vertex.}
\label{fig.Nndkinsing}
\end{figure}

\begin{figure}[h]
\begin{centering}
\includegraphics[width=7cm]{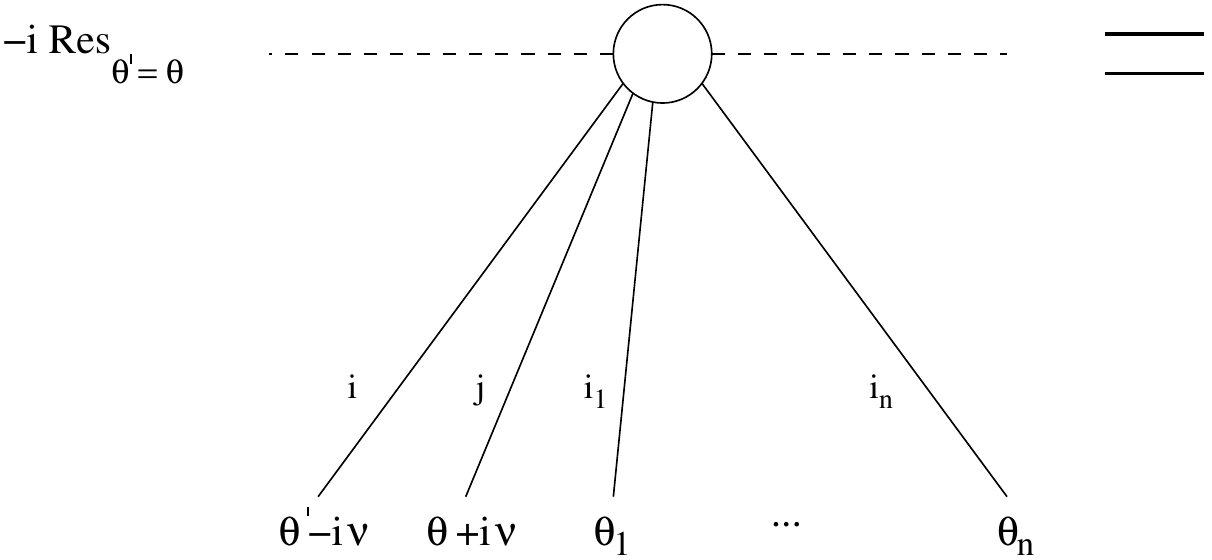}~~\includegraphics[width=4.5cm]{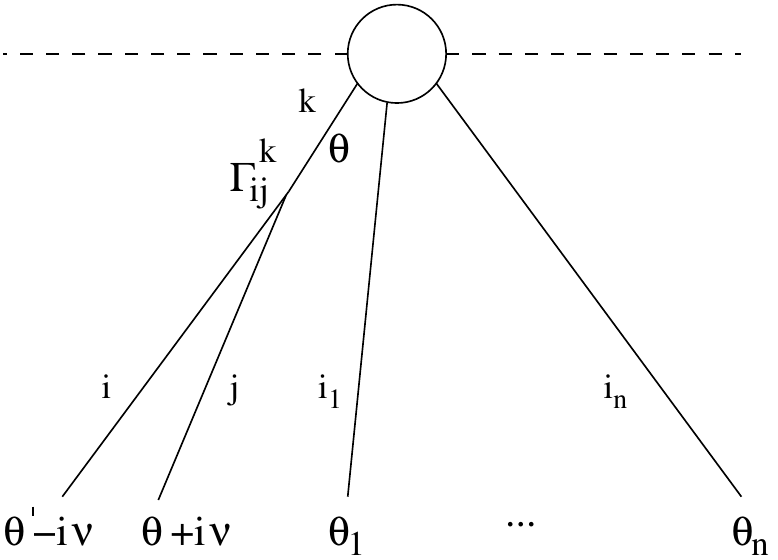}
\par\end{centering}
\caption{Dynamical singularity axiom for the SFT vertex.}
\label{fig.Nnddynsing}
\end{figure}

The crossing relation is valid if none of the incoming particles coincides
with any of the outgoing particles. Otherwise, the amplitude is singular,
but the residue of the pole is related to an onshell propagation of
the particle passing the reduced amplitude on both sides:
\eqn
-i\mbox{Res}_{\theta'=\theta}N_{\bullet,L}
(\theta'+i\pi,\theta,\theta_{1},\dots,\theta_{n})_{\bar{i},i,i_{1},\dots,i_{n}}\!\!&=&\!\!\!\!
(\delta_{i_{1}\dots i_{n}}^{j_{1}\dots j_{n}}-e^{ip(\theta)L}S_{ii_{1}}^{k_{1}j_{1}}
(\theta,\theta_{1})\dots S_{k_{n-1}i_{n}}^{ij_{n}}(\theta,\theta_{1}))\times\nonumber \\
&& N_{\bullet,L}(\theta_{1},\dots,\theta_{n})_{j_{1},\dots,j_{n}}
\eqnx
This axiom is called the kinematical singularity axioms, which connects
the amplitude with $n+2$ particles to an amplitude with $n$ particles.
This process is indicated on Figure \ref{fig.Nndkinsing}
Singularities of the SFT vertex always correspond to some kinematically
allowed onshell propagation of the particles. If, for example, two
particles with labels $i$ and $j$, with rapidities $\theta-i\nu$
and $\theta+i\nu$, can form a boundstate of type $k$ with rapidity
$\theta$, then the SFT vertex is singular and its residue is related
to the SFT vertex of the boundstate as 
\eq
-i\mbox{Res}_{\theta'=\theta}N_{\bullet,L}(\theta'-i\nu,\theta+i\nu,\theta_{1},\dots,\theta_{n})_{i,j,i_{1},\dots,i_{n}}=\Gamma_{ij}^{k}N_{\bullet,L}(\theta,\theta_{1},\dots,\theta_{n})_{k,j_{1},\dots,j_{n}}
\eqx
where $\Gamma_{ij}^{k}$ is the strength of the coupling, which is
related to the residue of the pole in the scattering matrix
\eq
-i\mbox{Res}_{\theta'=\theta}S_{ij}^{kl}(\theta'+i\nu,\theta-i\nu)=\Gamma_{ij}^{m}\Gamma_{m}^{kl}
\eqx
This singularity axiom is called the dynamical singularity axiom and
is represented graphically on Figure \ref{fig.Nnddynsing}. 
Generally, for any onshell propagation we have a singularity of the
SFT vertex. This is similar to how the singularities of the scattering
matrix can be explained by Coleman-Thun diagrams. 

In formulating the nondiagonal SFT vertex axioms we used rapidity 
parametrizations $\th$, with crossing transformations $\th \to \th \pm 
i \pi$, but we did not assume any relativistic invariance for the 
scattering matrix. The generalizations of these formulas for the AdS/CFT integrable model
can be obtained by using its rapidity parametrizations $\th \to z$
and its crossing transformations $z\to \pm \omega$. These axioms are very 
similar to the form factor axioms for world-sheet operators, \cite{KM1}, except the factor
$e^{ipL}$ appearing in the monodromy/periodicity equations.

\section{The program for the finite volume string vertex}
\label{s.program}

\begin{figure}
\begin{tabular}{ccccc}
\hspace{0cm}\includegraphics[height=3cm]{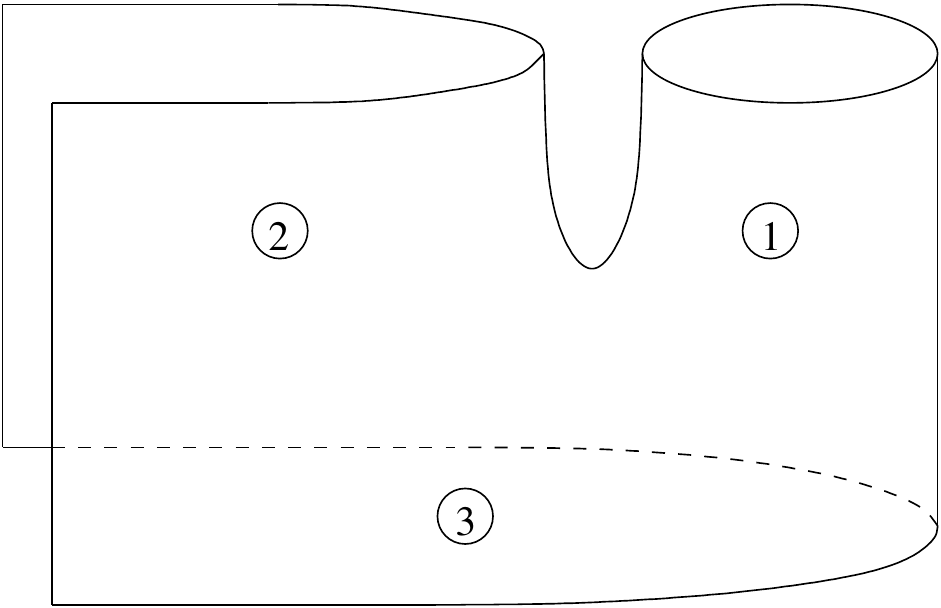} &  &  &  & \hspace{0cm}\includegraphics[height=3cm]{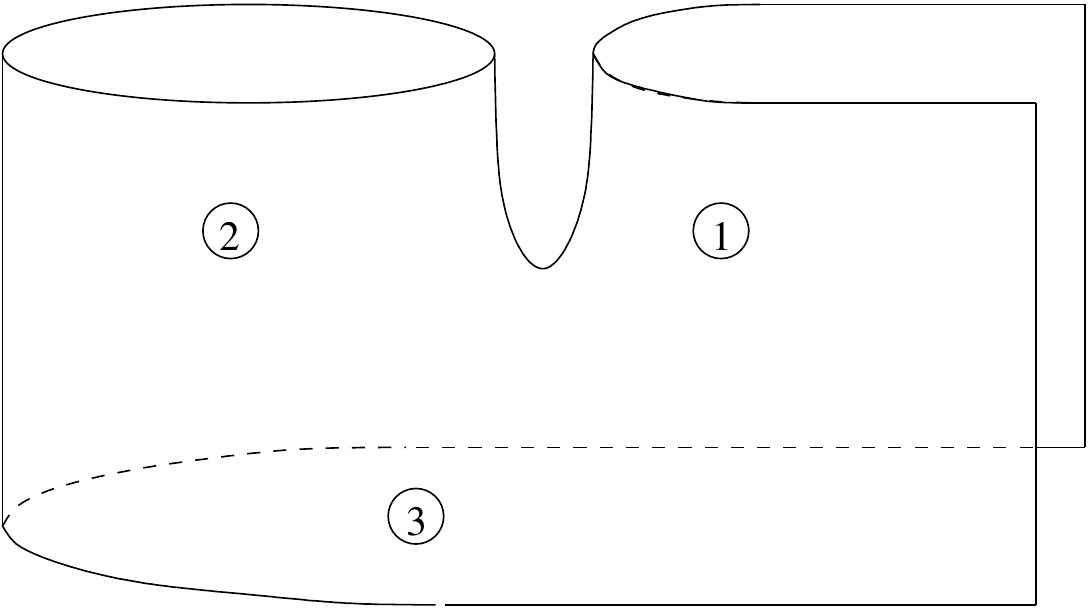}\tabularnewline
 & $\searrow$ &  & $\swarrow$ & \tabularnewline
 &  & \includegraphics[height=3cm]{sft1.pdf} &  & \tabularnewline
\end{tabular}
\caption{The program for obtaining the finite volume string field theory vertex 
up to wrapping corrections.\label{fig.program}}
\end{figure}

Let us now formulate our program for the general finite volume string field theory vertex
up to wrapping corrections. As explained before, we do not expect \emph{a-priori} an
exponential form of the vertex expressed in terms of some generalized Neumann coefficients\footnote{Although 
\emph{a-posteriori} a generalization might exist in analogy to similar structures for boundary states
in integrable relativistic QFT's \cite{Bndry}.} so what we are after is a general amplitude
with any prescribed multiparticle state on each string. In the limit that we are considering,
i.e. neglecting wrapping corrections, these states will be parametrized by momenta solving
Asymptotic Bethe Ansatz equations for each string individually\footnote{And of course the standard additional
Bethe roots associated to nesting. In the following, for brevity, we just explicitly indicate only the momenta.}.
We thus want to determine
\eq
\label{e.Nfin}
\Nfin^{3|2;1}_{L_3|L_2;L_1}\left( \{p^{(3)}_i\} \;\Bigl|\; \{p^{(2)}_j\}\,;\, \{p^{(1)}_k\} \right)
\eqx
where we explicitly indicated the sizes of the respective strings.

Our program consists of first solving the SFT functional equations derived in section~\ref{s.decomp} for
the two distinct decompactified versions of (\ref{e.Nfin}), namely
\eq
\label{e.decomp1}
\Nfin^{3|2}_{\{p^{(1)}_k\},L_1}\left( \{p^{(3)}_i\} \;\Bigl|\; \{p^{(2)}_j\}\right)
\eqx 
and
\eq
\label{e.decomp2}
\Nfin^{3|1}_{\{p^{(2)}_j\},L_2}\left( \{p^{(3)}_i\} \;\Bigl|\;  \{p^{(1)}_k\} \right)
\eqx 
It is important to note that in (\ref{e.decomp1}), the momenta $\{p^{(3)}_i\}$ and $\{p^{(2)}_j\}$
are assumed to be unconstrained and that the relevant SFT axioms involve explicitly \emph{only} these
momenta. There is no dependence in these axioms on the $\{p^{(1)}_k\}$, however there is certainly
a huge freedom in the choice of a particular solution which may depend on the $\{p^{(1)}_k\}$. 
This is in direct analogy to the case of form factors where the axioms do not depend on the particular
choice of local operator. They allow, however, for many solutions associated to different
choices of local operators. In the SFT vertex case we will soon show how to
strongly constrain this dependence.

Similarly, in the case of (\ref{e.decomp2}), the axioms involve explicitly only momenta
$\{p^{(3)}_i\}$ and $\{p^{(1)}_k\}$, and the particular solutions should \emph{a-priori}
depend on the $\{p^{(2)}_j\}$ this time.

Our main point is now that performing finite volume reduction from (\ref{e.decomp1}) and
(\ref{e.decomp2}) should yield exactly the same expression, which will be the original
quantity of interest (\ref{e.Nfin}). 

Since we are working only up to wrapping corrections, we should neglect\footnote{It is possible that
keeping the full decompactified solution in the following steps will yield lot of information
on some wrapping corrections for the OPE coefficients. However then the matching of
the finite volume reductions might require some care. We leave this possibility for future investigation.} wrapping
corrections (w.r.t. respectively $L_1$ and $L_2$) in the decompactified 
solutions (\ref{e.decomp1}) and (\ref{e.decomp2}). We denote the resulting asymptotic solutions
with a subscript ${}_{asympt}$ e.g.
\eq
\Nfin^{3|2}_{\{p^{(1)}_k\},L_1}\left( \{p^{(3)}_i\} \;\Bigl|\; \{p^{(2)}_j\}\right)_{\!\!asympt}
\eqx
This is the direct counterpart of the asymptotic formulas of section~\ref{s.asympt}
which were given for the case of the massive free boson (the pp-wave case).
Note that these expressions will typically still have some oscillatory $L_{1,2}$ dependence
in factors like $\sin \f{pL_1}{2}$.
Now again up to wrapping corrections, the finite volume reduction
should just amount to multiplying by the same factors as for finite volume form factors.
We thus get (up to wrapping corrections)
\eq
\Nfin^{3|2;1}_{L_3|L_2;L_1}\left( \{p^{(3)}_i\} \;\Bigl|\; \{p^{(2)}_j\}\,;\, \{p^{(1)}_k\} \right)
= \f{1}{\sqrt{\tilde{\rho}_3 \tilde{\rho}_2}} \Nfin^{3|2}_{\{p^{(1)}_k\},L_1}\left( \{p^{(3)}_i\} \;\Bigl|\; \{p^{(2)}_j\}\right)_{\!\!asympt}
\eqx
where $\tilde{\rho}$ is the same factor as in (\ref{e.fvff}), involving the Gaudin norm together with
a product of S-matrices.
It is important to note, that we could have equally well obtained the same expression from the second decompactification
\eq
\Nfin^{3|2;1}_{L_3|L_2;L_1}\left( \{p^{(3)}_i\} \;\Bigl|\; \{p^{(2)}_j\}\,;\, \{p^{(1)}_k\} \right)
= \f{1}{\sqrt{\tilde{\rho}_3 \tilde{\rho}_1}} \Nfin^{3|1}_{\{p^{(2)}_j\},L_2}\left( \{p^{(3)}_i\} \;\Bigl|\;  \{p^{(1)}_k\} \right)_{\!\!asympt}
\eqx
The consistency of these two expressions provides for us the key final equation
\eq
\label{e.consistency}
\f{1}{\sqrt{ \tilde \rho_2}} \Nfin^{3|2}_{\{p^{(1)}_k\},L_1}\left( \{p^{(3)}_i\} \;\Bigl|\; \{p^{(2)}_j\}\right)_{\!\!asympt}
\!=
\f{1}{\sqrt{ \tilde \rho_1}} \Nfin^{3|1}_{\{p^{(2)}_j\},L_2}\left( \{p^{(3)}_i\} \;\Bigl|\;  \{p^{(1)}_k\} \right)_{\!\!asympt}
\eqx
which should very strongly constrain the $\{p^{(1)}_k\}$-dependent choice of particular
solution of the SFT axioms for (\ref{e.decomp1}) and the $\{p^{(2)}_j\}$-dependent
choice of solution of the SFT axioms for (\ref{e.decomp2}).
The resulting expression yields the final finite volume amplitude (\ref{e.Nfin}).
We illustrate pictorially this strategy in figure~\ref{fig.program}.

\subsection{The program for the simplest plane wave SFT vertex\\ }

Let us see how to implement the program above for the simplest pp-wave SFT vertex, namely for the asymptotic value of 
$N^{33}(\th,\th ')$, which is denoted by 
\eq
N_{L_3\vert L_2,L_1}^{3\vert 2,1}\left(p,p'\bigl\vert \varnothing;\varnothing
\right)_{asympt}
\eqx
We use the previously calculated results but put into the context of the general program. 

As a first step we decompactify $L_2$ and $L_3$ and determine 
$N_{\varnothing \vert L_1}^{3\vert 2}(\th,\th '\vert \varnothing)$ from our axioms as
\eq
N_{\varnothing, L_1}^{3\vert 2}
(\th,\th '\vert \varnothing)=n(L_{1  })\sin \f{p L_1}{2}
 \sin \frac{p'L_1}{2}
 \frac{\sinh\frac{\theta}{2}\sinh\frac{\theta'}{2}}{\pi^2 \cosh\frac{1}{2}(\theta-\theta')}
 \cdot
\tilde{\Gamma}_{\f{ML_1}{2\pi}}(\theta) 
\tilde{\Gamma}_{\f{ML_1}{2\pi}}(\theta')  
\cdot e^{-\f{L_1}{2\pi}(p\th+p'\th')}
\label{N32L1}
\eqx
 where $n(L_1)$ is a normalization factor depending on the state in the finite
 string \#1.  

In the next step we take the $L_1\to \infty $ limit and neglect all exponentially small, $ e^{-mL_1}$ corrections:
\eq
N_{\varnothing,L_1}
^{3\vert 2}(\th,\th '\vert \varnothing)_{asympt}=n(L_{1  })\sin \f{p L_1}{2}
 \sin \frac{p'L_1}{2}\frac{1}{ \cosh\frac{1}{2}(\theta-\theta')}
\eqx

Now we repeat the same calculations for $L_2$. We decompactify $L_1$ and
 $L_3$ and solve the functional equations for 
$N_{\varnothing \vert L_2}^{3\vert 1}(\th,\th '\vert \varnothing) $. 
The result is the same as (\ref{N32L1}) but $L_1$ is exchanged with $L_2$. After taking the $L_2 \to \infty$ limit and neglecting wrapping corrections we obtain the asymptotic form: 
\eq
N_{\varnothing, L_2}^{3\vert 1}
(\th,\th '\vert \varnothing)_{asympt}=n(L_{2  })\sin \f{p L_2}{2}
 \sin \frac{p'L_2}{2}\frac{1}{ \cosh\frac{1}{2}(\theta-\theta')}
\eqx
Using that there are no particles in strings \#1  and \#2, thus $\rho_1=
\rho_2=1$, we demand that 
\eq
n(L_{1  })\sin \f{p L_1}{2}
 \sin \frac{p'L_1}{2}\frac{1}{ \cosh\frac{1}{2}(\theta-\theta')}=
n(L_{2  })\sin \f{p L_2}{2}
 \sin \frac{p'L_2}{2}\frac{1}{ \cosh\frac{1}{2}(\theta-\theta')} 
\eqx
Recall that $L_3=L_1+L_2$ and that both $p$ and $p'$ satisfy the 
asymptotic BA equations $e^{ipL_3}=1=e^{ip'L_3}$ together with the 
level matching condition $p=-p'$. This implies that 
\eq
\sin \f{p L_1}{2}
 \sin \frac{p'L_1}{2}=\sin \f{p (L_3-L_2)}{2}
 \sin \frac{p'(L_3-L_2)}{2}=
 \sin \f{p L_2}{2}
 \sin \frac{p'L_2}{2}
\eqx
and we are forced to take $n(L_1)=n(L_2)=n$. The asymptotic SFT vertex 
is then
\eq
N_{L_3\vert L_2,L_1}^{3\vert 2,1}\left(p,p'\bigl\vert \varnothing;\varnothing
\right)_{asympt}=
\frac{1}{\sqrt{\rho_1\rho_3}}N_{\varnothing,L_1}
^{3\vert 2}(\th,\th '\vert \varnothing)_{asympt}=\frac{n'}{\sqrt{\cosh\th\cosh\th'}}
\frac{\sin \f{p L_1}{2}
 \sin \frac{p'L_1}{2}}
{ \cosh\frac{1}{2}(\theta-\theta')}
\eqx
where we used that $\rho_1=1\ ; \ \rho_3=M^{2}L_1^{2}\cosh\th\cosh\th' $ and absorved $1/ML_1$
into the normalization $n'$. Clearly this answer agrees with 
the asymptotic form of the relevant pp-wave \emph{finite volume} Neumann coefficient.

\section{Weak coupling cross-checks with OPE coefficients}

In this section we  comment on  how the kinematical singularity axiom is
satisfied at weak coupling directly for the OPE coefficient.
More details can be found in Appendix C. We investigate
the 3-point functions in the $su(1\vert1)$ and $su(2)$ sectors
up to 1-loop based on the available explicit results obtained from
direct gauge theory calculations in \cite{Caetano,GVtheta}. As we do not know the exact relation between the 3-point
functions and the SFT vertex we  neglect the proper infinite volume normalization factors and 
 check the axioms only up to some proportionality factor. 

Even this comparision is not straightforward in two respects: Firstly, at weak coupling the kinematical domains of the crossed
amplitudes are infinitely far from each other and no longer  connected analytically. Secondly, the weak
coupling results were calculated for operators of finite sizes ($L_{i}$),
however our axioms are valid when two volumes $(L_{3}$ and $L_{2}$)
were sent  to infinity
by keeping the third volume ($L_{1}=L_{3}-L_{2}$)
finite. 

We can circumvent
these problems: first, by formulating the
kinematical singularity axioms directly for the crossed process, that
is, when we have the same type of particles in the initial state (operator
$\mathcal{O}_{3}$) and in the final state (operator $\mathcal{O}_{2}$),   and second, by   taking a careful
limit of the finite volume formulas. To spell out the details let us denote the momenta of the initial
state by $p_{j}^{(3)}$ and those of   the final state by $p_{i}^{(2)}$. 
The reformulation of the kinematical singularity axiom  into this  setting means that 
the OPE coefficient $C_{123}(\{p\})$ must have a pole whenever
$p_{n}^{(3)}=p_{m}^{(2)}$ with the residue
\begin{eqnarray}
-i\;\mbox{Res}\,C_{123}(\{p\})\propto
(1-e^{ip_{n}^{(3)}L_{1}}\prod_{j}^{N_{2}}S(p_{j}^{(2)},p_{n}^{(3)})
\prod_{k}^{N_{3}}S(p_{n}^{(3)},p_{k}^{(3)}))\times \nonumber \\
C_{123}(\{p\}
\setminus\{p_{n}^{(3)},p_{m}^{(2)}\})\label{eq:weaksingax}
\end{eqnarray}
where $S(p_{1},p_{2})$ is the scattering matrix, which we choose
to be diagonal. This motivates us to analyze the $su(2)$ and $su(1\vert1)$
closed diagonal subsectors of the theory. Observe that by sending $L_{3}$
and $L_{2}$ to infinity we went off-shell with the momenta, which
is crucial to have the singularity. For finite $L_{3}$ and $L_{2}$
volumes the BA equations 
\eq
e^{ip_{n}^{(3)}L_{3}}\prod_{j:j\neq n}S(p_{n}^{(3)},p_{j}^{(3)})=1\quad;
\qquad e^{ip_{m}^{(2)}L_{2}}\prod_{k:k\neq m}S(p_{m}^{(2)},p_{k}^{(2)})=1
\eqx
kill the singularity. Let us  emphasize that the power of the clear
analytical properties shows up only in the infinite volume limit. 

In the following we explain how to extract the infinite volume limit of the 3-point functions in the $su(1\vert 1)$  and $su(2)$ cases. In both cases
the explicit volume $(L_2$ or $L_3$) dependence comes from a term of the form
\eq
1-e^{-ip_{n}^{(3)}L_{2}}
\prod_{j:j\neq n }^{N_{2}}S(p_{j}^{(2)},p_{n}^{(3)})
\eqx
which prevents us to take the $L_2\to \infty$ limit. The idea is to use the Bethe Ansatz equation for  $p_{n}^{(3)}$ to replace the expression above with 
\eq
1- e^{ip_{n}^{(3)}L_{1}}\prod_{j}^{N_{2}}S(p_{j}^{(2)},p_{n}^{(3)})
\prod_{k}^{N_{3}}S(p_{n}^{(3)},p_{k}^{(3)})
\eqx
where $L_1=L_3-L_2$ is kept finite in the required limit, which now exists. Using this procedure and some renormalization of the states we obtained 
the infinite volume 3-point function in the two sectors as follows. 

\subsubsection*{The $su(1\vert1)$ sector}

The infinite volume limit of the OPE coefficient\footnote{More precisely of its complex conjugate w.r.t. the form written 
in~\cite{Caetano}.} in the $su(1\vert1)$
sector up to 1-loop takes the form:
\begin{equation}
C_{123}(\{p\})\propto\frac
{\prod_{r=1}^{2}\prod_{i<j}^{N_{r}}f(p_{i}^{(r)},
p_{j}^{(r)})}{
\prod_{i}^{N_{3}}\prod_{j}^{N_{2}}f(p_{i}^{(3)},p_{j}^{(2)})} 
 \prod_{l}^{N_{3}}
\left(1-
e^{
ip_{l}^{(3)}L_{1}}\prod_{i}^{N_{2}}S(p_{i}^{(2)},p_{l}^{(3)})
\prod_{j}^{N_{3}}S(p_{l}^{(3)},p_{j}^{(3)})\right)
\end{equation}
where the momenta $p^{(3)}$ and $p^{(2)}$ no longer satisfy any
quantization condition, $S$ denotes the scattering matrix in the $su(1\vert1)$
sector and $f$ is a known function, (displayed in Appendix~C),  whose explicit form is not relevant for us now, except that it vanishes at coinciding arguments.

In order to check the 
kinematical residue axiom at $p_{k}^{(3)}=p_{j}^{(2)}$
we calculate
\eq
\frac{-i\mbox{Res}_{p_{k}^{(3)}=p_{j}^{(2)}}C_{123}(\{p\})}{C_{123}(\{p\}
\setminus\{p_{k}^{(3)},p_{j}^{(2)}\})}
\eqx
The prefactor containing $f$ provides
 the required pole and its remaining part nicely cancels in the
ratio. The  product for $l$ not agreeing with $k$ contains
two extra terms in the numerator compared to the same term in 
the denominator, however they cancel by the unitarity of the scattering 
matrix. The $l=k$'th factor in the product exactly reproduces the 
term required by the kinematical singularity axiom.
More details are given in Appendix~C.

\subsubsection*{The $su(2)$ sector}

Here we explain the kinematical residue axioms at the tree level for the
$su(2)$ 3-pointfunction.  The 1-loop calculation is relegated to 
Appendix C. It is convenient to parametrize the momenta by the rapidities, $p(u)$, in terms of which the infinite volume limit 
 of the 3-point function takes 
the form
\eq
C_{123}(\{u\})\propto=
\frac{\prod_{j=1}^{N_{2}}\frac{Q^{(3)}(u_{j}^{(2)})}{(u_{j}^{(2)}-
\frac{i}{2})^{N_{1}}}\prod_{j=1}^{N_{3}}(u_{j}^{(3)}-\frac{i}{2})^{N_{1}}}{
\prod_{j<k}^{N_{3}}(u_{j}^{(3)}-u_{k}^{(3)}+i)\prod_{j<k}^{N_{2}}(u_{j}^{(2)}
-u_{k}^{(2)}+i)}D^{[0,1]}
\label{su2tree}
\eqx
where $D^{[0,1]}$ is given by an $N_3\times(N_2+N_1)$ determinant
\eq
D^{[0,1]}=\left|\begin{array}{ccccccc}
\partial_{u_{1}^{(3)}}T^{(3)}(u_{1}^{(2)}) & \dots & 
\partial_{u_{1}^{(3)}}T^{(3)}(u_{N_{2}}^{(2)}) & q_{2}(u_{1}^{(3)}) &
\dots & q_{N_{1}}(u_{1}^{(3)}) & q_{N_{1}+1}(u_{1}^{(3)})\\
\vdots & \vdots & \vdots & \vdots & \vdots & \vdots & \vdots\\
\partial_{u_{N_{3}}^{(3)}}T^{(3)}(u_{1}^{(2)}) & \dots &
\partial_{u_{N_{3}}^{(3)}}T^{(3)}(u_{N_{2}}^{(2)}) & q_{2}(u_{N_{1}}) &
\dots & q_{N_1} (u_{N_{3}}^{(3)})
& q_{N_{1}+1}(u_{N_{3}}^{(3)})
\end{array}\right|
\eqx
and we used  Baxter's  $Q$-function: 
\eq
Q^{(i)}(u)=\prod_{j=1}^{N_{i}}(u-u_{j}^{(i)})
\eqx
The derivative of the transfer matrix can be written in terms of the 
$su(2)$ scattering matrix as 
\begin{eqnarray*}
\partial_{u_{k}^{(3)}}T^{(3)}(u_{j}^{(2)}) & = & 
\frac{e^{ip(u_j^{(2)})L_3}}{u_{j}^{(2)}-u_{k}^{(3)}}
\frac{i}{u_{j}^{(2)}-u_{k}^{(3)}+i}\frac{Q^{(3)}(u_{j}^{(2)}+i)}{Q^{(3)}(u_{j}^{(2)})}\times\\
 &  & \left[1-e^{ip(u_{j}^{(2)})L_{1}}\prod_{m:m\neq j}S(u_{m}^{(2)},u_{j}^{(2)})
 \prod_{l:l\neq k}S(u_{j}^{(2)},u_{l}^{(3)})\right]
\end{eqnarray*}
In this formula  it is legitimate
to send $L_{3}$ and $L_{2}$ to infinity and keep
$L_{1}=L_{3}-L_{2}$ finite such that the rapidities no longer satisfy the BA equations. As a consequence,
the expression is singular for $u_{j}^{(2)}=u_{k}^{(3)}$ and we can
calculate its residue to check the kinematical singularity axiom.
One non-trivial requirement is that the result should be proportional
to the 3-pointfunction with two particles less, when $u_{j}^{(2)}=u_{k}^{(3)}$
were removed. Clearly the overall factors, which depend only on the
particles $u_{j}^{(1)}$ factor out. 
The only singularity at $u_{j}^{(2)}=u_{k}^{(3)}$ comes from
the matrix element
$\partial_{u_{k}^{(3)}}T^{(3)}(u_{j}^{(2)})$.  When we expand the 
determinant wrt. this element
the subdeterminant is nothing but the determinant, which appears in
the reduced 3-point functions and the prefactor is exactly the
required one:
\eq
\frac{-i\mbox{Res}_{u_{j}^{(2)}=u_{k}^{(3)}}C_{123}(\{u\})}{C_{123}(\{u\}
\setminus\{u_{j}^{(2)},u_{k}^{(3)}\})}\propto\left[
1-e^{ip(u_{j}^{(2)})L_{1}}\prod_{m:m\neq j}S(u_{m}^{(2)},u_{j}^{(2)})
 \prod_{l:l\neq k}S(u_{j}^{(2)},u_{l}^{(3)})\right]
\eqx

Finally, let us mention that the $su(2)$ OPE coefficients exhibit also the bound state
pole singularities required by the dynamical singularity axiom. 

\section{Conclusions}

In the paper we developed a new framework to determine the (light cone) SFT vertex 
for integrable worldsheet theories, including as a key special case 
the $AdS_5 \times S^5$ background. 

Our main idea was to use the integrable bootstrap approach to formulate 
functional relations for the SFT amplitudes incorporating crossing 
properties and including the scattering matrix of the theory. To achieve
this aim we decompactified the worldsheet of the process in which one 
big string (big J-charge) splits into two smaller ones in two alternative
ways. By sending to infinity the sizes of the big string together with any
of the other smaller strings allowed to define asymptotic states.  
The remaining finite string served as a nonlocal operator insertion 
appearing in the crossing equation in a nontrivial way. The solutions
of these functional equations contain exponentially small (wrapping)
corrections in the finite string size. After carefully getting rid 
of these wrapping corrections and applying a straightforward 
finite size reductions in all of the string sizes we would arrive at a formula 
valid for any coupling, which incorporates all finite size corrections which
are polynomial in the inverse powers of the sizes, but neglects 
exponentially small corrections. The fact that we can perform this finite size reduction 
procedure in two different ways, by starting with a finite size for any of the two small strings, 
gives very strong restrictions for the solutions of the functional relations. 

The feasibility of the program was demostrated by reproducing the results 
for the pp-wave SFT vertex. There,  to fix the analytical structure of the solution, 
we had to assume that the amplitude vanishes for those rapidities of the big string, 
which were the allowed finite volume states for the undecompactified small string. 
We do not have a physical explanation for this property but expect similar features for 
interactive theories. In the asymptotic limit of the SFT vertex we found some effective crossing 
formulas which bear striking resemblance to the recently 
discovered modification of crossing in Chern-Simons theories. 

Although we do not have control of the overall normalization of the 
SFT vertex, nor do we know the precise general relation between the SFT
vertex and OPE coefficients, nevertheless
we could check some of our diagonal  functional equations for the 
3-point functions in the weak coupling limit of the 
AdS/CFT correspondence by comparing them to explicit gauge theory calculations
of the OPE coefficients.

We also formulated the axioms for the generic non-diagonal case. 
For simplicity, and for 
being in accordance with the rest of the paper, we used rapidity 
parametrizations $\th$, with crossing transformations $\th \to \th \pm 
i \pi$, however, we did not assume any relativistic invariance for the 
scattering matrix. The generic formulas for the AdS/CFT integrable model
can be obtained simply by using its rapidity parametrizations $\th \to z$
and together with its crossing transformations $z\to z \pm \omega_2$. These axioms are very 
similar to the form factor axioms for world-sheet operators except the factor
$e^{ipL}$ appearing in the monodromy/periodicity equations and the
kinematical singularity axiom. Thus our 
axioms for the $L=0$ case reduces to them. 

An interesting dual line of investigation coming from the weak coupling
gauge theory side develops the concept of spin vertex \cite{KOMATSU,SPINVERTEX,YUNFENG} 
(but see also \cite{alday2}). One qualitative difference between that concept and the functional
equations presented in the present paper for the SFT amplitude
is that in the worldsheet formulation the wavefunctions of the
external states are in a natural way already incorporated into the SFT
amplitude (as it deals with multiparticle asymptotic states).
On the other hand, the weak coupling approach through the spin vertex
concentrates on contractions and loops in the interaction region
leaving aside the scalar product with the external states.

There are numerous directions for further research.
It would be very important to understand precisely the freedom
of choosing a particular solution to the SFT axioms. In particular,
even in the pp-wave limit, what are the features of the solution incorporating
the correct prefactor. Of course, the solution to the SFT axioms
remains an outstanding problem, which is not even solved for the 
ordinary form factor case i.e. in the $L\to0$ limit.

It would be furthermore important to understand the general relation between
the SFT vertex and OPE coefficients away from the pp-wave limit
(c.f. \cite{DY,SHIMADA,KMSFT}).

Let us emphasize once more that our approach would ultimately provide formulas for the 
SFT vertex, which are valid for any value of the 't Hooft coupling neglecting 
wrapping effects. But, even these exponentially 
small finite size effects are expected to be described in terms of the
asymptotic SFT vertex and the scattering matrix of the theory. 
Their systematic study should also be a direction of future research.

\bigskip

{\bf Acknowledgements.} We would like to thank Benjamin Basso, Shota Komatsu, Ivan Kostov, 
Didina Serban for interesting discussions.
We thank Simons Center for Geometry and Physics, Stony Brook for hospitality where a 
major part of this work was done.
RJ was supported by NCN grant 2012/06/A/ST2/00396 and ZB by a Lend\"ulet Grant. 
This work was supported by a Polish-Hungarian Academy of Science
cooperation.

\appendix

\section{The decompactified vertex formulation and solution}
\label{s.appdecomp}

\begin{figure}
\hfill\includegraphics[height=5cm]{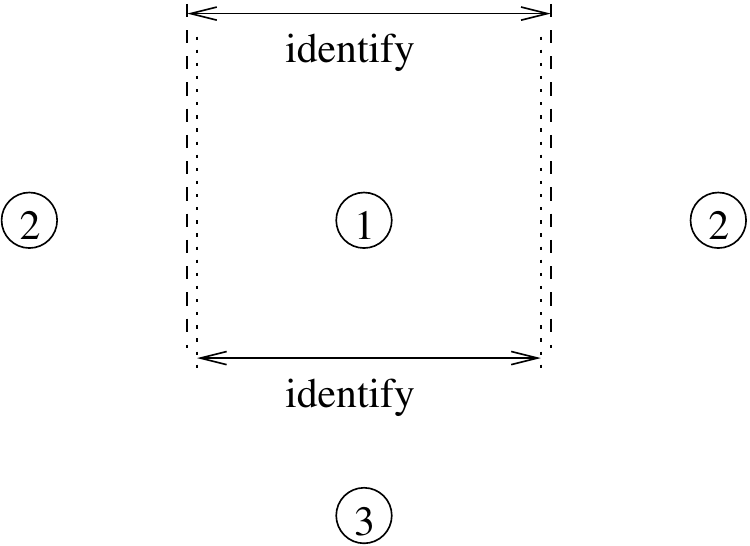}\hfill{}
\caption{The decompactified string field theory vertex \label{fig.decomp2}}
\end{figure}

The elementary exponential modes for the three strings in the decompactified
vertex shown in figure~\ref{fig.decomp2} are
\begin{align}
\phi^{(1)}_n(x) &= \f{2\pi}{L} e^{i\f{2\pi n x}{L}}  &\qq& x\in\left(-{L}/{2},{L}/{2}\right) \nonumber\\
\phi^{(2)}_k(x) &=  e^{ik \left(x+\f{L}{2}\right)}   &\qq&  x\in \left(-\infty, -{L}/{2}\right) \nonumber\\
                &= e^{ik \left(x-\f{L}{2}\right)} &\qq&  x\in \left({L}/{2},+\infty \right) \nonumber\\
\phi^{(3)}_k(x)&=  e^{ikx} &\qq&  x\in \left(-\infty,+\infty\right)
\end{align}
The `exponential' overlap matrices are defined through
\eq
\tilde{X}^r_{pk}=\f{1}{2\pi} \int_{-\infty}^\infty e^{-ipx} \phi^{(r)}_k(x) dx
\eqx
For the case at hand, we obtain
\begin{align}
X^1_{pn} & = \f{2}{L}  \f{(-1)^n \sin \f{p L}{2}}{p- \f{2\pi n}{L}} \\
X^2_{pk} &= \cos \f{pL}{2} \dl(p-k) -\f{1}{\pi} \sin \f{pL}{2} P_{\f{1}{p-k}} \\
X^3_{pk} &= \dl(p-k)
\end{align}
The overlap element for string \#2 was obtained using appropriate $e^{\mp\eps x}$ regularization on
the two half-lines and the use of
\eq
\f{1}{x \pm i\eps} = \mp i\pi \dl(x)+ P_{\f{1}{x}}
\eqx
where $P$ stands for the principle value.
Since there exist nice factorization properties of the Neumann coefficients when expressed in terms
of cosine and sine modes, it is convenient to define (for cosine modes)
\eq
X^r_{pk} \equiv \f{1}{2} \left( \tilde{X}^r_{pk}+ \tilde{X}^r_{-pk}+\tilde{X}^r_{p\,-k}+ \tilde{X}^r_{-p\,-k} \right)
\eqx
for positive $p$ and $k$. As reviewed in section~\ref{s.lsnsreview}, the Neumann coefficients for negative modes (sine-modes)
can be directly obtained from the cosine mode answer. So below we will concentrate exclusively on the
positive modes only.
Using such modes it has been shown in \cite{Scomments} that the Neumann coefficients
can be expressed as
\eq
\bar{N}^{rs}_{nm} = - \f{m n \al}{1-4\mu \al K} \f{ \bar{N}^r_m \bar{N}^s_n}{\al_s \om^r_n +\al_r \om^s_m}
\eqx
In order to provide formulas for $K$ and $\bar{N}^r_m$ it is customary to introduce quite a lot of notation.
We would like to adopt similar notation in the present decompactified case as in the standard
finite volume case as used in \cite{Scomments}, since we would like to make use of the factorizability proof of \cite{Scomments}
which is purely algebraic. 
To this end let us introduce
\eq
C=m\, \dl_{mn} \quad\text{or}\quad C=p\,\dl(p-k) 
\eqx
Note that in this appendix the meaning of $\mu$ and $\al_r$ will be {\bf different}
from the usage in the pp-wave case. However all algebraic relations employed in the factorizability proof
will still hold. We will use\footnote{Note that in this appendix $\al_1+\al_2+\al_3 \neq 0$.}
\eq
\mu \equiv M \qq \al_1=\f{L}{2\pi} \quad \al_2=1 \quad \al_3=-1 \quad \al=\al_1\al_2\al_3= -\f{L}{2\pi}
\eqx
and
\begin{align}
A^r_{pk} \equiv \sqrt{\f{k}{p}} X^r_{pk} \qq & C_r=\sqrt{C^2+\mu^2\al_r^2} \\
U_{(r)} = C^{-1} (C_r-\mu\al_r \id) \qq & U_{(r)}^{-1} = C^{-1} (C_r+\mu\al_r \id)
\end{align}
The key object neccessary for finding the Neuman vectors is the infinite matrix
\eq
\label{e.gmplusdef}
\Gm_+ =\sum_{r=1}^3 A^{r} U_{(r)} A^{rT}
\eqx
and an infinite vector $B$ (to be defined below). Then we have
\eqn
\bar{N}^r_m &=& - \left[ (C^{-1} C_r)^{1/2} U_{(r)}^{-1} A^{rT} Y \right]_m \qqqq Y \equiv \Gm_+^{-1} B \\
K &=& \f{1}{4} B^T \Gm_+^{-1} B
\eqnx

The proof of factorizability uses the following properties of the overlap matrices defined above
\begin{align}
\label{e.orthog}
A^{rT} C^{-1} A^{s} & = -\f{\al_r}{\al_3} C^{-1} \dl^{rs} \qq & r,s=1,2 \\
\label{e.beq}
\sum_{r=1}^3 \al_r A^r C^{-1} A^{rT} & = \f{\al}{2} BB^T &
\end{align}
The last equation defines for us the vector $B$ (which is related to overlaps with a constant mode).
These formulas seem at first glance to be quite convoluted but essentially reflect just the joint completness and 
mutual orthogonality of modes of strings \#1 and \#2. 

The matrices $A^r_{pk}$ in the decompactified case are given by
\begin{align}
A^1_{pn} & = \f{4}{L} \sqrt{\f{n}{p}} \f{(-1)^n p \sin \f{pL}{2}}{p^2- \f{4\pi^2 n^2}{L^2}} \\
A^2_{pk} & = \sqrt{\f{k}{p}} \left( \cos \f{pL}{2}\, \dl(p-k) -\f{1}{\pi} \sin \f{pL}{2}\, 
P_{\f{2p}{p^2-k^2}}\right ) \\
A^3_{pk} & = \dl(p-k)
\end{align}
One can check that indeed these matrices satisfy (\ref{e.orthog}) although this is quite involved and we
performed some checks for parts of the formulas only numerically. The equation (\ref{e.beq}) provides
for us the expression for the vector $B$ (which in the decompactified case is in fact a function on
the positive real line):
\eq
B_p =\f{4}{L}\, p^{-\f{3}{2}} \sin \f{pL}{2}
\eqx
In the decompactified case that we are considering here, the infinite matrix $\Gm_+$ becomes an integral kernel 
and the vector(function) $Y$ is defined through the integral equation
\eq
\label{e.inteq}
\int_0^\infty \Gm_+(p,p') Y(p')\, dp' = \f{4}{L}\, p^{-\f{3}{2}} \sin \f{pL}{2}
\eqx
It turns out, however, that even getting an explicit form for the kernel $\Gm_+(p,p')$ is quite involved.

Using the definition (\ref{e.gmplusdef}) and (\ref{e.beq}) we get
\begin{align}
\label{e.gmplusii}
\Gm_+ &=\sum_{r=1}^3 A^r C^{-1} C_r A^{rT}-\f{\al\mu}{2} B B^T= \nonumber\\
&= \sum_{r=1}^3 A^r C^{-1} C_r A^{rT}+ 
\f{4M}{\pi L} \left(p^{-\f{3}{2}}  \sin \f{pL}{2}\right)\left({p'}^{-\f{3}{2}}  \sin \f{p'L}{2}\right)
\end{align}
Let us now give the contributions of the three strings to the above formula:
\eq
A^3 C^{-1} C_3 A^{3T} =\f{1}{p} \sqrt{p^2+M^2}\, \dl(p-p')
\eqx
For string \#2 the result is quite messy:
\begin{align}
&\int_0^\infty A^2_{pk} \f{1}{k} \sqrt{k^2+M^2} A^2_{p'k} dk = \nonumber\\
&=\f{1}{\sqrt{pp'}} \left[ \sqrt{p^2+M^2}\, \dl(p-p') +Junk(p,p')
+\f{4}{\pi^2} \sin \f{pL}{2} \sin \f{p'L}{2} \f{pF(p')-p' F(p)}{p^2-{p'}^2} \right]
\end{align}
where
\eq
F(p)=\sqrt{p^2+M^2} \arctanh \f{p}{\sqrt{p^2+M^2}}
\eqx
and
\eq
Junk(p,p') =-\f{1}{\pi} \sqrt{p^2+M^2}  \cos \f{pL}{2} \sin \f{p'L}{2} P_{\f{2p'}{{p'}^2-p^2}} +\bigl(p \longleftrightarrow p'\bigr)
\eqx
The contribution of string \#1 involves a nontrivial infinite sum which can be handled using
the techniques of Appendix E in \cite{HSSV}.
\begin{align}
&\sum_{n=1}^\infty A^1_{pn} \f{1}{n} \sqrt{n^2+\f{M^2 L^2}{4\pi^2}} A^1_{p'n} =
\f{1}{\sqrt{pp'}} \f{16}{L^2} \sum_{n=1}^\infty \f{pp' \sin \f{pL}{2} \sin \f{p'L}{2} \sqrt{n^2+\f{M^2 L^2}{4\pi^2}}}{ 
\left(p^2-4n^2 \f{\pi^2}{L^2} \right) \left({p'}^2-4n^2 \f{\pi^2}{L^2} \right) }  \nonumber\\
&=\f{1}{\sqrt{pp'}} \left[ -Junk(p,p') +\f{1}{\pi^2} \sin \f{pL}{2} \sin \f{p'L}{2} G_L(p,p')-
\f{4M \sin \f{pL}{2} \sin \f{p'L}{2}}{\pi L p p'} \right]
\end{align}
where
\eq
G_L(p,p') = -4 pp' \int_M^\infty \f{\sqrt{\kap^2-M^2} \coth \f{\kap L}{2} d\kap}{(p^2+\kap^2)({p'}^2+\kap^2)}
\eqx
When we add all the above ingredients into the formula for $\Gm_+$, we see that the terms which involve $\cos \f{pL}{2}$
cancel out as well as the $BB^T$ term in (\ref{e.gmplusii}). The final formula for $\Gm_+$ is thus
\eq
\Gm_+(p,p') = \f{2}{p} \sqrt{p^2+M^2} \dl(p-p')+\f{1}{\pi^2 \sqrt{pp'}} \sin \f{pL}{2} \sin \f{p'L}{2} \bigl[ G_L(p,p')+
G_\infty(p,p') \bigr]
\eqx
and $G_\infty(p,p')$ can be evaluated to be\footnote{It is obtained from $G_L(p,p')$ by replacing the $\coth$ by $1$ and thus differs by exponential mutliple wrapping terms $e^{-nML}$.}
\eq
G_\infty(p,p')=4\f{pF(p')-p' F(p)}{p^2-{p'}^2}
\eqx
The Neumann vectors are now determined by the equation (\ref{e.inteq}).

Some comments are in order here. We see here the $\sin \f{pL}{2}$ factors
which eventually make their appearance in the Neumann coefficients even in the infinite $L$ limit.
These terms thus arise directly from the decompactified string vertex and are not associated to
some subtlety in finite volume reduction. These factors are also quite surprising in that they have
a highly oscillatory $L$ behaviour. This leads also to the fact that the large $L$ limit
of the equation (\ref{e.inteq}) is quite subtle. Indeed we cannot directly
take the large $L$ limit of (\ref{e.inteq}) but write
\eq
\f{2}{p} \sqrt{p^2+M^2} Y_\infty(p)+\f{2}{\pi^2\sqrt{p}} \sin \f{pL}{2} \int_0^\infty G_\infty(p,p') \sin \f{p'L}{2}  Y_\infty(p') \f{dp'}{\sqrt{p'}}
=\f{4}{L}\, p^{-\f{3}{2}} \sin \f{pL}{2}
\eqx
In the large $L$ limit, a consistent solution will have a $\sin \f{pL}{2}$ factor. After extracting this factor
and taking into account that under the integral we may substitute $\sin^2 \f{p' L}{2}$ by $\f{1}{2}$ up to
$e^{-ML}$ corrections, we can obtain an integral equation with no explicit $L$ dependence.
It is also convenient to pass to rapidity variables.
Indeed writing
\eq
Y_\infty(\th)=\f{2}{L M^{\f{3}{2}}} \sin \f{pL}{2} \cdot \f{1}{\sqrt{\sinh\th} \cosh\th} h(\th) 
\eqx
we get the following integral equation
\eq
h(\th) +\f{2}{\pi^2} \sinh \th \int_0^\infty \f{\th' \sinh\th  \cosh\th'-\th \cosh\th \sinh\th'}{
\sinh^2\th-\sinh^2\th'} \f{h(\th') d\th'}{\sinh\th'}=1
\eqx

One can check numerically that this is solved by $h(\th)=\f{1}{\sqrt{2}}\sqrt{1+\cosh\th}$. Thus the large $L$
solution becomes 
\eq
Y_\infty(\th)= \f{\sqrt{2}}{L M^{\f{3}{2}}} \sin \f{pL}{2} \f{\sqrt{1+\cosh\th}}{\cosh\th \sqrt{\sinh\th}}
\eqx
which  coincides\footnote{Upto a $(-1)^m$ sign factor.} with the large $L$ limit of the decompactified LSNS solution 
$f^{(3)}(\th)$ following from (4.3) in \cite{LSNS}.
Indeed we also verified numerically that its finite $L$ version  solves
the exact finite $L$ version of the integral equation (\ref{e.inteq}).
We also checked numerically that the $f^{(2)}(\th)$ vector is also reproduced.
The point of the above exercise was to ascertain that we have full control over the decompactified
string vertex solution and make sure that the puzzling  $\sin \f{pL}{2}$ factors indeed arise
already for the decompactified vertex and do not come from some complications in subsequent finite
volume reduction.

\section{Properties of the $\hat{\Gm}_\mu(\th)$ functions}
\label{s.appGammaMu}

In this appendix we summarize the properties of the deformed $\Gamma$
functions. We recall the original definition from \cite{LSNS} and renormalize
them in order to simplify the formulas for the Neumann coefficients and
to have simpler asymptotic behaviour. 

Following \cite{LSNS} we define the $\Gamma_{\mu}(z)$ functions as 
\eq
\Gamma_{\mu}(z)=\frac{e^{-\gamma\omega_{z}}}{z}\prod_{n=1}^{\infty}\frac{n}{\omega_{n}
+\omega_{z}}e^{\frac{\omega_{z}}{n}}\quad;\qquad\omega_{z}=\sqrt{z^{2}+\mu^{2}}
\eqx
where $\gamma$ is the Euler constant. The authors choose the finite
branch cut for the square root $\omega_{-z}=-\omega_{z}$ such that
$\Gamma_{\mu}(z)$ satisfies the functional equation:
\eq
\Gamma_{\mu}(z)\Gamma_{\mu}(-z)=-\frac{\pi}{z\sin\pi z}
\eqx
The large $z$ asymptotic has been calculated to be 
\eq
\log\Gamma_{\mu}(z)=\omega_{z}\log\frac{\mu}{2e}+z\log\frac{\omega_{z}+z}{\mu}+\log\frac{\sqrt{2\pi}
\sqrt{\omega_{z}+\mu}}{z}+O(e^{-\mu})
\eqx
or alternatively
\eq
\Gamma_{\mu}(z)=\frac{\sqrt{2\pi}\sqrt{\omega_{z}+\mu}}{z}\left(\frac{\mu}{2e}\right)^{\omega_{z}}
\left(\frac{\omega_{z}+z}{\mu}\right)^{z}+\dots
\eqx
The above equations apply only for $|\mathrm{arg}\, z|<\pi$.
For our purposes we renormalize this functions as
\eq
\hat{\Gamma}_{\mu}(z)=e^{-\omega_{z}\log\frac{\mu}{2e}}\Gamma_{\mu}(z)
\eqx
In order to resolve the branch cut it is convenient do introduce the
rapidity parametrization 
\eq
z=\mu\sinh\theta\qquad;\qquad\omega_{z}=\mu\cosh\theta
\eqx
and consider $\hat{\Gamma}_{\mu}$ as a functions of $\theta$ and
$\mu$
\eq
\hat{\Gamma}_{\mu}(z)\equiv\tilde{\Gamma}_{\mu}(\theta)
\eqx
In the $\theta$ variable the two sides of the branch cut are mapped
to $\theta$ and $\theta+i\pi$ and the function satisfies
\eq
\tilde{\Gamma}_{\mu}(\theta)=-\tilde{\Gamma}_{\mu}(-\theta)\quad;\qquad\tilde{\Gamma}_{\mu}
(\theta)\tilde{\Gamma}_{\mu}(\theta+i\pi)=-\frac{\pi}{\mu\sinh\theta\sin(\pi\mu\sinh\theta)}
\eqx
Consequently, it is $2\pi i$ periodic: 
\eq
\tilde{\Gamma}_{\mu}(\theta+2\pi i)=\tilde{\Gamma}_{\mu}(\theta)
\eqx
Moreover, it has the large $\mu$ asymptotical behaviour: 
\eq
\label{e.Gammamuas}
\tilde{\Gamma}_{\mu}(\theta)=\sqrt{\frac{\pi}{\mu}}\frac{e^{\theta\mu\sinh\theta}}
{\sinh\frac{\theta}{2}}+\dots
\eqx
where $|\Im m (\th)|<\pi$.

\section{Details on the \sutt and \suii OPE coefficients}

In this Appendix we check how the kinematical singularity axiom is
satisfied at weak coupling directly for the OPE coefficients. We investigate
the 3-point coefficients in the $su(1\vert1)$ and $su(2)$ sectors
up to 1-loop based on the available explicit results obtained from
direct gauge theory calculations in \cite{Caetano,GVtheta}. 

As we explained in the main text we have to extract  carefully the 
infinite volume limit of the OPE coefficient from the available finite 
volume 3-point functions. As we do not know the exact relation
between the OPE coeffcients and the SFT vertex we do not 
bother with the correct normalization of the infinite volume 
3-point functions and check  the axiom (\ref{eq:weaksingax})
only up to proportionality.
We start with the simpler $su(1\vert1)$ sector first and proceed with the
more complicated $su(2)$ sector afterwards.

\subsubsection*{The $su(1\vert1)$ sector}

In \cite{Caetano} the authors calculated the 3-point function in the $su(1\vert1)$
sector up to 1-loop. They parametrized the operators as closed spin
chain Bethe states satisfying the BA equation%
\footnote{Here we write the BA equation into our convention. In the AdS convention
the BA equation takes usually the form $e^{ip_{j}L}=\prod_{k\neq j}S_{su(1\vert1)}(p_{j},p_{k})$.
In \cite{Caetano} $S_{su(1\vert1)}$ was denoted by $S$. Strictly
speeking there is also a phase factor difference between $S_{su(1\vert 1)}$ and our $S$,
$S_{su(1\vert 1)}(p,k)=e^{i(p-k)/2}S(p,k)^{-1} $, but as it basically shifts the volume
we do not keep track of it.
} 
\eq
e^{ip_{j}^{(r)}L_{r}}\prod_{k:k\neq j}^{N_{r}}S(p_{j}^{(r)},p_{k}^{(r)})=1
\eqx
where the scattering matrix in our notation is
\eq
S(p_{1},p_{2})=-S_{su(1\vert1)}(p_{2},p_{1})^{-1}=1+8ig^{2}\sin\frac{p_{1}}{2}\,
\sin\frac{p_{2}}{2}\,\sin\frac{p_{1}-p_{2}}{2}+O(g^{4})
\eqx
The OPE coefficients (here we write the complex conjugate of the expression from \cite{Caetano}) up to 1-loop can be written as
\eq
C_{123}=\frac{C}{\sqrt{\rho_{N_{1}}\rho_{N_{2}}\rho_{N_{3}}}}\frac{\prod_{r=1}^{3}
\prod_{i<j}^{N_{r}}f(p_{i}^{(r)},p_{j}^{(r)})}{\prod_{i}^{N_{3}}
\prod_{j}^{N_{2}}f(p_{i}^{(3)},p_{j}^{(2)})}\prod_{k}^{N_{3}}
\left(1-e^{-ip_k^{(3)}L_2}
\prod_{i}^{N_{2}}S(p_{i}^{(2)},p_{k}^{(3)})\right)
\eqx
where 
\eq
f(p_{1},p_{2})=(e^{ip_{1}}-e^{ip_{2}})\left[1-g^{2}\left(1+\cos(p_{1}-p_{2})
-\cos p_{1}-\cos p_{2}\right)\right]+O(g^{4})
\eqx
and the normalization factor $C$ will not be relevant for our discussion.
The original expression in \cite{Caetano} contained factors of the form
$\left (1-e^{ip_k^{(3)}L_2}\prod_{i}S(p_{k}^{(3)},p_{i}^{(2)})\right ) $, i.e. the 
complex conjugate of the expression we wrote. 
As three point functions seem to be 
real we could just take the complex conjugate of their result. In any case
this does not modify the physics.
Let us also note that the appearance of the complex conjugate
expression might be related to the conjugate process in which two smaller strings join into a bigger one.

Recalling the  
general remarks in the main text we can observe that the
would be pole of~$C_{123} $ at $p_{i}^{(3)}=p_{j}^{(2)}$ is absent due to the zero
coming from the BA quantization condition of $p_{j}^{(2)}.$ As this
OPE coefficient does not have a direct $L_{2}\to\infty$ limit
we use the BA equations for $p_{k}^{(3)}$ to reformulate it following
\eq
e^{-ip_k^{(3)}L_2}
\prod_{i}^{N_{2}}S(p_{i}^{(2)},p_{k}^{(3)})\quad
\longrightarrow\quad e^{
ip_{k}^{(3)}L_{1}}\prod_{i}^{N_{2}}S(p_{i}^{(2)},p_{k}^{(3)})
\prod_{j}^{N_{3}}S(p_{k}^{(3)},p_{j}^{(3)})
\eqx
We now can safely take  the $L_{3},L_{2}\to\infty$
limit and obtain the infinite volume 3-point function
\eq
C_{123}(\{p\})\propto\frac{\prod_{r=1}^{2}\prod_{i<j}^{N_{r}}f(p_{i}^{(r)},
p_{j}^{(r)})}{\prod_{i}^{N_{3}}\prod_{j}^{N_{2}}f(p_{i}^{(3)},p_{j}^{(2)})}
\prod_{k}^{N_{3}}\left(1-e^{
ip_{k}^{(3)}L_{1}}\prod_{i}^{N_{2}}S(p_{i}^{(2)},p_{k}^{(3)})
\prod_{j}^{N_{3}}S(p_{k}^{(3)},p_{j}^{(3)})\right)
\eqx
where the momenta $p^{(3)}$ and $p^{(2)}$ no longer satisfy any
quantization condition. 

In order to check the kinematical residue axiom we calculate

\eq
\frac{-i\mbox{Res}_{p_{k}^{(3)}=p_{j}^{(2)}}C_{123}(\{p\})}{C_{123}(\{p\}
\setminus\{p_{k}^{(3)},p_{j}^{(2)}\})}
\eqx
Using the unitarity of the scattering matrix we obtain the required
factor 
\eq
1-e^{
ip_{k}^{(3)}L_{1}}\prod_{i}^{N_{2}}S(p_{i}^{(2)},p_{k}^{(3)})
\prod_{j}^{N_{3}}S(p_{k}^{(3)},p_{j}^{(3)})
\eqx

\subsubsection*{The $su(2)$ sector}

Let us recall the 1-loop structure constant from the literature \cite{GVtheta},
written in to our conventions when $L_{3}=L_{1}+L_{2}$. The operators
are again parametrized by the momenta (more precisely by the spin chain rapidities) of the closed BA
states satisfying:

\eq
e^{ip(u_{k}^{(r)})L_{r}}\prod_{j:j\neq k}^{N_{r}}S(u_{k}^{(r)}-u_{j}^{(r)})=
1\quad;\qquad S(u)=S_{su(2)}(u)=\frac{u-i}{u+i}
\eqx
The momenta and the rapidities are related via 
\eq
e^{ip(u)}=\frac{x(u+\frac{i}{2})}{x(u-\frac{i}{2})}\quad;\qquad x(u)
+\frac{1}{x(u)}=\frac{u}{g}
\eqx
The OPE coefficient up to 1-loop takes the form 
\eq
C_{123}(\{u\})=\frac{C(1-g^{2}(\Gamma_{1}+\Gamma_{32}^{2}-
\alpha_{32}))}{\sqrt{\rho_{1}\rho_{2}\rho_{3}}}P_{23}S_{23}A_{1}
\eqx
where the index refers to the set of rapidities $u_{i}^{(r)}$ the
various terms depend on. More explicitly
\eq
\Gamma_{i}=\sum_{j=1}^{N_{i}}\frac{1}{(u_{j}^{(i)})^{2}+\frac{1}{4}}\quad;
\qquad\Gamma_{ij}=\frac{1}{2}(\Gamma_{i}-\Gamma_{j})
\eqx
and
\eq
\alpha_{32}=
\sum_{j=1}^{N_{i}}\frac{u_{j}^{(3)}}{(u_{j}^{(3)})^{2}+\frac{1}{4}}-
\sum_{j=1}^{N_{i}}\frac{u_{j}^{(2)}}{(u_{j}^{(2)})^{2}+\frac{1}{4}}
\eqx
Introducing Baxter's $Q$ functions and the transfer matrices
\eq
Q^{(i)}(u)=\prod_{j=1}^{N_{i}}(u-u_{j}^{(i)})\qquad;\qquad T^{(i)}(u)=
\frac{Q^{(i)}(u-i)}{Q^{(i)}(u)}+e^{-ipL_{i}}\frac{Q^{(i)}(u+i)}{Q^{(i)}(u)}
\eqx
 the next term can be written as 
\eq
P_{23}=\frac{\prod_{j=1}^{N_{2}}\frac{Q^{(3)}(u_{j}^{(2)})}{(gx(u_{j}^{(2)}-
\frac{i}{2}))^{N_{1}}}\prod_{j=1}^{N_{3}}(gx(u_{j}^{(3)}-\frac{i}{2}))^{N_{1}}}{
\prod_{j<k}^{N_{3}}(u_{j}^{(3)}-u_{k}^{(3)}+i)\prod_{j<k}^{N_{2}}(u_{j}^{(2)}-u_{k}^{(2)}+i)}
\eqx
Finally the most complicated term can be compactly written in terms of a sum 
of $N_3\times (N_1+N_2)$ determinants 
\eq
S_{23}=D^{[0,1]}+g^{2}\left((N_{1}+1)D^{[0,3]}+(N_{1}-1)D^{[1,2]}-2\alpha_{32}D^{[0,2]}\right)
\eqx
in which the upper index shows, how the argument of the last two columns
are shifted 
\eq
D^{[i,j]}=\left|\begin{array}{ccccccc}
\partial_{u_{1}^{(3)}}T^{(3)}(u_{1}^{(2)}) & \dots & 
\partial_{u_{1}^{(3)}}T^{(3)}(u_{N_{2}}^{(2)}) & q_{2}(u_{1}^{(3)}) &
\dots & q_{N_{1}+i}(u_{1}^{(3)}) & q_{N_{1}+j}(u_{1}^{(3)})\\
\vdots & \vdots & \vdots & \vdots & \vdots & \vdots & \vdots\\
\partial_{u_{N_{3}}^{(3)}}T^{(3)}(u_{1}^{(2)}) & \dots &
\partial_{u_{N_{3}}^{(3)}}T^{(3)}(u_{N_{2}}^{(2)}) & q_{2}(u_{N_{1}}) &
\dots & q_{N_{1}+i}(u_{N_{3}}^{(3)}) & q_{N_{1}+j}(u_{N_{3}}^{(3)})
\end{array}\right|
\eqx
and 
\eq
q_{n}(u)=\frac{1}{(u+\frac{i}{2})^{n-1}}-\frac{1}{(u-\frac{i}{2})^{n-1}}
\eqx
The index of $q$ in the $N_2+i$'th column is $i+1$ except the last two columns which are shifted. 
The expression for $A_{1}$ is quite involved, however, we do not need
its explicit form to check the kinematical singularity axiom. 

To prepare for the infinite volume limit we rewrite the transfer matrix
in terms of the scattering matrix as
\begin{eqnarray}
T^{(j)}(u) & = &e^{-ipL_{j}} \frac{Q^{(j)}(u+i)}{Q^{(j)}(u)}
\left(1+e^{ipL_{j}}\frac{Q^{(j)}(u-i)}{Q^{(j)}(u+i)}\right)\nonumber \\
 & = &e^{-ipL_{j}} \frac{Q^{(j)}(u+i)}{Q^{(j)}(u)}\left(1+e^{ipL_{j}}
 \prod_{k=1}^{N_{j}}S(u-u_{k}^{(j)})\right)
\end{eqnarray}
 The derivative of the transfer matrix appears in the matrix element
of the determinant as:
\begin{eqnarray}
\partial_{u_{k}^{(3)}}T^{(3)}(u_{j}^{(2)}) & = & \frac{e^{-ip(u_{j}^{(2)})L_3}}{u_{j}^{(2)}-u_{k}^{(3)}}
\frac{i}{u_{j}^{(2)}-u_{k}^{(3)}+i}\frac{Q^{(3)}(u_{j}^{(2)}+i)}{Q^{(3)}(u_{j}^{(2)})}\times
\nonumber \\
 &  & \left[1-e^{ip(u_{j}^{(2)})L_{3}}\prod_{k:k\neq j}S(u_{j}^{(2)},u_{k}^{(3)})\right]
\end{eqnarray}
Clearly the $u_{j}^{(2)}=u_{k}^{(3)}$ would be pole is annihilated
by the zero which manifests the BA equation. In order to have a well
defined $L_{3},L_{2}\to\infty$ limit we use the BA of $u_{j}^{(2)}$
to rewrite this expression as 

\begin{eqnarray}
\partial_{u_{k}^{(3)}}T^{(3)}(u_{j}^{(2)}) & = & \frac{e^{-ip(u_{j}^{(2)})L_3}}{u_{j}^{(2)}-u_{k}^{(3)}}
\frac{i}{u_{j}^{(2)}-u_{k}^{(3)}+i}\frac{Q^{(3)}(u_{j}^{(2)}+i)}{Q^{(3)}(u_{j}^{(2)})}\times
\nonumber \\
 &  & \left[1-e^{ip(u_{j}^{(2)})L_{3}}\prod_{m:m\neq j}S(u_{m}^{(2)},u_{j}^{(2)})\prod_{k:k\neq j}S(u_{j}^{(2)},u_{k}^{(3)})\right]
\end{eqnarray}
Now it is legitimate to send $L_{3}$ and $L_{2}$ to infinity and keep
$L_{1}=L_{3}-L_{2}$ finite. This procedure, together with the renormalization
of the BA states to the infinite volume scattering basis, result
in the infinite volume `OPE coefficient', where the rapidities
no longer satisfy the BA equation, i.e. they are off-shell. As a consequence,
the expression is singular for $u_{j}^{(2)}=u_{k}^{(3)}$ and we can
calculate its residue to check the kinematical singularity axiom.
One non-trivial requirement is that the result should be proportional
to a similar decompactified OPE coefficient with two particles less, when $u_{j}^{(2)}=u_{k}^{(3)}$
were removed. Clearly the overall factors, which depend only on the
particles $u_{j}^{(1)}$ factor out. Additionally $\Gamma_{32}$ and
$\alpha_{32}$ reduces to the analogous expression with two particles
less. The only singularity at $u_{j}^{(2)}=u_{k}^{(3)}$ comes from
the common matrix element $\partial_{u_{k}^{(3)}}T^{(3)}(u_{j}^{(2)})$
of all determinants. When we expand the determinants wrt. this element
the subdeterminant is nothing but the determinant, which appears in
the reduced OPE coefficient and the prefactor is exactly the
required one:
\eq
\frac{-i\mbox{Res}_{u_{j}^{(2)}=u_{k}^{(3)}}C_{123}(\{u\})}{C_{123}(\{u\}
\setminus\{u_{j}^{(2)},u_{k}^{(3)}\})}\propto\left[1-e^{ip(u_{j}^{(2)})L_{1}}
\prod_{m:m\neq j}S(u_{m}^{(2)},u_{j}^{(2)})\prod_{l:l\neq k}S(u_{j}^{(2)},u_{l}^{(3)})\right]
\eqx

\pagebreak

\end{document}